\documentclass[a4paper,11pt]{article}
\usepackage{jheppub} 
\usepackage{lineno}
\usepackage[x11names]{xcolor}
\usepackage{float}
\usepackage{booktabs}
\usepackage{comment}
\usepackage{float}
\usepackage{tikz}
\usetikzlibrary{decorations.pathreplacing,decorations.markings,arrows.meta}
\usepackage{subcaption}
\usepackage{dsfont}
\usepackage{comment}
\usepackage{enumerate}
\usetikzlibrary{calc,
    decorations.markings,
    decorations.pathmorphing
}

\title{\boldmath Large $N$ hadron scattering bootstrap: \\
massive pions and glueballs}

\author{Sebastiano Bocchia,}
\author{Andrea Guerrieri}

\affiliation{Department of Mathematics, City St. George’s, University of London,\\ Northampton Square, EC1V 0HB, London, UK}

\emailAdd{sebastiano.bocchia@citystgeorges.ac.uk}
\emailAdd{andrea.guerrieri@city.ac.uk}

\abstract{We extend the large $N$ bootstrap of pion scattering to nonzero pion mass, and apply the same formalism to glueball scattering. Our bounds apply to Taylor coefficients and on-shell couplings of the exact planar amplitude. They do not rely on a truncated chiral expansion, so we can follow them as the quark mass varies, and apply them to glueballs, for which no low-energy effective theory exists. For pions, we bound the $\mathcal{O}(p^4)$ coefficients and the $\pi\pi f_2$ coupling. As in the chiral limit, the bound on the $\pi\pi f_2$ coupling features a kink, where the extracted spectrum closely resembles the real-world meson spectrum. For glueballs, we bound the couplings of the lightest scalar glueball to heavier glueballs of higher spin, using the large $N$ lattice spectrum as input, and find extremal solutions with linear Regge trajectories.
}

\begin{document}

\maketitle
\flushbottom

\section{Introduction}

Can the spectrum and couplings of QCD be computed exactly? By this we do not necessarily mean a closed formula, but any procedure that computes them to arbitrary precision, as the 't~Hooft equation does for large $N$ QCD in two dimensions~\cite{tHooft:1974pnl}. In four dimensions no such description is known. The large $N$ limit largely simplifies the problem~\cite{tHooft:1973alw,Witten:1979kh}: mesons and glueballs become stable and do not mix, while baryons become parametrically heavy. At leading order in $1/N$, meson and glueball amplitudes are meromorphic functions, with poles at the masses of the exchanged hadrons and residues fixed by their three-point functions. These simplifications do not make the dynamics perturbative: even in this limit, determining hadron masses and couplings remains a nonperturbative problem.

The S-matrix bootstrap approaches this problem from a different angle. Requiring hadronic amplitudes to be consistent with analyticity, crossing, positivity of residues and a bound on their high-energy behaviour constrains the spectra and couplings they can contain. For large-$N$ pion scattering, this program has led to bounds on chiral low-energy constants and to spectra with an intriguing resemblance to the meson spectrum~\cite{Albert:2022oes, Fernandez:2022kzi, Albert:2023jtd, Ma:2023vgc, Li:2023qzs, Albert:2023seb}. These analyses, as well as recent primal constructions~\cite{Haring:2023zwu,Bocchia:2026kew} and extensions to pion form factors~\cite{Albert:2026xyz}, assume the chiral limit. Massive pions have been considered at finite $N$, both in positivity bounds on chiral perturbation theory \cite{Ananthanarayan:1994hf, Manohar:2008tc, Wang:2020jxr, Alvarez:2021kpq} and in the nonperturbative S-matrix bootstrap~\cite{Guerrieri:2018uew, He:2023lyy, He:2024nwd, Guerrieri:2024jkn}, but not in the large-$N$ bootstrap.

Finite pion mass is essential for following bounds as the quark
mass is varied and for comparing them directly with lattice calculations
at the same masses. Recent lattice determinations of the large $N$
meson spectrum make this comparison increasingly concrete
\cite{Bonanno:2025hzr,Bonanno:2026zcq}. They provide both input for the
lightest states and a route to testing how the heavier spectrum changes
away from the chiral limit. A finite mass bootstrap can therefore ask
whether the relation between extremal amplitudes and QCD survives
this deformation.

In this paper, we derive dual bootstrap bounds for the scattering of massive pions and glueballs at large $N$. We write dispersion relations for the meromorphic amplitudes at fixed $t$ and fixed $u$, expanded around the crossing-symmetric point $s=t=u=4m^2/3$, where $m$ is the mass of the external particle. Under the assumed Regge behaviour, pion amplitudes require one subtraction and glueball amplitudes two. The resulting sum rules express low-energy observables as linear functionals of the spectral densities, while crossing symmetry imposes null constraints on the same densities. This extends the construction of~\cite{Caron-Huot:2020cmc} to massive external states, building on the treatment of massive positivity bounds in~\cite{EliasMiro:2022xaa}. Crossing-symmetric dispersion relations, in which null constraints are replaced by locality constraints, provide a complementary framework~\cite{Sinha:2020win,Chowdhury:2022obp}.

    Our observables are defined directly from the planar amplitude: Taylor coefficients around the crossing-symmetric point, and on-shell couplings of the exchanged hadrons. Neither relies on a truncated chiral expansion. For pions, matching the Taylor coefficients to chiral low-energy constants is a separate step, which receives corrections of order $m_\pi^2/\Lambda^2$ away from the chiral limit -- see Section~\ref{subsec:low_energy_observables}. This allows us to follow the bounds up to pion masses where the leading-order identification is no longer accurate. For glueballs, the external state is itself the lightest exchanged particle and no low-energy effective theory exists, so observables defined from the amplitude are the natural choice.

We apply this construction to three problems. For pions, we bound the normalized low-energy coefficients $\tilde g_2$ and $\tilde g_2'$, corresponding to $\mathcal{O}(p^4)$ operators in Chiral Perturbation Theory ($\chi$PT), and the normalized coupling of the $f_2$ to two pions. For glueballs, we bound the coupling of the higher spin glueballs to two scalar glueballs. Whenever we decide to input in the bootstrap the known low-energy states, we take their masses from large-$N$ lattice simulations: the $\rho$ mass as a function of the pion mass from~\cite{Bonanno:2025hzr}, and the glueball spectrum from~\cite{Athenodorou:2021qvs}.

For the $\rho$ we use the large $N$ lattice result of~\cite{Bonanno:2025hzr}, whose chiral-continuum fit is linear in $m_\pi^2$, with $m_{\rho,0}/\sqrt{\sigma_{\mathrm{str}}}=1.783(47)$ in the chiral limit and slope $0.1902(65)$ in units of the string tension $\sigma_{\mathrm{str}}$. Here $m_{\rho,0}$ denotes the $\rho$ mass in the chiral limit. Squaring the central values and keeping the leading dependence on $m_\pi^2$, we adopt
\begin{equation}\label{eq:rho_mass}
    \frac{m_\rho^2}{m_{\rho,0}^2}
    =1+0.678\,\frac{m_\pi^2}{m_{\rho,0}^2}.
\end{equation}

No large $N$ lattice determination of the $f_2$ mass is available. We therefore assume that the $\rho$ and the $f_2$ lie on a common linear Regge trajectory, whose slope is fixed by the string tension, $1/\alpha'=2\pi\sigma_{\mathrm{str}}$. At $N=\infty$ quark loops are absent, so $\sigma_{\mathrm{str}}$, and hence $\alpha'$, does not depend on the quark mass. The squared-mass spacing of the two states is then constant,
\begin{equation}\label{eq:f2_mass}
    \frac{m_{f_2}^2}{m_{\rho,0}^2}
    =1+\frac{1}{\alpha'm_{\rho,0}^2}
      +0.678\,\frac{m_\pi^2}{m_{\rho,0}^2}.
\end{equation}
The chiral-limit value $m_{\rho,0}/\sqrt{\sigma_{\mathrm{str}}}=1.783$ gives $1/(\alpha'm_{\rho,0}^2)\simeq1.98$, and hence $m_{f_2}^2/m_{\rho,0}^2\simeq2.98$ at $m_\pi=0$. Equation~\eqref{eq:f2_mass} is an input to the bootstrap, not a lattice determination of the tensor mass. It is a simple model, adopted in the absence of such a determination. 

In summary, the bounds on the pion low-energy coefficients keep the main features of the chiral limit as the pion mass increases, but do not simply rescale. Maximizing the $f_2$ coupling produces a kink at all pion masses studied, whose spectrum resembles the real-world mesons near the physical point. For glueballs, the extremal solutions are given by approximate linear Regge trajectories.

Section~\ref{sec:chiralEFT_largeN} introduces the massive pion amplitude and its low-energy expansion. Section~\ref{sec:bootstrap} states the bootstrap assumptions and derives the dispersion relations, null constraints and sum rules for pion scattering. Section~\ref{sec:bounds} presents the pion bounds and extremal spectra. Sections~\ref{sec:glueballs} and~\ref{sec:glueballbounds} discuss glueball scattering, and section~\ref{sec:discussion} concludes.

\paragraph{Note added.}
While this work was being completed, we became aware of ongoing work by Henriette Elvang and Aditi Chandra on the scattering of massive pions~\cite{ElvangChandra:toappear}, and by Alessandro Vichi and Long-Qi Shao on glueball scattering~\cite{Vichi:toappear}. We thank them for informing us about their work.

\section{Chiral EFT for large $N$ QCD with a mass}
\label{sec:chiralEFT_largeN}
We consider QCD with $N_f$ degenerate light quarks of mass $m_q$ in the 't Hooft limit, $N\to\infty$ with $N_f$ and $g^2_{YM}N$ held fixed. Assuming the standard pattern of chiral symmetry breaking
\begin{equation}
    SU(N_f)_L\times SU(N_f)_R\longrightarrow SU(N_f)_V ,
\end{equation}
the low-energy dynamics is described by the chiral effective field theory \cite{Gasser:1983yg,Gasser:1984gg}. In the massless limit, this is a theory of $N_f^2-1$ Goldstone bosons, the pions, which transform in the adjoint representation of the unbroken vector flavor symmetry $SU(N_f)_V$, while the broken axial symmetry is realized non-linearly.
The explicit breaking of the axial symmetry due to the quark masses is incorporated in the EFT through the spurion field $\chi(x)$, that in the end we set to its physical background value\footnote{The field $\chi$ transforms as $R(x)\chi(x)L^\dagger(x)$ with $R$ and $L$ right and left chiral rotations. In this way the quark mass terms $\bar q m_q q \to \bar q \chi q$ are invariant under chiral transformation. In the IR, we can use this field to construct invariant terms, and only in the end fix its value to the physical quark masses.
}
\begin{equation}
    \chi=2 B_0 \mathcal{M}_q,\qquad \mathcal{M}_q=m_q \mathds{1}.
\end{equation}
Throughout this work we set all external sources to zero, except for the scalar one through this constant quark-mass background.

\subsection{Low energy expansion and large \texorpdfstring{$N$}{N} limit}\label{sec:chiralEFT}
The leading-order Lagrangian of degree two in the number of derivatives is
\begin{equation}\label{eq:L2}
    \mathcal{L}_2=\frac{F^2}{4}\langle \partial_\mu U \partial^\mu U^\dagger\rangle+\frac{F^2}{4}\langle \chi U^\dagger+U \chi^\dagger\rangle, \qquad U=e^{2i\Phi(x)/F},
\end{equation}
with $\Phi(x)=\pi^a(x) T^a$, $a=1,\dots,N_f^2-1$, the Goldstone field. The symbol $\langle\,\cdots\rangle$ denotes the trace over flavour indices and the generators are normalized by $\langle T^a T^b\rangle=\tfrac{1}{2}\delta^{ab}$.
Expanding to quadratic order in the pion field
\begin{equation}
\mathcal{L}_2 =\frac{1}{2}\partial_\mu \pi^a\partial^\mu \pi^a-\frac{M^2}{2}\pi^a\pi^a+\dots, \qquad M^2=2B_0m_q.
\end{equation}
Both $F$ and $M$ are leading-order chiral parameters related to physical quantities by
\begin{equation}
    m_\pi^2=M^2+\mathcal{O}(p^4),\qquad f_\pi=F+\mathcal{O}(p^2).
\end{equation}
Unlike the decay constant $F\sim \mathcal{O}(\sqrt{N})$, the mass parameter $M$ does not scale with $N$, but is an order one number at fixed quark mass $m_q$.\footnote{The constant $B_0$ is related to the chiral condensate $\langle 0| \bar q q| 0\rangle_0$ by 
\begin{equation}
B_0=-\frac{\langle 0| \bar q q| 0\rangle_0}{F^2}\qquad  \implies \qquad F^2M^2=-2m_q \langle 0|\bar q q|0\rangle_0 ,\nonumber
\end{equation}
which is the leading order Gell-Mann--Oakes--Renner relation \cite{Gell-Mann:1968hlm}. 
Since $F^2=\mathcal{O}( N)$ and the chiral condensate $\langle 0|\bar q q|0\rangle_0=\mathcal{O}(N)$, one has $B_0=\mathcal{O}(1)$, and hence $M^2=\mathcal{O}(1)$ at fixed quark mass $m_q$.} 

In the strict planar limit, the $U(1)_A$ anomaly is suppressed by $1/N$, so the chiral symmetry is enlarged to
\begin{equation}
    U(N_f)_L\times U(N_f)_R\longrightarrow U(N_f)_V.
\end{equation}
The pseudoscalar singlet $\eta^\prime$ therefore joins the pions in the same $U(N_f)$ multiplet and, for degenerate quark masses, is degenerate with them at leading order in $1/N$. It can be incorporated by including the generator $T^0=\tfrac{\mathds{1}}{\sqrt{2N_f}}$. As discussed in \cite{Albert:2022oes}, amplitudes involving the $\eta^\prime$ are fixed in terms of the same flavor-ordered amplitudes that describe the scattering of non-singlet pions, and their inclusion does not produce additional constraints. Therefore, we restrict the scattering analysis below to the $SU(N_f)$ pion sector only.

The large $N$ and the chiral expansions are distinct \cite{Kaiser:2000gs}. The chiral expansion organizes the effective low energy Lagrangian according to the number of derivatives and insertions of the quark-mass spurion,
\begin{equation}
\mathcal{L}_\chi=\mathcal{L}_2+\mathcal{L}_4+\mathcal{L}_6+\dots,
\end{equation}
while the large $N$ counting is encoded in the flavour-trace structures of each interaction term. The coefficients of single-trace operators are leading, while terms containing multiple flavour traces are suppressed by powers of $1/N$. Meson loops also are suppressed by powers of $1/N$. Hence, in the planar limit the low-energy meson scattering is generated by tree-level diagrams of the single-trace EFT operators.

These two expansions play different roles. For a fixed four-meson amplitude, the coefficient of the leading single-trace operator at any order in the chiral expansion scales as $\mathcal{O}(N)$. Expanding $U=e^{2i\Phi/F}$ to four meson fields generates four powers of $1/F\sim 1/\sqrt{N}$. Every planar tree-level contribution to $2\to 2 $ scattering therefore scales uniformly as $1/N$. 
Instead, in the low-energy expansion its position is controlled by powers of momenta and quark masses over an $N$-independent hadronic scale that we call $\Lambda$. Schematically,
\begin{equation}
    M_{4\pi}^\text{planar}\sim \frac{p^2}{F^2}\left(1+\mathcal{O}\left(\frac{p^2}{\Lambda^2},\frac{M^2}{\Lambda^2}\right)+\dots\right).
\end{equation}
By constrast, each additional meson loop introduces an extra factor of $p^2/(16\pi^2 F^2)$, being therefore suppressed at large $N$.

Let's focus now on the planar $2\to 2$ pion amplitude. 
In the underlying large $N$ theory, the leading diagrams have the topology of a disk bounded by a single quark loop \cite{tHooft:1973alw, Witten:1979kh}. Their flavour dependence is therefore encoded in a single trace of generators ordered along the boundary. Following Albert and Rastelli and using their normalization \cite{Albert:2022oes}, we write the amplitude for the process $\pi^a(p_1)\pi^b(p_2)\to\pi^c(p_3)\pi^d(p_4)$ as
\begin{equation}\label{eq:flavor_decomposition}
\begin{aligned}
\mathcal{T}_{abcd} = \;& 4\left[\langle T_aT_bT_cT_d\rangle + \langle T_aT_dT_cT_b\rangle\right] M(s,t) + 4\left[\langle T_aT_cT_dT_b\rangle + \langle T_aT_bT_dT_c\rangle\right] M(s,u) \\
&+ 4\left[\langle T_aT_dT_bT_c\rangle + \langle T_aT_cT_bT_d\rangle\right] M(t,u),
\end{aligned}
\end{equation}
where
$s=(p_1+p_2)^2$, $t=(p_1-p_4)^2$, $u=(p_1-p_3)^2$ and $s+t+u=4m_\pi^2$. 
It is useful to define the combination $A(s|t,u)=M(s,t)+M(s,u)-M(t,u)$.\footnote{For $N_f=2$, where $d^{abc}=0$, this is the coefficient of $\delta_{ab}\delta^{cd}$ isospin term that at leading reduces to Weinberg's result \cite{PhysRevLett.17.616}
\begin{equation}
    A(s|t,u)=\frac{s-m_\pi^2}{f_\pi^2},\nonumber
\end{equation}
    that shows the Adler zero at $s=m_\pi^2$ \cite{PhysRev.137.B1022}.} The flavor-ordered amplitude $M(s,t)$ depends cyclically on two adjacent channels. Crossing symmetry implies the relation
\begin{equation}
    M(s,t)=M(t,s),
\end{equation}
but $M(s,t)$ is not fully symmetric under permutations of $s,t,u$.
The cyclic ordering of the planar quark boundary, drawn in Figure \ref{fig:planar-disk}, distinguishes the third channel. The ordered amplitude $M(s,t)$ has singularities only in the $s-$ and $t-$channels.
At leading order in the chiral expansion the flavor-ordered amplitude is
\begin{equation}\label{eq:M2}
    M^{(2)}(s,t)=\frac{s+t-2M^2}{2F^2}=\frac{s+t-2m_\pi^2}{2f_\pi^2}+\mathcal{O}(p^4),
\end{equation}
where in the second equality we have replaced the leading order parameters $M$ and $F$ with their physical values, producing an $\mathcal{O}(p^4)$ correction. In the $m_\pi\to0$ chiral limit $M^{(2)}$ vanishes when $u=0$, as expected from the pion soft theorem. At nonzero quark mass, the zero is shifted to $u=2m_\pi^2$. This zero is not protected by chiral symmetry and higher-order corrections can shift it.

\usetikzlibrary{calc,decorations.markings}

\begin{figure}[t!]
\centering
\begin{tikzpicture}[
  line cap=round,
  line join=round,
  quark/.style={draw=black, line width=0.75pt},
  disk/.style={fill=black!6},
  plain/.style={},
  flow/.style={
    postaction={decorate},
    decoration={
      markings,
      mark=at position 0.5 with
        {\arrow{Latex[length=1.8mm,width=1.25mm]}}
    }
  }
]

\def\R{1.45}
\def\opening{12}
\def\gap{0.060}
\def\leg{0.66}
\def\neck{0.20}
\def\bend{0.19}
\def\join{0.16}

\pgfmathsetmacro{\xa}{\R*cos(\opening)}
\pgfmathsetmacro{\ya}{\R*sin(\opening)}
\pgfmathsetmacro{\xc}{\xa+\bend*sin(\opening)}
\pgfmathsetmacro{\yc}{\ya-\bend*cos(\opening)}
\pgfmathsetmacro{\xb}{\R+\neck}
\pgfmathsetmacro{\xt}{\R+\leg}

\fill[disk] (0,0) circle[radius=\R];

\foreach \ang/\lab in {135/a,225/b,45/c,315/d} {
  \begin{scope}[rotate=\ang]
    \fill[disk]
      (\xa,-\ya)
      .. controls (\xc,-\yc) and ({\xb-\join},-\gap)
      .. (\xb,-\gap)
      -- (\xt,-\gap) -- (\xt,\gap) -- (\xb,\gap)
      .. controls ({\xb-\join},\gap) and (\xc,\yc)
      .. (\xa,\ya) -- cycle;

    \foreach \sgn in {-1,1} {
      \draw[quark]
        (\xa,{\sgn*\ya})
        .. controls (\xc,{\sgn*\yc})
        and ({\xb-\join},{\sgn*\gap})
        .. (\xb,{\sgn*\gap})
        -- (\xt,{\sgn*\gap});
    }
  \end{scope}
  \node at (\ang:{\xt+0.28}) {$\pi^{\lab}$};
}

\foreach \ang/\arcstyle in {45/flow,135/plain,225/flow,315/plain} {
  \draw[quark,\arcstyle]
    ({\ang+\opening}:\R)
    arc[start angle={\ang+\opening},
        end angle={\ang+90-\opening},
        radius=\R];
}

\end{tikzpicture}
\caption{Leading large-$N$ disk topology for four-pion scattering.
The oriented boundary represents a single quark loop, with the
external pions in cyclic order $(a,b,d,c)$.
The shaded region denotes planar gluon dressings.}
\label{fig:planar-disk}
\end{figure}
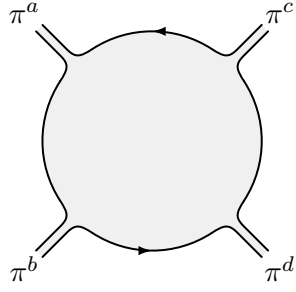

Next, we consider the first non-universal corrections. At order $p^4$, the single-trace operators relevant for pion scattering are~\footnote{At finite $N$ and generic $N_f$ there are four independent four derivative operators: the two single-trace ones in \eqref{eq:L4} with coefficients $L_3^\prime$ and $L_4^\prime$ in the notation of \cite{Albert:2022oes}, and two double-trace ones $L_1^\prime$ and $L_2^\prime$ whose couplings are $1/N$ suppressed. For $N_f=3$, the Cayley-Hamilton theorem relates them leaving only $L_1=L_1^\prime+\tfrac{L_4^\prime}{2}$, $L_2=L_2^\prime+L_4^\prime$, $L_3=L_3^\prime-2L_4^\prime$ in the convention of \cite{Gasser:1984gg}. At leading order in $1/N$ we can set $L_1^\prime=L_2^\prime=0$. 
For $N_f=2$, a second relation leaves only two low energy constants: the $\ell_1, \ell_2$ of \cite{Gasser:1983yg}, that at leading order in $1/N$ are given by $\ell_1=2(L_3^\prime-L_4^\prime)$, and $\ell_2=4L_4^\prime$.}
\begin{equation}\label{eq:L4}
\begin{aligned}
    \mathcal{L}_4 = &\; L_3' \langle \partial_\mu U^\dagger \partial^\mu U \partial_\nu U^\dagger \partial^\nu U\rangle + L_4' \langle \partial_\mu U^\dagger \partial_\nu U \partial^\mu U^\dagger \partial^\nu U\rangle \\
    &+ L_5 \langle \partial_\mu U^\dagger \partial^\mu U (\chi^\dagger U + U^\dagger \chi)\rangle + L_8 \langle \chi^\dagger U \chi^\dagger U + U^\dagger \chi U^\dagger \chi\rangle.
\end{aligned}
\end{equation}
The four couplings $L_3^\prime, L_4^\prime, L_5, L_8$ are all of $\mathcal{O}(N)$.
The mass-dependent operators proportional to $L_5$ and $L_8$ generate four-pion interactions and modify the relation between $M,F$ and the physical pion mass and decay constant
\begin{equation}\label{eq:mpi_fpi}
    m_\pi^2 = M^2\left[1 + \frac{8 M^2}{F^2}\,(2L_8 - L_5)\right] + \mathcal{O}(p^6), \qquad f_\pi = F\left[1 + \frac{4 M^2}{F^2}\, L_5\right] + \mathcal{O}(p^4).
\end{equation} 
Expanding the $p^4$ operators to four pion fields, putting the external particles on the physical mass-shell, and expressing the result in terms of $m_\pi$ and $f_\pi$ gives the next-order amplitude
\begin{equation}\label{eq:M4}
    M^{(4)}(s,t)=\frac{1}{f_\pi^4}\Big(c_{2,0}(s^2+t^2)+2c_{1,1}s t+c_{1,0}\,m_\pi^2(s+t)+c_{0,0}\,m_\pi^4\Big),
\end{equation}
with
\begin{equation}\label{eq:c_vs_L}
\begin{aligned}
    c_{2,0}&=2L_3'+4L_4', &\qquad c_{1,0}&=4\left(L_5-2L_3'-4L_4'\right),\\
    c_{1,1}&=4L_4', & c_{0,0}&=16\left(L_3'+L_4'-L_5+L_8\right) .
\end{aligned}
\end{equation}
The coefficients of the quadratic structures, $c_{2,0}$ and $c_{1,1}$, are the same as in the chiral limit $m_\pi=0$. 
This has a simple origin.
Trading $M$, $F$ for $m_\pi$, $f_\pi$ in \eqref{eq:M2}, together with the associated wave-function renormalization, can only generate terms proportional to $m_\pi^2(s+t)$ and $m_\pi^4$. At order $p^4$ the dependence on the quark mass therefore only affects $c_{1,0}$ and $c_{0,0}$, which also receive contributions from the mass-dependent couplings $L_5$ and $L_8$. The details are given in Appendix \ref{app:matching}.

\subsection{Nonperturbative low-energy observables and matching conditions}\label{subsec:low_energy_observables}

In the planar limit, the flavor-ordered $2\to 2$ meson amplitude $M(s,t)$ is a meromorphic function with isolated poles associated to meson resonances. It is therefore analytic in a neighbourhood of any point that does not contain a meson pole, and its Taylor expansion around such point has a finite radius of convergence determined by the nearest singularity. This statement concerns the Taylor expansion of the on-shell amplitude and it has to be distinguished from the chiral EFT expansion.

For massless pions the natural expansion point is $s=t=u=0$. For massive pions we instead expand around the crossing symmetric point in the spirit of \cite{EliasMiro:2026utl}
\begin{equation}
    s=t=u=\sigma,\qquad \sigma=\frac{4}{3}m_\pi^2.
\end{equation}
For convenience, the same point will be later used as subtraction point for the dispersion relations in section~\ref{sec:dispersion}. Defining $\hat s=s-\sigma$, $\hat t=t-\sigma$, and $\hat u=-\hat s-\hat t=u-\sigma$, we write the crossing-symmetric expansion
\begin{equation}\label{eq:Mlow}
\begin{aligned}
M(\hat s,\hat t) &= M_0 + \sum_{n=1}^{\infty}\sum_{k=0}^{\lfloor n/2\rfloor} g_{n,k}\left(\hat{s}^{n-k}\hat{t}^{k}+\hat{t}^{n-k}\hat{s}^{k}\right)= \\
&= M_0 + g_{1,0}(\hat{s}+\hat{t}) + g_{2,0}(\hat{s}^2+\hat{t}^2) + 2g_{2,1}\hat{s}\hat{t} + g_{3,0}(\hat{s}^3+\hat{t}^3) + g_{3,1}(\hat{s}^2\hat{t}+\hat{t}^2\hat{s}) + \cdots.
\end{aligned}
\end{equation}
The quantities $M_0$ and $g_{n,k}$ are extracted directly from the exact planar amplitude $M(s,t)$, and do not rely on the truncation of the chiral expansion. They are the observables we will bound in the next sections. Their interpretation in terms of EFT couplings is a separate matching problem.

The constant $M_0=M(\sigma,\sigma)$ vanishes in the chiral limit when the expansion point coincides with the Adler zero. For massive pions no symmetry fixes its value, and at the leading order in chiral perturbation theory it reads
\begin{equation}\label{eq:matching}
    M_0=\frac{m_\pi^2}{3f_\pi^2}\left(1+\mathcal{O}\left(\frac{m_\pi^2}{\Lambda^2}\right)\right).
\end{equation}
In the once-subtracted dispersions we use below, $M_0$ plays the role of the subtraction constant. It is not constrained by positivity, and it will not play any role in the bounds we will discuss below. This differs from the nonperturbative S-matrix bootstrap where $M_0$ can be bounded using full unitarity \cite{ LOPEZ1977426, Paulos:2017fhb, He:2021eqn, Guerrieri:2021tak, Chen:2022nym, EliasMiro:2023fqi, deRham:2025vaq}. 

We can now match the observables defined in \eqref{eq:Mlow} with the chiral EFT couplings. Expanding the perturbative amplitudes at order $p^2$ \eqref{eq:M2} and $p^4$ \eqref{eq:M4} around the crossing symmetric point, we get
\begin{equation}\label{eq:gtilde}
    g_{1,0}=\frac{1}{2f_\pi^2}\,(1+2\epsilon)\left[1+\mathcal{O}\!\left(\frac{m_\pi^4}{\Lambda^4}\right)\right], \qquad \text{with} \qquad \epsilon\equiv\frac{m_\pi^2}{f_\pi^2}\Big[c_{1,0}+\tfrac{8}{3}\big(c_{2,0}+c_{1,1}\big)\Big]
\end{equation}
and
\begin{equation}
    g_{2,0}=\frac{c_{2,0}}{f_\pi^4}\left[1+\mathcal{O}\!\left(\frac{m_\pi^2}{\Lambda^2}\right)\right],\quad
    g_{2,1}=\frac{c_{1,1}}{f_\pi^4}\left[1+\mathcal{O}\!\left(\frac{m_\pi^2}{\Lambda^2}\right)\right].
\label{eq:g2defs}
\end{equation}
Since the $\mathcal{O}(p^4)$ coefficients satisfy the scaling relation $c_{n,k}/f_\pi^2 \sim 1/\Lambda^2$, we have that $\epsilon=\mathcal{O}(m_\pi^2/\Lambda^2)$. We omit the corrections coming from $\mathcal{O}(p^6)$. Notice the important conceptual difference with the chiral limit: since the expansion point itself is of order $m_\pi^2$, higher derivative operators will contribute not only to the higher order coefficients but also the lower order ones $g_{n,k}$. The matching between the amplitude observables and the chiral low energy constants (LECs) is only perturbative in $m_\pi^2/\Lambda^2$. In the context of the Bootstrap, this point has been explained in \cite{EliasMiro:2023fqi}.

Following \cite{Caron-Huot:2020cmc}, we normalize the coefficients $g_{n,k}$ in units of the leading dispersive coefficient $g_{1,0}$, which is positive,  and it reduces to $1/2f_\pi^2$ at leading order in the chiral expansion. At the next order in the Taylor expansion we define the dimensionless couplings
\begin{equation}
\label{eq:g2eg2prime}
\tilde g_2=\Lambda^2\frac{g_{2,0}}{g_{1,0}}, \qquad
\tilde g_2^\prime=2\Lambda^2\frac{g_{2,1}}{g_{1,0}}.
\end{equation}
At order $p^4$ in the chiral expansion they are given by
\begin{equation}
    \tilde g_2=\frac{2\Lambda^2}{f_\pi^2}c_{2,0}\left(1-2\epsilon+\mathcal{O}\left(\frac{m_\pi^2}{\Lambda^2}\right)\right), \quad \tilde g_2^\prime=\frac{4\Lambda^2}{f_\pi^2}c_{1,1}\left(1-2\epsilon+\mathcal{O}\left(\frac{m_\pi^2}{\Lambda^2}\right)\right).
\end{equation}
The explicit factor $1-2\epsilon$ comes from the common normalization by $g_{1,0}$. In the chiral $\epsilon\to 0$ limit they coincide with the observables studied in \cite{Albert:2022oes}
\begin{equation}
    \tilde g_2=(4L_3'+8L_4')\,\frac{\Lambda^2}{f_\pi^2},\quad \tilde g_2^\prime=16L_4'\,\frac{\Lambda^2}{f_\pi^2}.
\end{equation}
What is the typical size of the corrections to the Taylor expansion coefficients? For physical pion masses $\epsilon\simeq0.03$, as estimated in appendix~\ref{app:exp}, while for the largest pion masses studied in this paper $\epsilon\sim0.3$. This shows explicitly that the identification of $g_{2,0}$ and $g_{2,1}$ with the order-$p^4$ chiral couplings works well at the physical point, but becomes unreliable for large pion masses.

The distinction between amplitude observables and EFT interpretation is central in our analysis. The bounds we derive apply directly to the Taylor coefficients $g_{n,k}$ of the planar amplitude and do not assume that the amplitude itself can be approximated by a finite order truncation of chiral perturbation theory. Chiral perturbation theory is only needed to translate these bounds in terms of the more familiar LECs $L_i$. At finite pion mass this dictionary receives higher order corrections, and for the heaviest pion masses considered below it should be regarded qualitative and taken with a grain of salt.

The Taylor coefficients remain well defined as long as no pole sits at the crossing symmetric point (in which case one can always move the subtraction point elsewhere -- see Section \ref{sec:glueballs}). If the nearest singularity is the $\rho$ pole, then the typical scale is $R_\rho=m^2_\rho-\sigma$, with $\sigma=\tfrac{4}{3}m_\pi^2$.
Their magnitude is expected to be controlled by inverse powers of $R_\rho$, schematically
\begin{equation}
    \frac{g_{n,k}}{g_{1,0}}\sim \frac{1}{R_\rho^{n-1}}.
\end{equation}
Thus, when measured in the same units, for instance $m_{\rho,0}$ in the exact chiral limit, higher order coefficients are expected to be smaller for light pions, and grow as $m_\pi^2$ increases and $R_\rho$ becomes smaller. This expectation is consequence of the analytic structure of the planar amplitude and of our Taylor expansion and is independent of whether the EFT expansion converges or not. For the case of pion scattering, the subtraction point
remains below the $\rho$ pole throughout the range studied in
section~\ref{sec:bounds}, including when the $\rho$ lies below the
two-pion threshold. The glueball case is different and will be addressed in Section \ref{sec:glueballs}.

\section{Meson scattering}\label{sec:bootstrap}

 We will initially focus on the meson sector and study the flavour-ordered amplitude $M(s,t)$ of section~\ref{sec:chiralEFT} in the planar limit.
 Below, we review the general properties that this amplitude must satisfy.
 Large $N$ QCD at a given pion mass is one point in the space of functions with these properties, and the bounds we derive apply to every point of that space. 

\subsection{General properties}
\label{subsec:genprop}
\paragraph{Analyticity.}
At leading order in $1/N$ the amplitude $M(s,t)$ is a meromorphic function of two complex variables.
It has simple poles at the squared masses of the exchanged states.
It is hermitian analytic, $M(s^*,t)=M(s,t)^*$, so that its imaginary part on the real axis is supported at the poles and coincides with the discontinuity $\text{Im}M(s,t)=\tfrac{1}{2i}\text{Disc}_s M(s,t)$.

Planarity fixes which channels can contain such poles. For instance, in the case of $M(s,t)$ a pole in the $u$ channel would require either an exotic four-quark intermediate state or a process involving quarks annihilations violating the OZI rule, both suppressed at $N=\infty$. Since the argument only depends on the flavour tolopogy, it is not affected by the quark mass. Thus, $M(s,t)$ has only poles in the $s$ and $t$ channels. Consequently, if we fix $t$ this function has only $s$-channel poles for positive $s$, while at  fixed $u$, the amplitude evaluated at $M(s,4m_\pi^2-s-u)$, has poles on both sides: if $s_R$ is a pole, then $4m_\pi^2-s-s_R$ is the crossed-channel $t$-pole. The two analytic structures are shown in figure~\ref{fig:dispersion_elations}.

We denote by $\Lambda$ the mass of the lightest exchanged meson, so that all the singularities of $M(s,t)$ lie at $s\geq\Lambda^2$ or $t\geq\Lambda^2$. In QCD, we identify $\Lambda=m_\rho$ with the mass of the $\rho$ meson. 
Since $m_\rho^2$ and $m_\pi^2$ depend differently on the quark mass, nothing forbids $\Lambda<2m_\pi$, in which case the lightest meson is a true bound state below the two-pion threshold. We shall mostly assume $\Lambda\geq 2m_\pi$, and return to the bound-state case in section~\ref{app:Bound_state}.

\paragraph{Crossing.}
The ordered amplitude is symmetric under the exchange of the two adjacent channels, $M(s,t)=M(t,s)$, with $s+t+u=4m_\pi^2$, but there is no symmetry involving $u$. As discussed in section~\ref{sec:chiralEFT}, we expand around the point $s=t=u=\sigma=\tfrac43 m_\pi^2$, which is invariant under $s\leftrightarrow t$ and reduces in the chiral limit to the origin.

\paragraph{Unitarity.}
For $s\geq 4m_\pi^2$ the imaginary part of the amplitude admits the $s$-channel partial-wave decomposition
\begin{equation}\label{eq:partial_waves}
    \mathrm{Im}\,M(s,t)=16\pi\sum_{J=0}^{\infty}(2J+1)\,\rho_J(s)\,P_J\big(x_t(s)\big),\qquad x_t(s)\equiv 1+\frac{2t}{s-4m_\pi^2}=\cos\theta_s,
\end{equation}
where $\theta_s$ is the scattering angle in the $s$-channel and $P_J$ are the Legendre polynomials. Unitarity of the full amplitude $\mathcal{T}_{abcd}$, in each flavour channel, implies positivity of the corresponding spectral densities.
The absence of exotic states implies that at fixed-$s$, the $M(t,u)$ amplitude has no poles for $s\geq 4m_\pi^2$.\footnote{Of course it has poles for negative $s$ at $s=4m_\pi^2-u-t$, but those do not correspond to physical exchanges.} 
Under this assumption, the singlet and antisymmetric adjoint $s$-channel imaginary parts read as \cite{Albert:2022oes}
\begin{align}
\label{eq:nice_channels}
    \mathrm{Im}\,M^{\text{singlet}}(s,t,u)&=\frac{2(N_f^2-1)}{N_f}(\mathrm{Im}\,M(s,t)+\mathrm{Im}\,M(s,u))\\
    \mathrm{Im}\,M^{\text{Adj}_A}(s,t,u)&=N_f(\mathrm{Im}\,M(s,t)-\mathrm{Im}\,M(s,u)).
\end{align}
Recalling that $x_u(s)=-x_t(s)$, we have that
\begin{equation}
\mathrm{Im}\,M(s,u)=16\pi\sum_{J=0}^{\infty}(-1)^J(2J+1)\,\rho_J(s)\,P_J\big(x_t(s)\big).
\end{equation}
From this immediately follows that $M^\text{singlet}$ contains only even spins, while $M^{\text{Adj}_A}$ the odd spins.
Positivity of their corresponding imaginary parts implies
\begin{equation}\label{eq:positivity}
    \rho_J(s)\geq 0,\qquad s\geq 4m_\pi^2,\quad J=0,1,2,\dots,
\end{equation}
where 
\begin{equation}
    \rho_{J, \text{even}}(s)=\tfrac{N_f}{4(N_f^2-1)}\rho_J^{\text{singlet}}(s),\qquad \rho_{J,\text{odd}}(s)=\tfrac{1}{2N_f}\rho^{\text{Adj}_A}(s).
    \label{eq:proj_nice}
\end{equation}
Unitarity also bounds the $\rho_J$ from above, but meson couplings scale as $1/\sqrt N$ and the upper bound is irrelevant at leading order. For a meromorphic amplitude the $\rho_J$ are sums of delta functions. Around each pole the amplitude behaves as,
\begin{equation}\label{eq:coupling_def}
    M(s,t)\;\xrightarrow[s\to m_X^2]{}\;\frac{g^2_{\pi\pi X}\,m_X^2\,P_{J_X}\big(x_t(m_X^2)\big)}{m_X^2-s},
\end{equation}
so one has
\begin{equation}\label{eq:rho_meromorphic}
    16(2J+1)\,\rho_J(s)=\sum_{X:\,J_X=J} g^2_{\pi\pi X}\,m_X^2\,\delta(s-m_X^2),
\end{equation}
and \eqref{eq:positivity} is the positivity of the squared couplings. In principle, we shall not assume that the spectrum is discrete: we parametrize our ignorance of the physics above the cutoff by a generic positive measure $\rho_J(s)$ supported on $s\geq\Lambda^2$. Every bound derived in this way holds also for a discrete spectrum.

The positivity of the residues holds for bound states as well, but it has to be stated in terms of the physical couplings. The residue of a spin-$J$ exchange is the coupling squared times the polarization sum, which is proportional to $(|\vec p\,|^2)^{J}P_J(\cos\theta)$ with $|\vec p\,|^2=(s-4m_\pi^2)/4$; absorbing positive constants in the coupling,
\begin{equation}\label{eq:physical_coupling}
g^2_{\pi\pi X}\,m_X^2=\hat g^2_X\,\big(m_X^2-4m_\pi^2\big)^{J_X},\qquad \hat g^2_X\geq0 ,
\end{equation}
where $\hat g_X$ is the coupling that unitarity, through the completeness relation, makes the square of a real number, whether $m_X$ lies above or below the two-pion threshold. Above threshold \eqref{eq:physical_coupling} is just \eqref{eq:positivity}. For a bound state of odd spin, instead, the coefficient $g^2_{\pi\pi X}$ of \eqref{eq:coupling_def} is negative. This is the case of a $\rho$ lighter than two pions, to which we come back in section~\ref{app:Bound_state}.

\paragraph{High-energy behaviour.}
Regge theory relates the high-energy behaviour of the amplitude to the leading Regge trajectories exchanged. The pomeron, which is responsible for the Froissart growth of total cross sections in QCD, carries vacuum quantum numbers and, like the glueballs, decouples from the mesons at large $N$. The leading meson trajectories for odd and even spins are those of the $\rho$ and the $f_2$. In the planar limit they are degenerate, with intercept $\alpha(0)\simeq0.5$.\footnote{At $N=3$ this is the value extracted from fits to total cross sections~\cite{Donnachie:1992ny}, and it agrees with the Chew--Frautschi plot of the $\rho$, $a_2$, $\rho_3$ and $a_4$ masses. Since the low-lying large $N$ spectrum is close to the physical one~\cite{Bonanno:2026zcq}, we assume that the same holds for the intercept.} We therefore assume
\begin{equation}\label{eq:regge}
    \lim_{|s|\to\infty}\frac{M(s,t)}{s}=0\quad\text{at fixed }t,\qquad
    \lim_{|s|\to\infty}\frac{M(s,4m_\pi^2-s-u)}{s}=0\quad\text{at fixed }u,
\end{equation}
for $t$ and $u$ in a neighbourhood of the crossing-symmetric point $\sigma$. At fixed $t$, the limit is controlled by the exchange of the $\rho$ and $f_2$ trajectories in the $t$-channel, $M\sim s^{\alpha(t)}$, and \eqref{eq:regge} requires $\alpha(t)<1$, i.e.\ $t$ below the $\rho$ mass. At fixed $u$, both $s$ and $t$ are large, and the asymptotics is controlled by the $u$-channel, which carries no planar exchanges, so the amplitude is much softer.\footnote{For instance, for the Lovelace--Shapiro amplitude at fixed $u$, $\alpha(s)+\alpha(t)$ is constant and the amplitude decays exponentially away from the real axis. It would be interesting to 
impose stronger high-energy constraints in this channel as done in \cite{Haring:2023zwu} for stringy amplitudes, and check their effect on the bounds studied in Section~\ref{sec:bounds}.} The two conditions are independent because the amplitude is symmetric only under $s\leftrightarrow t$. They allow us to write once-subtracted dispersion relations both at fixed $t$ and at fixed $u$ (Section~\ref{sec:dispersion}).~\footnote{Note that~\cite{Albert:2023seb} denotes ordered amplitudes by $M(s,u)$: their fixed-$u$ limit corresponds to our fixed-$t$ one.}

\subsection{Once-subtracted Dispersion Relations}\label{sec:dispersion}

A convenient way to package the information coming from analyticity and crossing is through dispersion relations. They are the starting point of the dual S-matrix Bootstrap formulation both in the weakly-coupled limit \cite{Caron-Huot:2020cmc} and nonperturbatively \cite{Guerrieri:2021tak}.
Since for the case of $\pi\pi$ scattering we will always subtract at the crossing-symmetric point, the value of $x_t$ at $t=\sigma=4/3m_\pi^2$,
\begin{equation}\label{eq:x0}
x_0(z)\equiv x_\sigma(z)=1+\frac{2\sigma}{z-4m_\pi^2},
\end{equation}
appears repeatedly. Note that $x_0(z)>1$ for $z>4m_\pi^2$, so that $P_J(x_0)>0$ and $P_J'(x_0)>0$.

\label{subsec:fixed_t}\label{subsec:fixed_u}

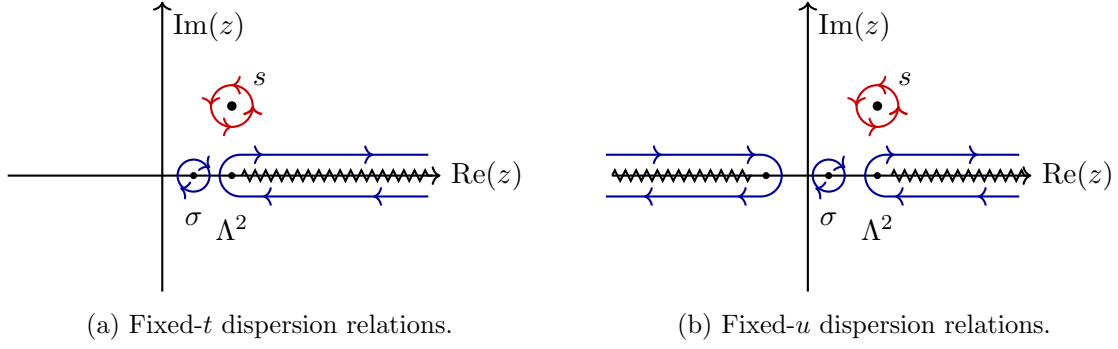
\begin{figure}[t!]
    \centering

    \begin{subfigure}[b]{0.48\textwidth}
        \centering

        \def\xr{4}
        \def\yr{2.5}

        \begin{tikzpicture}[thick, scale=0.92]

            \draw[->] (-\xr/1.8,0) -- (\xr,0)
                node[right] {$\text{Re}(z)$};

            \draw[->] (0,-\yr/1.5) -- (0,\yr)
                node[below right] {$\text{Im}(z)$};

            \draw[
                xshift=-5,
                blue!60!black,
                decoration={
                    markings,
                    mark=between positions 0.125 and 0.875 step 0.25
                    with {\arrow{>}}
                },
                postaction={decorate}
            ]
            (\xr,-0.3)
            -- (1.3,-0.3)
            arc (-90:-270:0.3)
            (1.3,0.3)
            -- (\xr,0.3);

            \draw[
                decorate,
                decoration={zigzag,segment length=4,amplitude=2}
            ]
            (\xr/3.5,0)
            -- (\xr/1.03,0);

            \draw[fill]
            (1,1) circle (1.5pt)
            node[above right=3.8] {$s$};

            \draw[
                red!80!black,
                decoration={
                    markings,
                    mark=between positions 0 and 1 step 0.25
                    with {\arrow{>}}
                },
                postaction={decorate}
            ]
            (1,1) circle (0.3);

            \draw[fill] (1,0) circle (1pt) node[below=11pt] {$\Lambda^2$};
            \draw[fill] (0.45,0) circle (1pt) node[below=10pt] {$\sigma$};

            \draw[
                blue!60!black,
                decoration={
                    markings,
                    mark=between positions 0.15 and 0.85 step 0.55
                    with {\arrow{<}}
                },
                postaction={decorate}
            ]
            (0.45,0) circle (0.23);

        \end{tikzpicture}

        \caption{Fixed-$t$ dispersion relations.}
        \label{fig:fixed_t}

    \end{subfigure}
    \hfill
    \begin{subfigure}[b]{0.48\textwidth}
        \centering

        \def\xr{4.5}
        \def\yr{2.5}

        \begin{tikzpicture}[thick, scale=0.92]

            \draw[->] (-\xr/1.6,0) -- (\xr/1.4,0)
                node[right] {$\text{Re}(z)$};

            \draw[->] (0,-\yr/1.5) -- (0,\yr)
                node[below right] {$\text{Im}(z)$};

            \draw[
                xshift=-5,
                blue!60!black,
                decoration={
                    markings,
                    mark=between positions 0.125 and 0.875 step 0.25
                    with {\arrow{>}}
                },
                postaction={decorate}
            ]
            (\xr/1.4,-0.3)
            -- (1.3,-0.3)
            arc (-90:-270:0.3)
            (1.3,0.3)
            -- (\xr/1.4,0.3);

            \draw[
                xshift=-5,
                blue!60!black,
                decoration={
                    markings,
                    mark=between positions 0.125 and 0.875 step 0.25
                    with {\arrow{<}}
                },
                postaction={decorate}
            ]
            (-\xr/1.65,-0.3)
            -- (-0.5,-0.3)
            arc (-90:90:0.3)
            (-0.5,0.3)
            -- (-\xr/1.65,0.3);

            \draw[
                decorate,
                decoration={zigzag,segment length=4,amplitude=2}
            ]
            (1.2,0)
            -- (\xr/1.43,0);

            \draw[
                decorate,
                decoration={zigzag,segment length=4,amplitude=2}
            ]
            (-\xr/1.6,0)
            -- (-0.8,0);

            \draw[fill]
            (1,1) circle (1.5pt)
            node[above right=3.8] {$s$};

            \draw[
                red!80!black,
                decoration={
                    markings,
                    mark=between positions 0 and 1 step 0.25
                    with {\arrow{>}}
                },
                postaction={decorate}
            ]
            (1,1) circle (0.3);

            \draw[fill] (1,0) circle (1pt) node[below=11pt] {$\Lambda^2$};
            \draw[fill] (-0.6,0) circle (1pt);
            \draw[fill] (0.3,0) circle (1pt) node[below=10pt] {$\sigma$};

            \draw[
                blue!60!black,
                decoration={
                    markings,
                    mark=between positions 0.15 and 0.85 step 0.55
                    with {\arrow{<}}
                },
                postaction={decorate}
            ]
            (0.3,0) circle (0.23);

        \end{tikzpicture}

        \caption{Fixed-$u$ dispersion relations.}
        \label{fig:fixed_u}

    \end{subfigure}

    \caption{Analytic structure of the flavour-ordered amplitude in the complex plane of its first argument, $z$, at fixed $t$ (left) and at fixed $u$ (right), together with the contours used for the once-subtracted dispersion relations of section~\ref{sec:dispersion}. At large $N$ the only singularities are poles on the real axis at the masses of the exchanged mesons, whose union is drawn as a cut starting at $z=\Lambda^2$; at fixed $u$ the $t$-channel poles appear as well, on the negative real axis at $z=4m_\pi^2-u-m_X^2$. The subtraction point $\sigma$ and the point $z=s$ where the amplitude is evaluated are also shown.}
    \label{fig:dispersion_elations}

\end{figure}

Let $f(z)$ be a real-analytic function of the complex variable $z$, with singularities only on the real axis and such that $f(z)/z\to0$ at infinity. For an analytic point $s$ and a subtraction point $a$, Cauchy's theorem on a small circle around $z=s$ gives
\begin{equation}\label{eq:cauchy}
f(s)=\frac{1}{2\pi i}\oint dz\,\frac{f(z)}{z-s}\,\frac{s-a}{z-a}.
\end{equation}
The factor $(s-a)/(z-a)$ equals one at $z=s$ and makes the integrand decay faster than $1/z$, so that the contour can be blown up as in figure~\ref{fig:dispersion_elations}: the circle at infinity does not contribute, the pole at $z=a$ produces the subtraction constant $f(a)$, and the singularities contribute with their discontinuity, $2i\,\mathrm{Im}f(z+i\epsilon)$. Hence
\begin{equation}\label{eq:DR_general}
f(s)-f(a)=\frac{1}{\pi}\int dz\,\mathrm{Im}f(z+i\epsilon)\,K_1(z;s,a),\qquad K_1(z;a,b)\equiv\frac{1}{z-a}-\frac{1}{z-b},
\end{equation}
where the integral runs over the support of the singularities. By the Regge assumptions of section~\ref{sec:bootstrap}, \eqref{eq:DR_general} applies to the amplitude both at fixed $t$ and at fixed $u$, for $t$ or $u$ in a neighbourhood of $\sigma$. In both cases we subtract at the crossing-symmetric point, $a=\sigma$, which lies below the two-pion threshold and hence below every singularity. 

\paragraph{Fixed $t$ dispersions.}
The function $f(z)=M(z,t)$ is singular only at $z\geq\Lambda^2$, figure~\ref{fig:fixed_t}, so that
\begin{equation}\label{eq:DR_fixed_t_a}
M(s,t)=M(\sigma,t)+\frac{1}{\pi}\int_{\Lambda^2}^\infty dz\,\mathrm{Im}\,M(z,t)\,K_1(z;s,\sigma).
\end{equation}
The subtraction function still depends on $t$, but crossing, $M(\sigma,t)=M(t,\sigma)$, identifies it with the amplitude at fixed second argument $\sigma$, to which \eqref{eq:DR_fixed_t_a} applies again in the variable $t$: $M(\sigma,t)=M_0+\frac{1}{\pi}\int_{\Lambda^2}^\infty dz\,\mathrm{Im}\,M(z,\sigma)\,K_1(z;t,\sigma)$, with $M_0\equiv M(\sigma,\sigma)$. Inserting the partial-wave expansion \eqref{eq:partial_waves} of the absorptive parts,\footnote{For $t$ outside the physical region, \eqref{eq:partial_waves} is understood as the analytic continuation in $t$ of the partial-wave expansion of the absorptive part, which converges inside the Lehmann--Martin ellipse; the neighbourhood of $\sigma$ that we need lies well inside it.} we obtain the fixed-$t$ representation
\begin{equation}\label{eq:DR_fixed_t}
M(s,t) = M_0 + \Big\langle P_J\big(x_t(z)\big)\,K_1(z;s,\sigma) + P_J\big(x_0(z)\big)\,K_1(z;t,\sigma)\Big\rangle ,
\end{equation}
where we have introduced the shorthand
\begin{equation}\label{eq:average}
\langle\, h\,\rangle \equiv \sum_{J}\int_{\Lambda^2}^{\infty} dz\, 16(2J+1)\,\rho_J(z)\, h(z,J)
\end{equation}
for the linear functionals of the spectral densities, which we use throughout the paper; by \eqref{eq:positivity}, $\langle h\rangle\geq0$ whenever $h\geq0$ on the support of the densities. The amplitude is entirely determined by $M_0$ and by the spectral densities. 
The $s\leftrightarrow t$ symmetry is not manifest, and imposing it will produce the null constraints of section~\ref{subsec:null_constraints}.

\paragraph{Fixed $u$ dispersions.}
The function $f(z)=M(z,4m_\pi^2-z-u)$ is singular on both sides of the real axis, figure~\ref{fig:fixed_u}: at the $s$-channel poles $z\geq\Lambda^2$ and at the $t$-channel poles $z=4m_\pi^2-u-z'$ with $z'\geq\Lambda^2$. Since $f(z)=M(4m_\pi^2-z-u,z)$ by crossing, the discontinuity across the left cut is again the $s$-channel absorptive part, evaluated at the reflected point $z'$ and with the opposite orientation. Changing variable to $z'$ folds the left cut onto the right one, and \eqref{eq:DR_general} becomes
\begin{equation}\label{eq:DR_fixed_u_raw}
M(s,4m_\pi^2-s-u) = M(\sigma,4m_\pi^2-\sigma-u) + \frac{1}{\pi}\int_{\Lambda^2}^{\infty} dz\, \mathrm{Im}\,M(z,4m_\pi^2-u-z)\, K_{\rm fold}(z;s,u,\sigma),
\end{equation}
with
\begin{equation}\label{eq:Kfold}
\begin{aligned}
K_{\rm fold}(z;s,u,\sigma)
&\equiv K_1(z;s,\sigma)-K_1(4m_\pi^2-u-z;\,s,\sigma)\\
&=\frac{1}{z-s}+\frac{1}{z-4m_\pi^2+s+u}-\frac{1}{z-\sigma}-\frac{1}{z-4m_\pi^2+u+\sigma}.
\end{aligned}
\end{equation}
The absorptive part is now evaluated at $t=4m_\pi^2-u-z$, that is at $x_t(z)=-x_u(z)$, and its partial-wave expansion acquires a signature factor
\begin{equation}\label{eq:signature}
\mathrm{Im}\,M(z,4m_\pi^2-u-z) = 16\pi\sum_{J} (2J+1)(-1)^J \rho_J(z)\, P_J\big(x_u(z)\big),
\end{equation}
the only place where the parity of the spin enters. The subtraction function $M(\sigma,4m_\pi^2-\sigma-u)$ is the fixed-$t$ amplitude at $s=\sigma$, for which only the second term of \eqref{eq:DR_fixed_t} survives since $K_1(z;\sigma,\sigma)=0$. Writing everything in terms of $s$ and $t$, with $u=4m_\pi^2-s-t$ and $4m_\pi^2-\sigma-u=s+t-\sigma$, the fixed-$u$ representation reads
\begin{equation}\label{eq:DR_fixed_u}
M(s,t) = M_0 + \Big\langle P_J(x_0)\, K_1\big(z;\,s+t-\sigma,\,\sigma\big) + (-1)^J P_J(x_u)\, K_{\rm fold}(z;s,u,\sigma)\Big\rangle .
\end{equation}
Unlike \eqref{eq:DR_fixed_t}, it is manifestly symmetric under $s\leftrightarrow t$: the exchange swaps the first two terms of $K_{\rm fold}$ and leaves $x_u$, the subtraction and the densities untouched.

\subsection{Null constraints from crossing}\label{subsec:null_constraints}

The dispersion relations of section~\ref{subsec:fixed_t} reconstruct the amplitude from the $s$-channel absorptive part along lines of fixed $t$ and fixed $u$ in the Mandelstam plane. For an arbitrary set of positive densities, however, the resulting amplitude need not be crossing symmetric. The null constraints \cite{Caron-Huot:2020cmc} follow from imposing crossing on these two families of lines,
\begin{equation}\label{eq:crossing_lines}
M(s,t)=M(t,s),\qquad M(s,4m_\pi^2-s-u)=M(4m_\pi^2-s-u,s),
\end{equation}
with the corresponding dispersion relations substituted on both sides. The result is a set of homogeneous linear constraints on the spectral densities $\rho_J$, in which the unknown constant $M_0$ cancels.

At fixed $t$, both sides of the first relation are given by the fixed-$t$ dispersion relation \eqref{eq:DR_fixed_t}, which is not manifestly symmetric under $s\leftrightarrow t$. We obtain
\begin{equation}\label{eq:null_F}
\big\langle \mathcal{F}_J(z;s,t)\big\rangle = 0,
\qquad \forall\,s,t,
\end{equation}
where
\begin{equation}\label{eq:Fdef}
\mathcal{F}_J(z;s,t)
\equiv
P_J(x_t)\,K_1(z;s,\sigma)
-
P_J(x_s)\,K_1(z;t,\sigma)
+
P_J(x_0)\left[
\frac{1}{z-t}-\frac{1}{z-s}
\right].
\end{equation}

At fixed $u$, the crossing transformation $s\leftrightarrow t$, with $t=4m_\pi^2-s-u$,
reflects the dispersion line onto itself. This symmetry was already used in deriving the fixed-$u$ dispersion relation \eqref{eq:DR_fixed_u}: the left-hand cut was folded onto the right-hand one, and the resulting fixed-$u$ representation is therefore manifestly symmetric under $s\leftrightarrow t$. Imposing crossing once again within this representation gives no additional constraint.
There is, however, a distinct consistency condition. The fixed-$t$ and fixed-$u$ dispersion relations provide two different dispersive reconstructions of the amplitude at the same point in the Mandelstam plane $(s,t)$.\footnote{The two representations converge in an open neighbourhood of the crossing symmetric point, which is all we need, but it would be interesting to characterize better the region that the two representation have in common.} For spectral densities that come from a crossing-symmetric amplitude, both reconstructions must return the same value
\begin{equation}
M_{\text{fixed-}t}(s,t)
=
M_{\text{fixed-}u}(s,t).
\end{equation}
Requiring their equality therefore gives a second family of null constraints
\begin{equation}\label{eq:null_G}
\big\langle \mathcal{G}_J(z;s,t)\big\rangle =0
\end{equation}
with
\begin{equation}\label{eq:Gdef}
\begin{aligned}
\mathcal{G}_J(z;s,t) \equiv\;&
P_J(x_t)\,K_1(z;s,\sigma)
-
(-1)^J P_J(x_u)\,
K_{\rm fold}(z;s,u,\sigma)
\\
&+
P_J(x_0)\Big[
K_1(z;t,\sigma)
-
K_1(z;s+t-\sigma,\sigma)
\Big].
\end{aligned}
\end{equation}

The two families are not independent. Since the fixed-$u$ representation is symmetric under $s\leftrightarrow t$, their antisymmetric parts satisfy
\begin{equation}\label{eq:G_minus_Gt}
\mathcal{G}_J(z;s,t)
-
\mathcal{G}_J(z;t,s)
=
\mathcal{F}_J(z;s,t),
\end{equation}
so the antisymmetric part of $\mathcal G$ reproduces the fixed-$t$ crossing constraint $\mathcal F$. The genuinely new content of \eqref{eq:null_G} is therefore its symmetric part, which is also the only place where the signature factor $(-1)^J$ survives. In the chiral limit the two families reduce to the null constraints of \cite{Albert:2022oes}. Ours is one choice among many: dispersion relations can be written along other lines through the crossing-symmetric point, for instance in $t$ along the line $s=u$, and crossing imposed on them \cite{Guerrieri:2023qbg}. We leave the exploration of these alternatives to future work.

Taylor expanding \eqref{eq:null_F} and the symmetric part of \eqref{eq:null_G} around the crossing-symmetric point turns each of them into a countable set of constraints. Let
\begin{equation}\label{eq:XY_def}
\mathcal{X}^{(n,k)}_J(z)
\equiv
\left.
\partial_s^n \partial_t^k\,
\mathcal{F}_J(z;s,t)
\right|_{s=t=\sigma},
\quad
\mathcal{Y}^{(n,k)}_J(z)
\equiv
\left.
\partial_s^n \partial_t^k\,
\Big[
\mathcal{G}_J(z;s,t)
+
\mathcal{G}_J(z;t,s)
\Big]
\right|_{s=t=\sigma}.
\end{equation}
Both $\mathcal{F}$ and the symmetric part of $\mathcal{G}$ vanish identically when either $s=\sigma$ or $t=\sigma$, so only $n,k\geq1$ contribute. Moreover,
\[
\mathcal{X}^{(n,k)}_J
=
-\mathcal{X}^{(k,n)}_J,
\qquad
\mathcal{Y}^{(n,k)}_J
=
\mathcal{Y}^{(k,n)}_J.
\]
An independent set of null constraints is therefore\footnote{At fixed total derivative order $N=n+k$,
the number of independent constraints is $N-1$. Indeed, for $N=2m$ the antisymmetric family contributes $m-1$ constraints and the symmetric family contributes $m$, while for $N=2m+1$ each family contributes $m$. Keeping all constraints with $n+k\leq n_{\max}$ therefore gives
$\sum_{N=2}^{n_{\max}}(N-1)
=
\frac{n_{\max}(n_{\max}-1)}{2}$
independent null constraints in total. 
Amusingly, the resulting closed-form expression can also be derived with a single prompt to your favourite large language model.
}
\begin{equation}\label{eq:null_XY}
\big\langle \mathcal{X}^{(n,k)}_J\big\rangle = 0,
\quad n>k\geq1,
\qquad\qquad
\big\langle \mathcal{Y}^{(n,k)}_J\big\rangle = 0,
\quad n\geq k\geq1.
\end{equation}

\subsection{Dispersive sum rules for the observables}\label{subsec:observables}

Expanding the fixed-$t$ dispersion relation \eqref{eq:DR_fixed_t} around the crossing-symmetric point and comparing with the Taylor expansion of the amplitude \eqref{eq:Mlow} expresses the $g_{n,k}$ coefficients as linear functionals \eqref{eq:average} of the spectral densities. 
We find that for the observables studied in the next section $g_{n,k}=\big\langle K_{n,k}\big\rangle$ with the kernels\footnote{The sum rules are easy to derive pen and paper. The $s$ dependence sits entirely in $K_1(z,s;\sigma)$, the $t$ dependence in $P_J(x_t)$ and $K_1(z,t;\sigma)$, and
\begin{equation}\label{eq:derivatives}
\partial_{s}^{\,n} K_1(z,s;\sigma)\Big|_{s=\sigma} = \frac{n!}{(z-\sigma)^{n+1}}, \qquad
\partial_{t}^{\,k} P_J\big(x_t(z)\big)\Big|_{t=\sigma} = \Big(\frac{2}{z-4m_\pi^2}\Big)^{k} P_J^{(k)}\big(x_0(z)\big).\nonumber
\end{equation}} 
\begin{equation}\label{eq:kernels}
K_{1,0}(z,J) = \frac{P_J(x_0)}{(z-\sigma)^2}, \qquad
K_{2,0}(z,J) = \frac{P_J(x_0)}{(z-\sigma)^3}, \qquad
K_{2,1}(z,J) = \frac{P_J'(x_0)}{(z-4m_\pi^2)(z-\sigma)^2}.
\end{equation}
In the chiral limit $\sigma\to0$, $x_0\to1$, and using $P_J(1)=1$, $P_J'(1)=\tfrac{J(J+1)}{2}$ they reduce to the massless sum rules of \cite{Albert:2022oes}.

Two simple consequences follow from the positivity of the kernels for $z>4m_\pi^2$. First, $g_{1,0}>0$ for any nonvanishing density
; section~\ref{app:Bound_state} shows that this remains true for a bound state. Second, since $K_{2,0}/K_{1,0}=1/(z-\sigma)$ is bounded by $1/(\Lambda^2-\sigma)$ on the support of the densities and $K_{2,1}\geq0$,
\begin{equation}\label{eq:simple_bounds}
0\leq \tilde g_2 \leq \frac{\Lambda^2}{\Lambda^2-\tfrac{4}{3} m_\pi^2},\qquad \tilde g_2'\geq 0 .
\end{equation}
The upper bound is saturated by the exchange of a scalar of mass $\Lambda$, and reduces to $\tilde g_2\leq1$ in the chiral limit \cite{Albert:2022oes, Fernandez:2022kzi}. These bounds do not use the null constraints: they delimit the region of figure~\ref{fig:g20g21nmax}, whose remaining boundary requires the full Bootstrap machinery.

\paragraph{Resonance couplings.}\label{subsubsec:couplings}
To extract the resonance couplings studied in section~\ref{sec:bounds} we resolve the two lightest states explicitly. The spectral density contains explicitly a spin-1 meson of mass $m_\rho$ and a spin-2 meson of mass $m_{f_2}$, with couplings normalized as in \eqref{eq:coupling_def}, and a generic positive density above a cutoff $\Lambda>m_{f_2}$:
\begin{equation}\label{eq:Im_rho_f2}
\begin{split}
\mathrm{Im}\,M(z,t) =\ & \pi\, g^2_{\pi\pi\rho}\, m_\rho^2\, \delta(z-m_\rho^2)\, P_1(x_t(z)) + \pi\, g^2_{\pi\pi f_2}\, m_{f_2}^2\, \delta(z-m_{f_2}^2)\, P_2(x_t(z)) \\
+\ & 16\pi\sum_{J} (2J+1)\, \rho_J(z)\, P_J(x_t(z))\, \theta(z-\Lambda^2).
\end{split}
\end{equation}
Every linear functional then acquires the two pole terms, evaluated at $(z,J)=(m_\rho^2,1)$ and $(m_{f_2}^2,2)$,
\begin{equation}\label{eq:average_poles}
\langle\, h\, \rangle' \equiv g^2_{\pi\pi\rho}\, m_\rho^2\, h(m_\rho^2,1)+g^2_{\pi\pi f_2}\, m_{f_2}^2\, h(m_{f_2}^2,2)+\sum_{J}\int_{\Lambda^2}^{\infty} dz\, 16(2J+1)\,\rho_J(z)\, h(z,J),
\end{equation}
so that the null constraints become
\begin{equation}\label{eq:null_const_rhof2}
\big\langle\mathcal{X}^{(n,k)}_J\big\rangle'=0,\qquad \big\langle\mathcal{Y}^{(n,k)}_J\big\rangle'=0,
\end{equation}
and the low-energy coefficients $g_{n,k}=\big\langle K_{n,k}\big\rangle'$.
The density is positive on $z\geq\Lambda^2$, and the two explicit couplings enter as parameters, non-negative when the corresponding states lie above the two-pion threshold (see section~\ref{app:Bound_state} otherwise).

\section{Bounds and Amplitudes for the scattering of Pions}\label{sec:bounds}

Once we have derived dispersion relations, null constraints, and sum-rules for the physical observables we are interested in, what remains is to formulate the corresponding dual problem and solve it! 
To put it in optimization dialect: the once-subtracted dispersion relations in Section~\ref{sec:dispersion} express $M(s,t)-M_0$ as a linear functional of the spectral densities $\rho_J(z)$, $z\geq\Lambda^2$; the Taylor coefficients $g_{n,k}$ of \eqref{eq:Mlow} become linear functionals of the $\rho_J$; crossing symmetry, which is not fully manifest in dispersive representations, becomes an infinite set of homogeneous linear constraints on the $\rho_J$, the null constraints of \cite{Caron-Huot:2020cmc}. 
The space of amplitudes is thus the set of positive measures $\{\rho_J\}$ satisfying the null constraints. It is convex, and so is its image in the space of the $g_{n,k}$. 

The precise dual formulation of our problems is derived in Appendix~\ref{sec:dual}. We explore two classes of problems.
First, we study the allowed region of the normalized Taylor coefficients
$\tilde g_2$ and $\tilde g_2'$ \eqref{eq:g2eg2prime} and extract the corresponding extremal amplitudes.
Then, we bound the normalized
$\pi\pi f_2$ coupling \eqref{eq:f2coupling}. 
We examine how these bounds and the spectra depend on the pion mass. The numerical
implementation is described in detail in Appendix~\ref{sec:numerics}.

The value of $m_\pi$ is related to the masses of the quarks in the QCD Lagrangian. At low energy, it introduces a new scale in the Bootstrap problem. The bounds in this section depend parametrically on this scale. 

\subsection{The \texorpdfstring{$\tilde g_2'$--$\tilde g_2$}{g2'-g2} region}
\label{subsec:g20_g21_plane}

\begin{figure}[t!]
    \centering
    \includegraphics[width=0.8\linewidth]{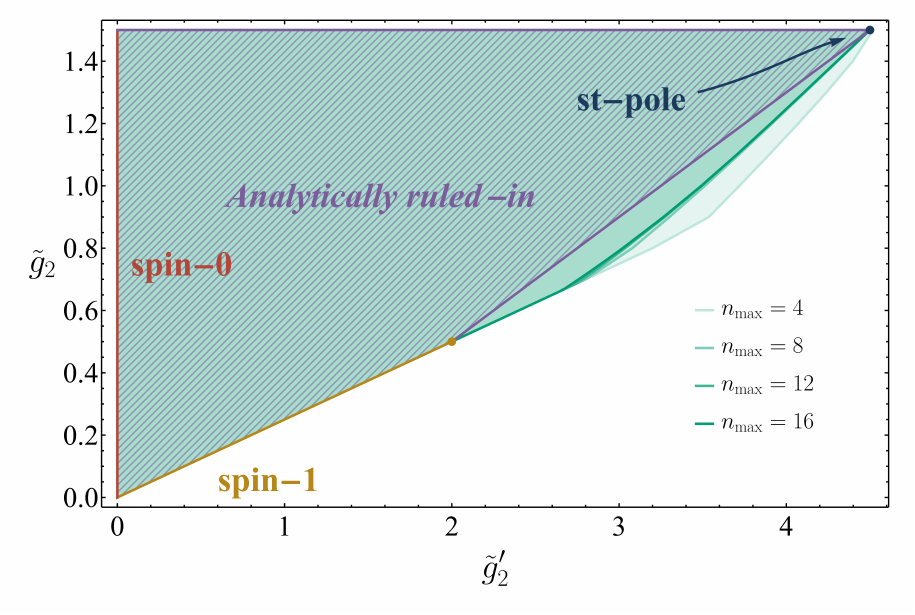}
    \caption{Exclusion plot in the $(\tilde g_2',\tilde g_2)$ plane
    for $m_\pi^2=1/4$ and $n_{\max}=4,8,12,16$, in units
    $\Lambda=1$. The yellow line marks the $\tilde g_2'=4\tilde g_2$ Skyrme line, which ends at the point $(2,1/2)$, given by
    \eqref{eq:spin1_point} with a vector resonance of mass  $m=1$. The red boundary is saturated by the exchange of a spin zero state, while the purple region is given by a convex combination of a spin-$0$ exchange \eqref{eq:scalar_exchange}, a spin-$1$ exchange \eqref{eq:vector_uv_completion} and the st-pole amplitude \eqref{eq:st_pole}.}
    \label{fig:g20g21nmax}
\end{figure}

For this problem, to set the units we choose an arbitrary cutoff $\Lambda$, writing $\Lambda=1$, 
and vary $m_\pi^2$.
If $m^2_\pi>1/4$, the cutoff goes below threshold and the discontinuity in the interval $(1,4m^2_\pi )$ is not necessarily positive. Hence, we always consider $0\leq m^2_\pi\leq 1/4$.
Figure~\ref{fig:g20g21nmax} shows the regions not ruled out at increasing $n_{\max}$ for the maximal value $m^2_\pi=1/4$, where the
two-pion threshold coincides with the cutoff. 
The bounds improve as
more null constraints are imposed and display the same qualitative
features as in the chiral limit case studied in ~\cite{Albert:2022oes, Fernandez:2022kzi}: a straight segment
on the right boundary ends at a kink, whose neighbourhood converges most
slowly with $n_{\max}$.

\paragraph{Analytic bounds.}
The left and top edges follow from the positivity bounds
\eqref{eq:simple_bounds} setting $\Lambda=1$,
\[
    \tilde g_2'\geq0,\qquad
    \tilde g_2\leq \frac{1}{1-\tfrac{4}{3}m^2_\pi}.
\]
The upper bound on $\tilde g_2$ diverges when $m_\pi^2=3/4$, which is outside the physical domain.\footnote{This divergence follows from the definition of Wilson coefficients in \eqref{eq:kernels}: all sum rules diverge when $\Lambda^2=4/3 m_\pi^2$. Had we chosen a different expansion point this divergence would have then shown up for some other value of $m_\pi^2$, even physical ones. For instance, if the expansion point in $s$ were $\sigma=2m_\pi^2-t_0/2$ and $0<t_0<4m_\pi^2$, then the bound would have diverged at $\Lambda^2=2m_\pi^2-t_0/2$.}
For $m^2_\pi=1/4$ used in Figure~\ref{fig:g20g21nmax}, the upper bound is $3/2$. The yellow line in Figure~\ref{fig:g20g21nmax} $\tilde g_2=\tilde g^\prime_2/4$ is fixed by
the lowest null constraint. Indeed, evaluating \eqref{eq:XY_def} at $n=k=1$ gives
\begin{equation}\label{eq:Y11}
    \mathcal Y^{(1,1)}_J(z)
    =4K_{2,1}(z,J)-4\big[1-(-1)^J\big]K_{2,0}(z,J).
\end{equation}
Then, taking the positive average $\langle  \mathcal Y^{(1,1)}\rangle=0$, and recalling the definitions of $g_{2,0}$ and $g_{2,1}$ \eqref{eq:kernels}, it yields
\begin{equation}\label{eq:rho_line}
    g_{2,1}=2\langle K_{2,0}\rangle_{\mathrm{odd}\ J}
    \leq2g_{2,0},\qquad \text{or}\qquad \tilde g_2'\leq4\tilde g_2.
\end{equation}
Equality requires the even-spin contributions to $g_{2,0}$ to
vanish. The inequality is
independent of the pion mass. Higher null constraints, however,
restrict how far the straight segment extends. In the numerical regions
shown, this line at each fixed $n_\text{max}$ ends at a kink whose position moves towards the left as
$n_{\max}$ increases.

This line has the same EFT interpretation as in the chiral
limit. The four-derivative term of the Skyrme model corresponds to
$L_3'=-L_4'=-1/(16e^2)$~\cite{Zahed:1986qz}. Equation~\eqref{eq:c_vs_L}
then gives $c_{2,0}=2L_4'$ and $c_{1,1}=4L_4'$. The pion-mass term changes
$c_{1,0}$ and $c_{0,0}$ but leaves the relation between the quadratic
coefficients unchanged at this order. We therefore continue to call
$\tilde g_2=\tilde g'_2/4$ the Skyrme line. 
Our bound applies to the exact
Taylor coefficients; its interpretation as a bound on the normalized
Skyrme coupling uses the EFT matching, which receives higher-order
corrections at nonzero pion mass as discussed in section \ref{subsec:low_energy_observables}.

Most of the boundary of the allowed region is saturated by simple meromorphic amplitudes belonging to the same three families found in the massless case \cite{Albert:2022oes}. Part of this similarity has a simple origin. In appendix~\ref{app:skyrme_endpoint} we show that every admissible amplitude with pion mass $m_\pi$ and gap $\Lambda$ is mapped into an admissible amplitude with any smaller pion mass, in particular the massless one, and shifted gap $\Lambda^2-\sigma$, with the same Taylor coefficients. Every bound obtained at a lighter mass, and in particular every massless bound, therefore implies a bound at heavier masses after rescaling, at any finite $n_{\max}$. Where a massive amplitude saturates the rescaled bound, as for the endpoint of the Skyrme segment, the massive boundary coincides with the rescaled massless one; elsewhere, for instance along the $st$-pole family, the two differ.

\paragraph{Spin-$0$ exchange amplitude.}
The red boundary $\tilde g_2^\prime=0$ in Figure~\ref{fig:g20g21nmax} is saturated by
a scalar
exchange with mass $m\geq1$ 
\begin{equation}\label{eq:scalar_exchange}
    M_{\mathrm{spin\text{-}0}}(s,t)
    =\frac{1}{1-s/m^2}+\frac{1}{1-t/m^2}.
\end{equation}
The ratio $s/m^2$ does not depend on the value of the cutoff $\Lambda$, and therefore we can think of this amplitude as universal or \emph{scale independent} -- for an extended discussion of this definition, see \cite{us_to_appear}.
It is manifestly crossing symmetric, it has residue $g^2_{\pi\pi0}=1$ at $s=m^2$ and
$J=0$, and satisfies the Regge behaviour imposed. 
This amplitude has
\begin{equation}\label{eq:spin0_point}
    \tilde g_2=\frac{1}{m^2-\tfrac{4}{3}m^2_\pi},\qquad
    \tilde g_2'=0.
\end{equation}
Varying $m$ from one to infinity covers the left edge of figure \ref{fig:g20g21nmax}, with the
origin reached as a decoupling limit $m\to \infty$.

\paragraph{Spin-$1$ exchange amplitude.}
The second family comprises a single-vector exchange
\[
    M_{\mathrm{spin\text{-}1}}(s,t)
    =
      \frac{x_t(m^2)}{1-s/m^2}+\frac{x_s(m^2)}{1-t/m^2}
    ,
\]
with $x_t(m^2)=1+2t/(m^2-4m^2_\pi)$ and similarly for $x_s(m^2)$, and with
\begin{equation}\label{eq:spin1_point}
    \tilde g_2=\frac{1/3
    }{m^2-\tfrac{4}{3}m_\pi^2},\qquad
    \tilde g_2'=\frac{4/3}{m^2-\tfrac{4}{3}m_\pi^2}.
\end{equation}
These values of $\tilde g_2$ and $\tilde g_2^\prime$ are extremal and parametrize the yellow line in Figure~\ref{fig:g20g21nmax} as function of $m$. However, the exchange of an isolated spin-$1$ resonance grows as $s$ at fixed $t$ and it is marginally excluded by the Regge behaviour imposed by the single-subtracted dispersion relations. As noticed in \cite{Albert:2022oes, Fernandez:2022kzi}, the simplest way out is to introduce an appropriate dressing factors.
For $m^2>4m_\pi^2$ and $M_\infty>m$, consider
\begin{equation}\label{eq:vector_uv_completion}
    M_{\mathrm{V,UV}}(s,t)
    =
      \frac{x_t(m^2)}{1- s/m^2}\frac{M_\infty^2}{M_\infty^2-t}
      +\frac{x_s(m^2)}{1- t/m^2}\frac{M_\infty^2}{M_\infty^2-s}
    .
\end{equation}
It has positive residue and satisfies the Regge bound for any $M_\infty$.\footnote{We invite the reader to ask their favourite LLM model to show positivity of the residues of \ref{eq:vector_uv_completion} for any $M_\infty$. The proof produced is simple and instructive.} Its low-energy
coefficients approach \eqref{eq:spin1_point} as $M_\infty\to\infty$.

In Appendix~\ref{app:skyrme_endpoint}, we prove that the Skyrme line ends exactly at
\begin{equation}\label{eq:massive_skyrme_endpoint_bound}
    \tilde g_2\leq
    \frac{1}{3-4m_\pi^2},
    \qquad
    0\leq m_\pi^2\leq\frac14,
\end{equation}
as $n_\text{max} \to \infty$, i.e. when the vector in \eqref{eq:vector_uv_completion} has mass 1, analogously to the massless framework \cite{Fernandez:2022kzi}.
At $m_\pi^2=1/4$, this gives the endpoint $(2,1/2)$ marked in
Figure~\ref{fig:g20g21nmax}.

\begin{figure}[t!]
    \centering
    \includegraphics[width=0.8\linewidth]{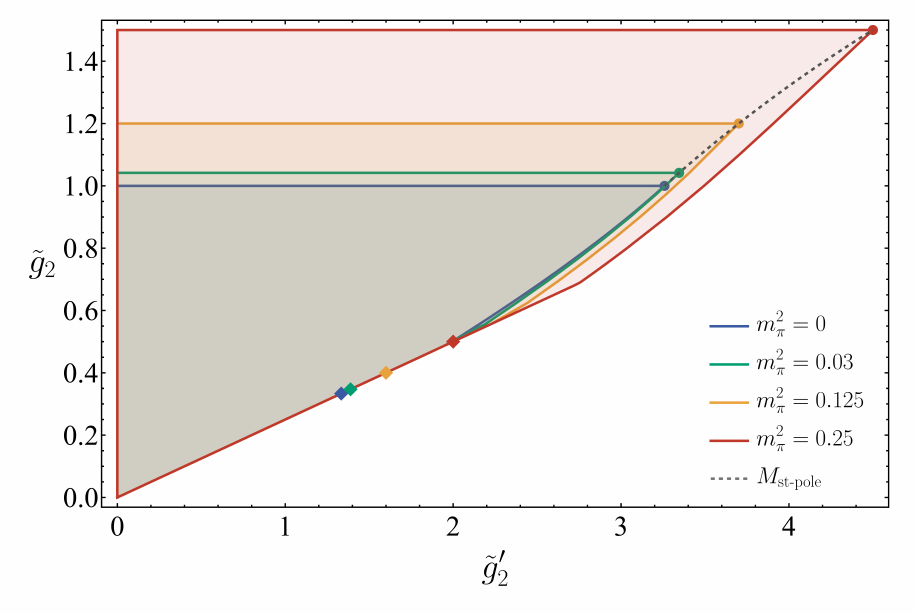}
    \caption{Allowed region at $n_{\max}=16$ for
    $m^2_\pi=0$, $0.03$ (physical), $0.125$ and $0.25$. The dashed curve is the $st$-pole point
    \eqref{eq:stpole_point} at $m=1$ as function of $m_\pi$. The diamonds track the end of the Skyrme line as function of $m_\pi$.}
    \label{fig:g20g21masses}
\end{figure}

\paragraph{The $st$-pole amplitude and convex combinations.}
The last family of analytic amplitudes is given by convex combinations of the spin-$0$ and spin-$1$ exchanges with a peculiar amplitude that has poles both in the $s$ and $t$ channels
\begin{equation}\label{eq:st_pole}
\begin{aligned}
    M_{st\text{-pole}}(s,t)
    &=\frac{1}{(1-s/m^2)(1- t/m^2)}
      -\alpha(m)M_{\mathrm{spin\text{-}0}}(s,t),\\
      \alpha(m)
    &=\frac{1}{1-4m^2_\pi/m^2}
      \log(2-4m^2_\pi/m^2).
\end{aligned}
\end{equation}
The subtraction removes the spin-zero pole and is the largest scalar
subtraction compatible with positivity. For $m^2>4m_\pi^2$, the partial-wave coefficients are positive,
proportional to
$Q_J\big(1+2/(1-4m_\pi^2/m^2)\big)>0$. In this case, we have
\begin{equation}\label{eq:stpole_point}
\begin{aligned}
    \tilde g_2&=\frac{1}{m^2-\tfrac{4}{3}m_\pi^2},\\
    \tilde g_2'
    &=\frac{m^2-4m_\pi^2}{
    (m^2-\tfrac{4}{3}m_\pi^2)
    \left[m^2-4m_\pi^2-
    (m^2-\tfrac{4}{3}m_\pi^2)\log(2-4m_\pi^2/m^2)\right]}.
\end{aligned}
\end{equation}
At $m=1$ and $m^2_\pi=0$ this gives
$(\tilde g_2',\tilde g_2)=(1/(1-\log2),1)$, as
in~\cite{Albert:2022oes}. For nonzero pion mass, setting $m=1$
gives the corresponding point $(9/2,3/2)$ on the top right edge of Figure \ref{fig:g20g21nmax} for $m_\pi^2=1/4$ and for generic values of $m_\pi^2$ as in Figure \ref{fig:g20g21masses}. 
Sending $m\to\infty$ the parametric curve $(\tilde g_2^\prime, \tilde g_2)$ traces a line that connects this vertex to the origin.

Finally, we can rule-in the extremal amplitudes taking convex combinations of the $st$-pole amplitude \eqref{eq:stpole_point} with the scalar and vector exchanges.
We explicitly show only the combination that saturates the upper horizontal purple edge of Figure~\ref{fig:g20g21nmax} 
\begin{equation}
\gamma M_{st\text{-pole}}( s, t) +(1-\gamma)M_{\mathrm{spin\text{-}0}}(s, t), \qquad \gamma=\frac{\tilde g_{2}^\prime}{\tilde g^\prime_2[M_{st\text{-pole}}]} 
\end{equation}
with fixed $m=1$, and $\tilde g^\prime_2[M_{st\text{-pole}}]$ given by equation \eqref{eq:stpole_point}.

\paragraph{Pion mass dependence.}
Figure~\ref{fig:g20g21masses} compares the allowed regions in the $\{\tilde g_2,\tilde g_2'\}$ plane as function of 
$m_\pi^2=0$, $0.03$, $0.125$ and $0.25$. The value $m_\pi^2=0.03$ is
approximately physical if we identify $\Lambda$  with the $\rho$ mass.
The displayed regions expand as the pion mass increases. Their top
edges are given by $(1-\tfrac{4}{3}m_\pi^2)^{-1}$.
 The growth of the allowed region can be accounted for by the decreasing distance $m^2-4m_\pi^2/3$ from the point where all sum rules explode.
Changing the pion mass does not amount to a universal rescaling: for instance, the $st$-pole
coordinate $\tilde g_2'$ contains a pion mass-dependent logarithm.
In contrast with the massless case, also higher-spin contributions introduce a non linear dependence on $m_\pi^2$.
To make this point explicit, the Taylor coefficients of a
crossing-symmetrized isolated spin-$J$ exchange give
\begin{equation}\label{eq:coeff_wilson_spin}
\begin{aligned}
    \tilde g_2[J]&=
    \frac{3[P_J(z_0)+2z_0^2P_J''(z_0)]}{
    (3m^2-4m_\pi^2)[P_J(z_0)+2z_0P_J'(z_0)]},\\
    \tilde g_2'[J]&=
    \frac{4 P_J'(z_0)}{
    (m^2-4m_\pi^2)[P_J(z_0)+2z_0P_J'(z_0)]},\\
\end{aligned}
\end{equation}
where $z_0=x_0(m^2)
    =\frac{3m^2-4 m_\pi^2}{3(m^2-4 m_\pi^2)}>1$.
For $J=0,1$ these expressions reproduce
\eqref{eq:spin0_point} and \eqref{eq:spin1_point}. 
Their corresponding ratios $\tilde g_2^\prime[J]/\tilde g_2[J]$ are constants and 
parametrize straight lines as function of $m$ at fixed $m_\pi^2$. But already starting at $J=2$ 
the same ratio varies nonlinearly with the pion mass:
\begin{equation}\label{eq:J2_ratio}
    \left.\frac{\tilde g_2'[J]}{\tilde g_2[J]}\right|_{J=2}
    =\frac{4(3m^2-4 m_\pi^2)^2}{
           21m^4-48 m^2 m_\pi^2+16 m_\pi^4}.
\end{equation}
These isolated exchanges illustrate the kinematic dependence of our observables, but they cannot be promoted to consistent amplitudes; for $J\geq1$ their growth at fixed $t$ violates our Regge assumptions in 
\eqref{eq:regge}. Therefore, the simple power-law scaling observed in the massless framework \cite{Berman:2024eid} holds only for scalars and vectors, but not for higher spins.

\begin{figure}[t!]
    \centering
    \includegraphics[width=0.85\linewidth]{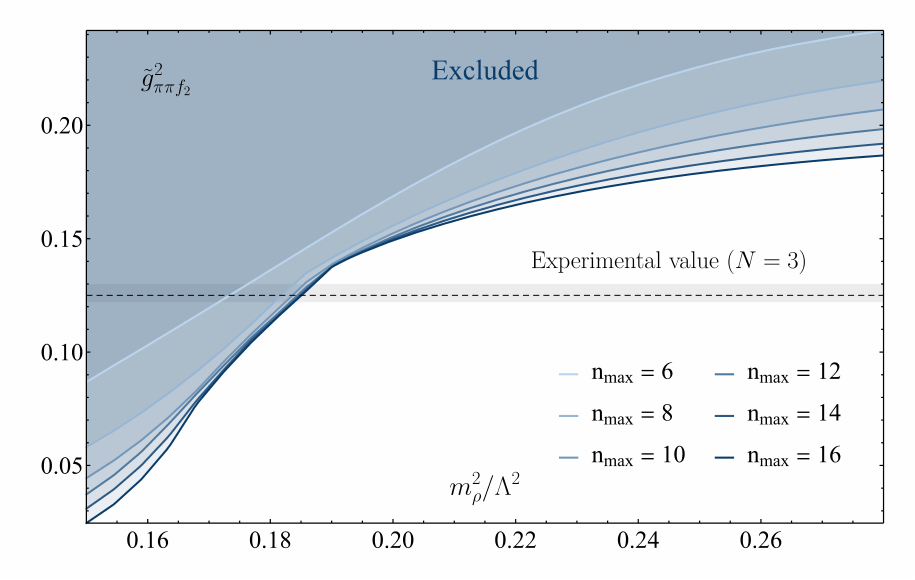}
    \caption{Upper bound on $\tilde g^2_{\pi\pi f_2}$ at the physical mass $m_\pi^2=0.03$, as a function of $m_\rho^2/\Lambda^2$, for increasing values of $n_{\max}$. The blue shaded area is excluded. The dashed line is the experimental value at $N=3$, the gray shaded width is the error uncertainty.}
    \label{fig:gpipi_nmax}
\end{figure}

\subsection{Bounds on the \texorpdfstring{$\pi\pi f_2$}{pi pi f2} coupling}\label{subsec:f2_coupling}

We now turn to the problem of maximizing the $\pi\pi f_2$ coupling in presence of a pion mass following \cite{Albert:2023seb}. 
The dual formulation of this problem is in Appendix~\ref{subsec:dual_coupling}.
We normalize the contribution of the $f_2$ pole by $g_{1,0}$ and define the quantity
\begin{equation}\label{eq:norm}
    \tilde{g}^2_{\pi\pi f_2}
    \equiv \frac{g_{\pi\pi f_2}^{2}\,m_{f_2}^2\,K_{1,0}(m_{f_2}^2,2)}{g_{1,0}}
    =\frac{g_{\pi\pi f_2}^{2}}{g_{1,0}}
    \frac{m_{f_2}^{2}\,P_2(x_0(m^2_{f_2}))}{\left(m_{f_2}^{2}-\tfrac{4}{3}m_\pi^{2}\right)^{2}},
\end{equation}
with $x_0=1+\tfrac{8m_\pi^2}{3\left(m_{f_2}^2-4m_\pi^2\right)}$.
This quantity is precisely the fraction of $g_{1,0}$ coming from the $f_2$ pole. For $m^2_{f_2}>4m_\pi^2$ it lies in the interval $0\leq \tilde g_{\pi\pi f_2}\leq 1$.

In the rest of this section, we work in units of the mass of the $\rho$ in the chiral limit, setting $m_{\rho,0}=1$.  
We then fix both the mass of the $\rho$ and $f_2$ in terms of the pion mass.
Their dependence on $m_\pi$ is our lattice input: $m_\rho$ is taken from \cite{Bonanno:2025hzr} and given explicitly in \eqref{eq:rho_mass}; 
$m_{f_2}$ instead is estimated in  \eqref{eq:f2_mass} using the large $N$ value of the string tension and $m_\rho$. The determination of the $f_2$ mass in the large $N$ limit is an interesting open problem and it would provide us a crucial missing input.
Finally, we maximize $\tilde g^2_{\pi\pi f_2}$ as in \eqref{eq:norm} as a function of the cutoff on the continuous spectral density $\Lambda>m_{f_2}$. 

We first consider the physical pion mass $m_\pi^2=0.03$.
Figure~\ref{fig:gpipi_nmax} shows the bound for $n_{\max}$ ranging from $6$ to $16$. 
For each fixed $\Lambda$, we bound the maximum value of $\tilde g_{\pi\pi f_2}^2$.
The curves at $n_{\max}=12,14,16$ 
converge faster in a region of the boundary that features the presence of a kink at
$m_\rho^2/\Lambda^2\simeq0.190$. Our kink lies to the left of the value $m_\rho^2/\Lambda^2\simeq0.2106$ in \cite{Albert:2023seb}: 
the difference is due to the masses used as inputs. 
The mass of the $f_2$ resonance we use in \eqref{eq:f2_mass} is derived from the large $N$ string tension of \cite{Bonanno:2025hzr}, assuming it does not depend on the pion mass. It predicts a larger $f_2$ mass than the finite $N=3$ input of \cite{Albert:2023seb}. The displacement therefore includes the effect of this choice, as well as that of the nonzero pion mass from the $m_\rho$ dependence in \eqref{eq:f2_mass}.

We compare the fraction of $g_{1,0}$ carried by the $f_2$ at the kink with the real-world value, extracted from the $f_2\to\pi\pi$ width as detailed in Appendix~\ref{app:exp}\footnote{The quoted uncertainty is experimental only; the leading-order identification $g_{1,0}=1/(2f_\pi^2)$ receives a relative correction $2\epsilon\simeq0.06$ at the physical point, estimated in Appendix~\ref{app:exp}.}
\begin{equation}
\label{eq:g2comparison}
\tilde g^{2\,({\rm exp})}_{\pi\pi f_2}(N=3)=0.125^{+0.005}_{-0.003}\,,\qquad
\tilde g^{2}_{\pi\pi f_2}\Big|_{\rm kink}\le 0.138\,.
\end{equation}
The real-world value lies about $10\%$ below the bound. This is expected: scalar exchanges contribute to $g_{1,0}$ but not to the null constraints, and the extremal spectrum contains no daughter trajectories. A similar gap was found in the chiral limit \cite{Albert:2023seb}.

\begin{figure}[t!]
    \centering
    \includegraphics[width=\linewidth]{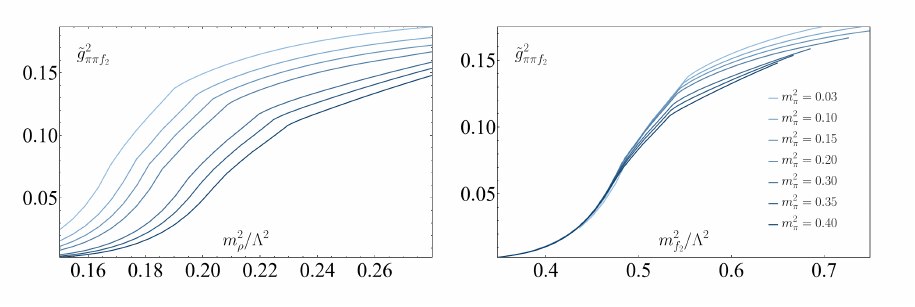}
    \caption{Upper bound on $\tilde g^2_{\pi\pi f_2}$ at $n_{\max}=16$ for $m_\pi^2=0.03$, $0.10$, $0.15$, $0.20$, $0.30$, $0.35$ and $0.40$, as a function of $m_\rho^2/\Lambda^2$ (left) and $m_{f_2}^2/\Lambda^2$ (right). The two largest pion masses have a bound-state $\rho$ and use the relaxed residue-sign condition of section~\ref{app:Bound_state}.}
    \label{fig:gpipi_masses}
\end{figure}

\paragraph{Dependence on the pion mass.}
Figure~\ref{fig:gpipi_masses}, left panel, shows the bound at $n_{\max}=16$ for different values of the pion mass $m_\pi^2$ between $0.03$ and $0.40$. For the two largest values used, the $\rho$ lies below the two-pion threshold, and becomes a genuine bound state. For those points we use the formulation of Appendix~\ref{app:Bound_state}, with the $\rho$-residue sign unrestricted, that give valid upper bounds on the coupling of the $f_2$ resonance even in that case. 
In figure~\ref{fig:gpipi_masses}, right panel, we show the same plot as function of $m^2_{f_2}/\Lambda^2$ instead.
This brings the curves closer together, especially for $m_{f_2}^2/\Lambda^2\lesssim0.45$, at fixed $n_\text{max}$. Their separation becomes more pronounced toward the upper end of the displayed range.

\begin{figure}[t!]
    \centering
    \begin{minipage}{0.495\linewidth}
        \centering
        \includegraphics[width=\linewidth]{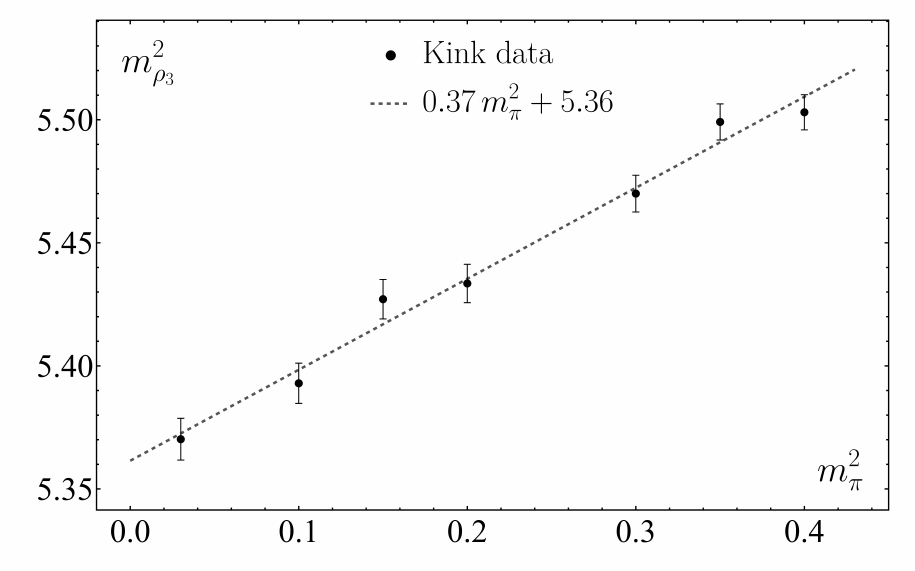}
    \end{minipage}
    \hfill
    \begin{minipage}{0.495\linewidth}
        \centering
        \includegraphics[width=\linewidth]{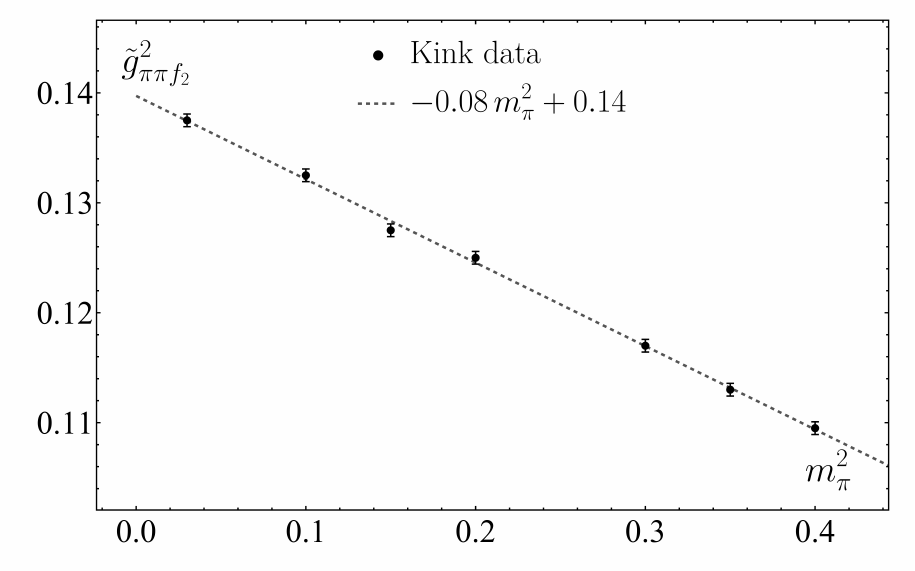}
    \end{minipage}
    \caption{
    Properties of the kink as functions of $m_\pi^2$.
    \textit{Left:} Candidate spin-three squared mass
    $m_{\rho_3}^2$, identified with the cutoff at the kink.
    The dashed line is the fit \eqref{eq:rho3_mass}.
    \textit{Right:} Value of $\tilde g^2_{\pi\pi f_2}$ at the kink,
    with the linear fit \eqref{eq:kinkheight}.
    The uncertainties in both panels estimate the sampling resolution.
    }
    \label{fig:kinkfits}
\end{figure}

At each sampled pion mass, the dual functional at the kink has a spin-3 zero at the cutoff, $z=\Lambda^2$. Identifying this zero with the candidate $\rho_3$ state, a linear fit of the kink position, shown in Figure \ref{fig:kinkfits} left panel, gives
\begin{equation}\label{eq:rho3_mass}
 m_{\rho_3}^2
    =(5.361\pm0.007)+(0.37\pm0.02)\,
    m_\pi^2.
\end{equation}
Here the quoted errors describe the uncertainty of the determination of the kink position using a grid in $\Lambda$.\footnote{The error in $n_{\mathrm{max}}$ is neglegible for the spin 3. Since the observed spin-3 sits at the threshold $z=\Lambda$, in principle there is no error in the extraction, except the uncertainty coming from the grid in $\Lambda$ itself used to determine Figure \ref{fig:gpipi_masses}. 
} 
The fit \eqref{eq:rho3_mass} should not be read as a prediction for the $\rho_3$ mass, since it depends on the prescriptions \eqref{eq:rho_mass} and \eqref{eq:f2_mass} for the $\rho$ and $f_2$ masses, which are partly assumed. Rather, it is a proof of principle: given lattice determinations of the low-lying masses as functions of $m_\pi$, the bootstrap turns them into predictions for the pion-mass dependence of the higher states.

If all states moved up together with the $\rho$ and the $f_2$, as we have indeed assumed, the $\rho_3$ would stay on their straight line and follow
\begin{equation}\label{eq:rho3_rigid_shift}
m_{\rho_3}^2 \simeq 1+\frac{2}{\alpha'}+0.678\,m_\pi^2 \simeq 4.96+0.678\,m_\pi^2\,.
\end{equation}
The extremal solution differs in two ways. At $m_\pi=0$ the $\rho_3$ lies above this line, because the extremal trajectory bends towards larger masses, as already found in \cite{Albert:2023seb} and confirmed in Section~\ref{subsec:extremal_spectra}. As $m_\pi$ grows, the $\rho$ becomes heavier while the $\rho$--$f_2$ spacing is kept fixed, and this bending decreases. The $\rho_3$ therefore becomes heavier more slowly than the $\rho$ and the $f_2$, with a slope about half the input one -- see also Appendix~\ref{app:kink_spectrum_mpi}.

Over the same pion-mass range, a linear fit summarizes the decrease in the height of the kink, shown in the right panel of Figure~\ref{fig:kinkfits}
\begin{equation}\label{eq:kinkheight}
\tilde g^2_{\pi\pi f_2}\Big|_{\rm kink}\simeq 0.14-0.08\,m_\pi^2\,.
\end{equation}
Extrapolated to $m_\pi=0$, it gives $\simeq 0.14$, for the ratio $m_{f_2}^2/m_\rho^2\simeq 2.98$ which is close to the value $\simeq 0.147$ of \cite{Albert:2023seb} obtained with a smaller ratio $m_{f_2}^2/m_\rho^2\simeq 2.72$. Unlike the masses, the height of the kink is not a function of $m_{f_2}^2/m_\rho^2$ alone: at $m_\pi^2=0.20$, our input gives $m_{f_2}^2/m_\rho^2\simeq 2.74$, comparable to the chiral limit of \cite{Albert:2023seb}, but the max coupling at the kink is $\simeq 0.125$, $15\%$ lower.

\subsection{Extremal spectra at the kink}\label{subsec:extremal_spectra}

The zeros of the dual extremal functional identify the spectrum of the amplitudes saturating the bounds.\footnote{Technically speaking, this is the KKT complementarity: when the duality gap closes, either the dual functional vanishes or the constraints are saturated. In our case, a nonzero spectral density forces the dual functional to vanish at optimality.} We extract candidate masses from those zeros using the procedure in Appendix~\ref{subsec:spectrum-extraction}, quoting the values at $n_{\max}=16$ and comparing them with $n_{\max}=14$ to check convergence and estimate systematic errors. To establish true convergence, however, we should look at the stability of the residues too. Residues are extracted from the output of the SDPB \cite{Simmons-Duffin:2015qma}, the solver we use in practice for our optimization problems. We have checked that unlike the masses which converge quite fast in the number of null constraints, the residues are quite unstable.
This problem has been discussed in several works already \cite{Albert:2022oes,Albert:2026xyz,Caron-Huot:2021rmr,us_to_appear}. We will not discuss residues further here, but comment on possible strategies to solve it in Section \ref{sec:discussion}.

\begin{figure}[t!]
    \centering
    \begin{minipage}{0.49\linewidth}
        \centering
        \includegraphics[width=\linewidth]{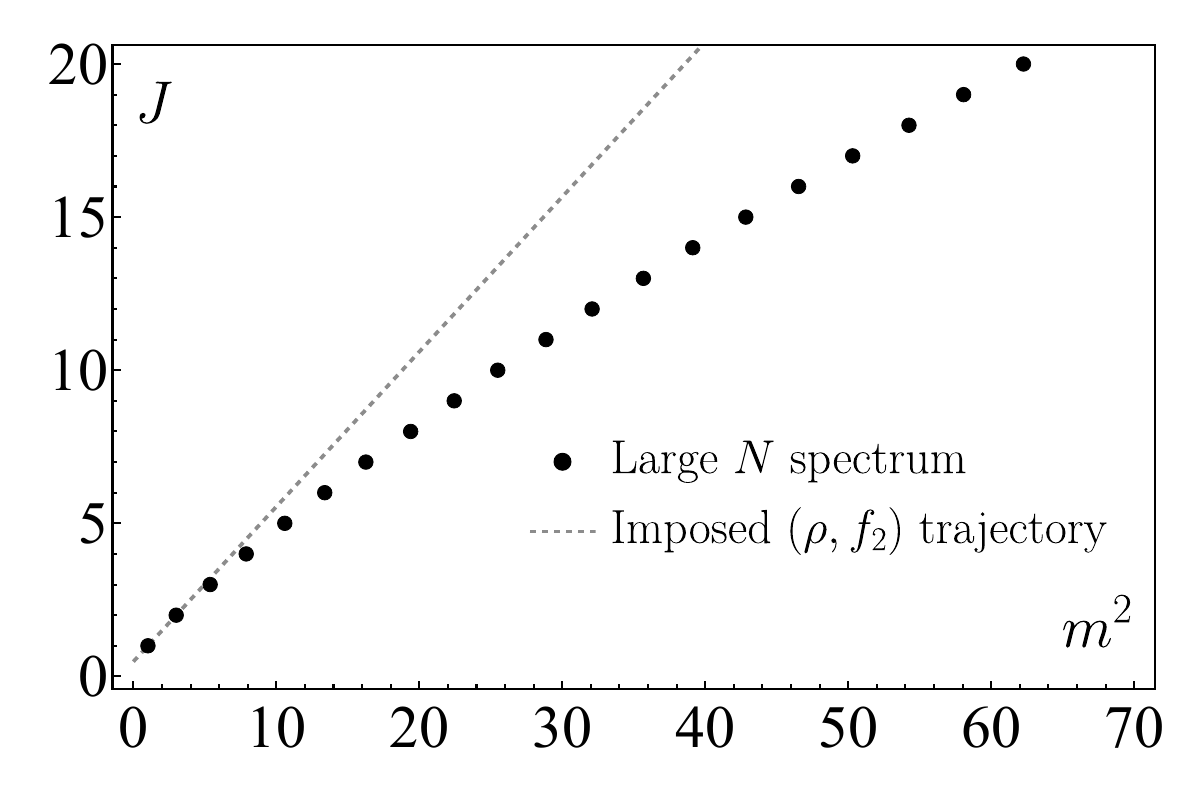}
    \end{minipage}
    \hfill
    \begin{minipage}{0.49\linewidth}
        \centering
        \includegraphics[width=\linewidth]{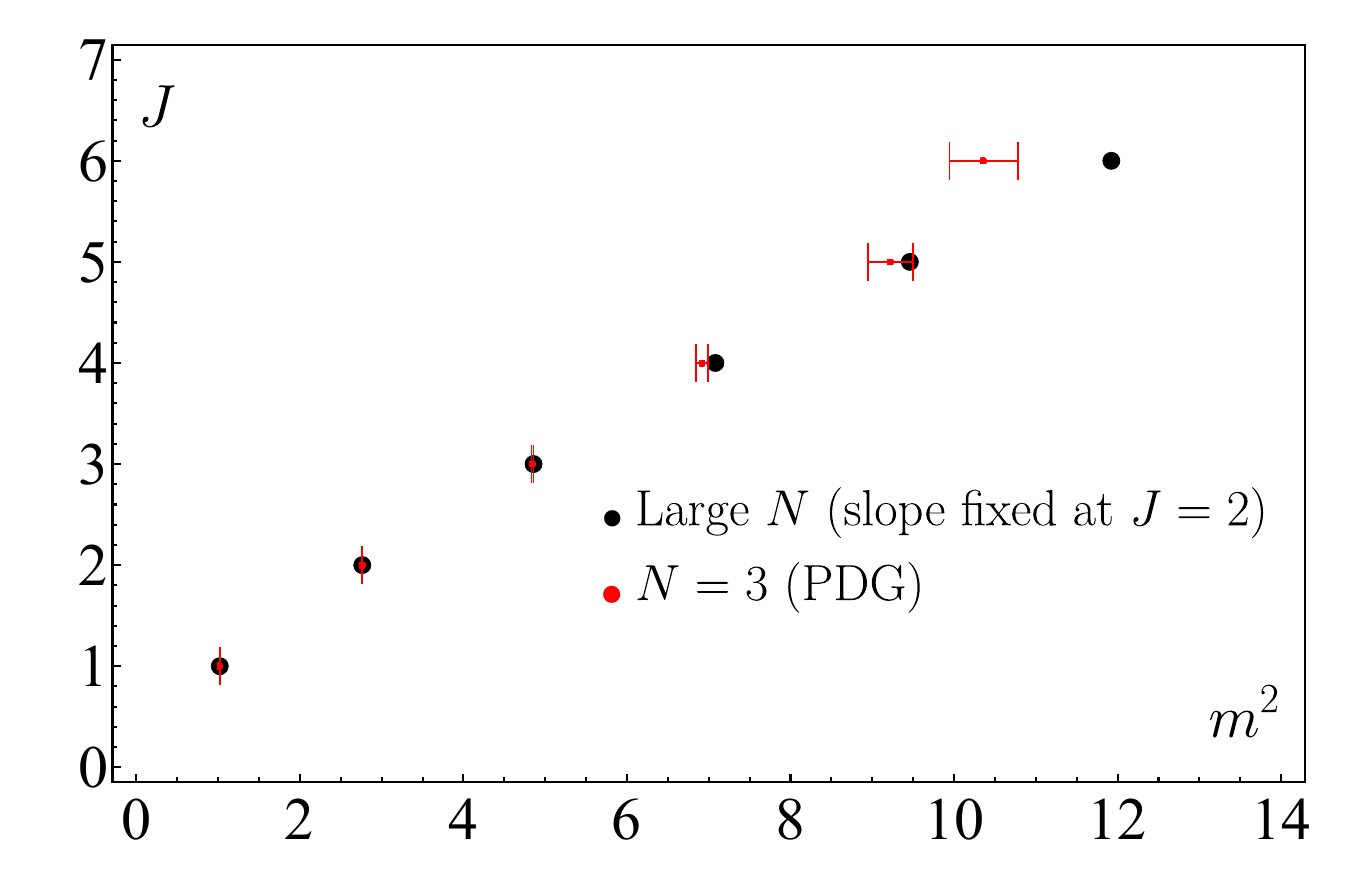}
    \end{minipage}
    \caption{
    Candidate spectrum at the kink for physical pion mass 
    $m_\pi^2/m_{\rho,0}^2=0.03$.
    \textit{Left:} Chew-Frautschi extracted from the extremal functional
    at $n_{\max}=16$.
    The dashed line is the input $\rho$--$f_2$ trajectory.
    \textit{Right:} Comparison with the experimental spectrum
    after the calibration \eqref{eq:spectrum_mass_calibration}.
    Black dots denote extremal-functional zeros, while 
    red errors bars denote the experimental states $\rho(770)$,
    $f_2(1270)$, $\rho_3(1690)$, $f_4(2050)$,
    $\rho_5(2350)$ and $f_6(2510)$, with their PDG mass
    uncertainties \cite{PDG2026}. The mass-squared spacing is rescaled by
    $\lambda\simeq0.88$ about the $\rho$ to align the $f_2$
    masses.
    }
    \label{fig:spectrum_at_the_kink}
\end{figure}

\paragraph{The physical pion mass.}
Figure~\ref{fig:spectrum_at_the_kink}, left panel, shows the candidate spectrum at the physical point $m_\pi^2=0.03$, together with the input $\rho$--$f_2$ trajectory denoted with a dashed line. The states with $J\geq3$ do not lie on the input trajectory, but bend to larger masses.
Their curvature and local slope in the $(m^2,J)$ plane are examined below. At this point, our spectrum qualitatively matches the one found in \cite{Albert:2023seb} for large $N$ pion amplitudes in the chiral limit, in supergravity \cite{Boisvert:2026sfh}, and for neutral Goldsone amplitudes \cite{Berman:2024eid, us_to_appear}: a single curved trajectory and no subleading trajectories, except for a number of spurious poles with small residues that we have already filtered out. The precise method of the extraction of the physical trajectory is discussed in appendix~\ref{subsec:spectrum-extraction}; in summary, we used the residues as a criterion to exclude unphysical poles to construct a reference trajectory, we have then looked directly at the dual functional to reconstruct the whole physical spectrum. We have then checked, \textit{a posteriori}, that forcing the spurious poles to disappear, by introducing a spin dependent cutoff, did not change the bound and the physical spectrum.

To compare our large $N$ spectrum with the experimental $N=3$, we calibrate the mass axis using the $\rho$ and the $f_2$. After expressing both spectra in common units of $m^2_{\rho,0}$, so that the $\rho$ mass at the physical point is fixed at $m_\rho^2\simeq 1.02$, we rescale the slope, using the spin-2 experimental mass:
\begin{equation}\label{eq:spectrum_mass_calibration}
    m^2\longmapsto m_\rho^2+\lambda\left(m^2-m_\rho^2\right),
    \qquad
    \lambda=\frac{m_{f_2,\mathrm{exp}}^2-m_\rho^2}
    {m_{f_2,\mathrm{input}}^2-m_\rho^2}\simeq0.88.
\end{equation}
This fixes the spacing between the $\rho$ and $f_2$ masses to be the same and coincide with their experimental one. Figure~\ref{fig:spectrum_at_the_kink}, right panel, then compares the remaining masses at $J\geq3$. 
This agreement should not be regarded as an independent test. As shown in appendix~\ref{app:kink_spectrum_mpi}, at fixed $r\equiv m_{f_2}^2/m_\rho^2$ the spectrum at the kink is approximately independent of the pion mass and coincides with the chiral-limit spectrum of \cite{Albert:2023seb}. The calibration \eqref{eq:spectrum_mass_calibration} moves $r$ close to its experimental value, 
so that the comparison above essentially reproduces the agreement already observed in \cite{Albert:2023seb}.

\paragraph{Dependence on the pion mass.}
Next, we study the dependence of the spectrum at the kink as function of the pion mass. To do so, we determine the position of the kink from the fit in figure \ref{fig:kinkfits}, left panel, then run the coupling maximization problem at fixed pion mass using the corresponding value of the kink as cutoff.

\begin{figure}[t!]
    \centering
    \includegraphics[width=0.9\linewidth]{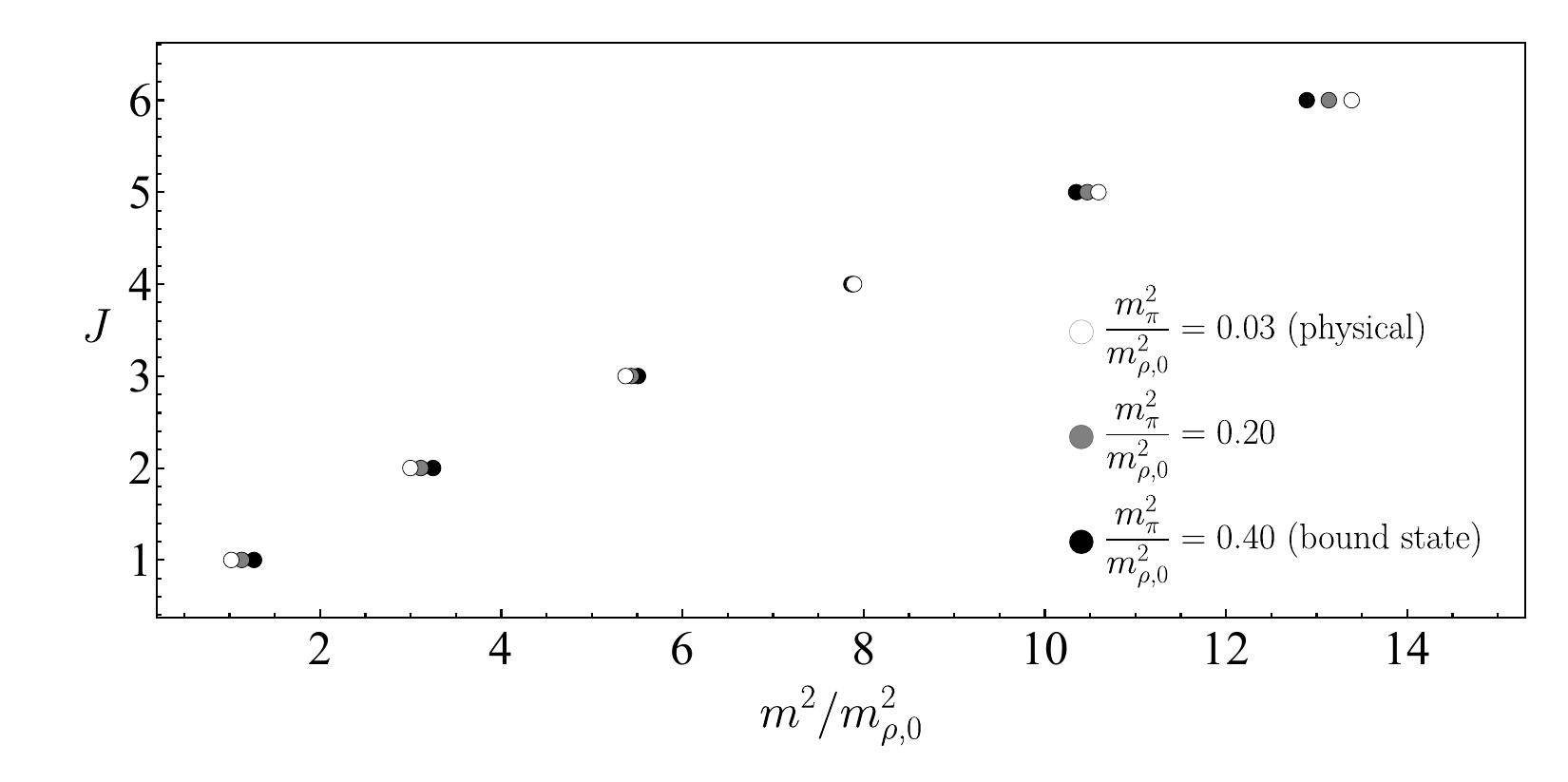}
    \caption{Candidate spectra at the kink for $m_\pi^2/m_{\rho,0}^2=0.03$, $0.20$ and $0.40$, in units of $m_{\rho,0}^2=1$. At the largest pion mass the $\rho$ is a bound state. The fitted masses with $J\geq4$ decrease as the pion mass increases.}
    \label{fig:spectrum_masses}
\end{figure}
 We use a simple fit function truncating the expansion at the leading order, inspired by the Lattice analysis \cite{Bonanno:2025hzr, Bonanno:2026zcq}\footnote{For the largest pion mass considered, the next order would give a correction of size $m_\pi^4/m_J^2 \sim 0.16/m_J^2$.}
\begin{equation}\label{eq:trajectory_pion_mass_fit}
    m_J^2=A_J+B_J x,
    \qquad x\equiv m_\pi^2.
\end{equation}
This fit model works extremely well for all resonances up to spin $J=40$.
The fits for the pion mass dependence of the resonances on the Regge trajectory are shown in fig~\ref{fig:quadratic_fits}, Appendix \ref{appsubsec:the_meson_trajectory}.

 We observe a peculiar behaviour. While the spin-three mass increases with the pion mass as the $\rho$ and $f_2$ do, for $J\geq4$ the masses decrease. Figure~\ref{fig:spectrum_masses} illustrates the trend. Its origin is kinematical rather than dynamical. The extremal problem depends only on $r=m^2_{f_2}/m^2_\rho$, on $m_\pi^2/m_\rho^2$ and on the cutoff, which is fixed by the kink; but in practice the dependence on $m_\pi^2/m_\rho^2$ is negligible for the spectrum. In fact, at $m_\pi^2/m_{\rho,0}^2=0.20$, i.e.\ $m_\pi\simeq 2.6\,m_\pi^{\rm (phys)}$, where our input gives $r\simeq2.74$, the spectrum at the kink agrees with the chiral-limit one of \cite{Albert:2023seb} at approximately the same  $r\simeq2.72$ -- see Appendix~\ref{app:kink_spectrum_mpi}. Along our input trajectory, $r$ decreases with $m_\pi$ by construction, and the lightening of the higher-spin states follows from this decrease. 
 The same is not true for the height of the kink: 
 the pion mass affects the bound but not the extremal spectrum.

\section{Glueball scattering}
\label{sec:glueballs}

The mass gap of pure Yang--Mills theory is set by the lightest glueball. At $N=3$, however, glueballs mix with flavour-singlet mesons, and their identification remains debated. Recent BESIII results support a glueball-dominated pseudoscalar, the $X(2370)$ \cite{BESIII:2026mvn}, while the assignment of the scalar glueball among its candidates 
is still open \cite{Chevalier:2026pwf}.
The planar limit, however, provides a clean laboratory to study Glueball properties since they decouple from the mesons and form a separate consistent sector \cite{tHooft:1973alw,Witten:1979kh}. With single-trace gluonic operators normalized like the action, so that $\mathcal{L}\sim N\,{\rm tr}(\dots)$, all connected correlators scale as $N^2$. After the LSZ reduction each of the four external legs is divided by the square root of the two-point function, a factor of order $N$, so the $2\to2$ glueball amplitude is of order $N^2/N^4=1/N^2$. Glueball scattering is therefore suppressed compared to meson scattering, which is of order $1/N$.

We focus on the scattering of the lightest scalar glueball, the $0^{++}$. Unlike the pion, it is not a Goldstone boson, and there is no separation of scales: nothing forbids a cubic self-coupling, so the lightest state exchanged in $2\to2$ scattering is the $0^{++}$ itself, with a pole at $s=m_{0^{++}}^2$. There is therefore no low-energy EFT with a small expansion parameter. We can nevertheless define observables directly from the amplitude, such as the on-shell couplings of the exchanged glueballs, or the Taylor coefficients around points away from the poles as in \eqref{eq:Mlow}, without relying on a truncated expansion. 
Non-perturbative bounds of this kind on glueball couplings were derived in \cite{Guerrieri:2023qbg} for SU(3) Yang--Mills theory, imposing full unitarity. Glueball--glueball scattering has also been studied on the lattice for SU(2) \cite{Yamanaka:2019yek} and in a dilaton effective model \cite{Giacosa:2021brl}.
Here we work at leading order in $1/N$, where the amplitude is meromorphic and positivity of the residues replaces unitarity.

The large $N$ glueball spectrum has been studied extensively on the lattice; see for instance \cite{Athenodorou:2021qvs, Lucini:2010nv, Lucini:2004my, Lucini:2001ej}. In the scattering of two identical $0^{++}$ glueballs, Bose symmetry restricts the exchanged states to even spin, and parity and charge conjugation then fix them to be $J^{++}$. We work in units of the lightest glueball mass, $m_{0^{++}}=1$, and use as input the large $N$ spectrum of~\cite{Athenodorou:2021qvs} up to $J=4$,
\begin{equation}
\label{eq:glueballspec}
m^2_{0^{++}} {=} 1, \quad
m^2_{2^{++}} {=} 2.241(25), \quad
m^2_{0^{++*}} {=} 3.62(7), \quad
m^2_{2^{++*}} {=} 4.591(65), \quad
m^2_{4^{++}} {=} 5.45(10).
\end{equation}
In the numerics we use the central values and neglect the uncertainties.

\subsection{Fully crossing symmetric amplitudes}

The amplitude $M(s,t,u)$, with $s+t+u=4$, is symmetric under all permutations of $s$, $t$ and $u$, and has poles in all three channels. A state $X$ of spin $J_X$ and mass $m_X$ contributes
\begin{equation}
\label{eq:glueballpole}
M(s,t,u)\supset g_X^2\,\frac{P_{J_X}\!\left(1+\frac{2t}{m_X^2-4}\right)}{m_X^2-s}+\text{crossed channels}\,,
\end{equation}
where, unlike in \eqref{eq:coupling_def}, we do not include a factor $m_X^2$ in the residue. Most of the properties of section~\ref{subsec:genprop} carry over, and we only stress the differences with pion scattering.

\paragraph{Poles below threshold.}
No symmetry forbids a cubic coupling of the lightest glueball, so the first singularity of the amplitude is the $0^{++}$ pole at $s=1$, below the two-particle threshold at $s=4$. According to the spectrum \eqref{eq:glueballspec}, the $2^{++}$ and the $0^{++*}$ also lie below threshold and are bound states, while the heavier states are resonances. At large $N$ this distinction is only kinematical. Since only even spins are exchanged, bound states and resonances have residues of the same sign as discussed in Section~\ref{sec:bootstrap}.

\paragraph{Regge behaviour and observables.}
The leading glueball Regge trajectory passes through the $2^{++}$, so the amplitude satisfies dispersion relations with two subtractions for $t<m^2_{2^{++}}$. Their explicit form is reviewed in appendix~\ref{app:stu}, where we also study a toy model without bound states; see also for the first positivity bounds for massive particles \cite{Guerrieri:2023qbg}. The $0^{++}$ pole and the bound states lie close to the crossing-symmetric point $s=t=u=4/3$. 
They limit the radius of convergence of the low-energy Taylor expansion there and spoil the positivity of its coefficients as we will see in Section \ref{subsec:spin4}, and cannot provide a positive normalization to the dual problem. We therefore bound on-shell couplings of the exchanged glueballs rather than Taylor coefficients.

\paragraph{Which coupling.}
With $k$ subtractions, a single exchange of spin $J\le k$ can be completed by states at arbitrarily high mass, while a state with $J>k$ forces an infinite tower of higher-spin states at finite mass \cite{Albert:2023seb}. For pions ($k=1$), this is why the $\rho$ alone gives no Regge trajectory and one maximizes the $f_2$ coupling. For glueballs ($k=2$, even spins only), the $2^{++}$ plays the role of the $\rho$, and the first nontrivial case is spin four. We therefore maximize the coupling of the $4^{++}$.

\subsection{The spin-four coupling}
\label{subsec:spin4}
The null constraints follow from writing the amplitude as a twice-subtracted dispersion relation at fixed $t$ and then imposing full crossing symmetry \cite{EliasMiro:2022xaa}, see also Appendix~\ref{app:stu}. The fixed-$t$ representation reconstructs the $s$- and $u$-channel poles explicitly, but not the $t$-channel ones. Each exchanged state therefore contributes a function $F_J(m_X^2;s,t)$, given in \eqref{eq:crossing_symmetric_ncs}, which measures by how much its dispersive reconstruction fails to be symmetric under $s\leftrightarrow t\leftrightarrow u$. Crossing symmetry of the full amplitude requires $\sum_X g_X^2\,F_{J_X}(m_X^2;s,t)=0$, and the null constraints $n_k$ are the Taylor coefficients of this condition around $s=t=u=4/3$. A spin-zero exchange,
\begin{equation}
\label{eq:scalarexch}
M_X(s,t,u)=g_X^2\left(\frac{1}{m_X^2-s}+\frac{1}{m_X^2-t}+\frac{1}{m_X^2-u}\right),
\end{equation}
is a consistent amplitude on its own. It is symmetric under all permutations of $s$, $t$ and $u$, and at fixed $t$ it tends to a constant as $|s|\to\infty$, so it satisfies the twice-subtracted Regge bound. 
Therefore, the null constraints vanish identically on spin zero
\begin{equation}
\label{eq:nullscalar}
F_0(m_X^2;s,t)\equiv0 \qquad\Longrightarrow\qquad n_k(m_X^2,0)=0 \quad \text{for every } k \text{ and every } m_X^2\,.
\end{equation}
Exchanges with $J\ge2$ behave differently: a single such exchange grows at least as $s^2$ at fixed $t$, so it is not consistent on its own and contributes non-trivially to the null constraints. As for pions~\cite{Albert:2023seb}, scalars are thus invisible to crossing and enter the problem only through the sum rules for physical quantities, such as $a_2(t)$ below.

\paragraph{Normalization.}
As for mesons, bounding the $4^{++}$ coupling requires a normalization that is positive for every allowed spectrum. To do so, we generalize the expansion in \eqref{eq:Mlow}. We expand at fixed $t$ around the $s\leftrightarrow u$ symmetric point $s=u=2-t/2$
\begin{equation}
\label{eq:a2expansion}
M(s,t,u)=a_0(t)+a_2(t)\,\frac{(s-u)^2}{4}+a_4(t)\,\frac{(s-u)^4}{16}+\dots
\end{equation}
For $t<m^2_{2^{++}}$ the twice-subtracted dispersion relation in $s$ gives
\begin{equation}
\label{eq:a2sumrule}
a_2(t)=\sum_X g_X^2\,K_t(m_X^2,J_X)+\sum_J\int_{\Lambda^2}^{\infty}\!dz\;16(2J+1)\,\rho_J(z)\,K_t(z,J),
\end{equation}
with $K_t(z,J)=\tfrac{2\,P_J(x_t(z))}{(z-2+t/2)^3}$.
Here $s_0=2-t/2$ is the expansion point, $x_t(z)=1+2t/(z-4)$, and the sum runs over the isolated states, whose residues are normalized as in \eqref{eq:glueballpole}.
The natural choice $t=4/3$ corresponds to the crossing-symmetric point $s=t=u=4/3$. There the two lightest states contribute with negative weights
\begin{equation}
\label{eq:negweights}
K_{4/3}(1,0)=\frac{2}{\left(1-\frac43\right)^3}=-54\,,\qquad
K_{4/3}(m^2_{2^{++}},2)=\frac{2\,P_2\!\left(1-\frac{8}{3(4-m^2_{2^{++}})}\right)}{\left(m^2_{2^{++}}-\frac43\right)^3}\simeq-0.27\,.
\end{equation}
The external $0^{++}$ lies below the expansion point. For the $2^{++}$ bound state, the argument of $P_2$ is $\simeq-0.52$, where $P_2<0$. By \eqref{eq:nullscalar}, the $0^{++}$ coupling can then be tuned to make $a_2(4/3)$ vanish, and no bound exists.

Every term in \eqref{eq:a2sumrule} is positive when two conditions hold. The denominator is positive for $t>2(2-m_X^2)$, which for the external $0^{++}$ requires $t>2$. For the $2^{++}$ bound state, the Legendre factor is positive when $|1+2t/(m^2_{2^{++}}-4)|>1/\sqrt3$, which holds throughout the window.\footnote{This wisdom follows from the explicit form of $P_2(x)=1/2(3x^2-1)$.} All states therefore contribute positively for $2<t<m^2_{2^{++}}=2.241$, and we take $t=2.1$, where for instance $K_{2.1}(2,m^2_{2^{++}})\simeq2.2$. The dependence of the bound on this choice is discussed in Section~\ref{sec:glueballbounds}.

\paragraph{The optimization problem.}
We define the fraction of $a_2(t)$ carried by the $4^{++}$ state
\begin{equation}
\label{eq:g4tilde}
 0\leq \tilde g^2_{4^{++}}=\frac{g^2_{4^{++}}\,K_t(m^2_{4^{++}},4)}{a_2(t)}\leq 1\,.
\end{equation}
We maximize it subject to crossing symmetry, imposed through the null constraints
\begin{equation}
\label{eq:glueballnull}
\sum_X g_X^2\,n_k(m_X^2,J_X)+\sum_J\int_{\Lambda^2}^{\infty}\!dz\;16(2J+1)\,\rho_J(z)\,n_k(z,J)=0\,,\qquad k=1,\dots,N(n_{\max})\,.
\end{equation}
Here $n_k(z,J)$ denotes a linearly independent subset of the derivatives $N^{(n,k)}_J(z)$ in \eqref{eq:ncsforreal}, obtained by expanding the crossing equations of appendix~\ref{app:stu} around $s=t=u=4/3$ up to order $n+k\le n_{\max}$, and $N(n_{\max})$ is their number. Scalar exchanges do not enter the null constraints, \eqref{eq:nullscalar}, and inside the window $2<t<m^2_{2^{++}}$ they contribute to $a_2(t)$ with $K_t(m_X^2,0)>0$. Increasing their couplings can therefore only lower the fraction \eqref{eq:g4tilde}, and the maximum is reached when they vanish. Removing the isolated scalars, the external $0^{++}$ at $m_X^2=1$ and the $0^{++*}$ at $m_X^2=3.62$, leaves the bound exactly unchanged.
In particular, the bound holds for any value of the cubic self-coupling of the lightest glueball. As a check, we keep spin-zero states in the continuum and verify that the extremal solution assigns them vanishing couplings.

\begin{figure}[t!]
    \centering
    \includegraphics[width=0.85\linewidth]{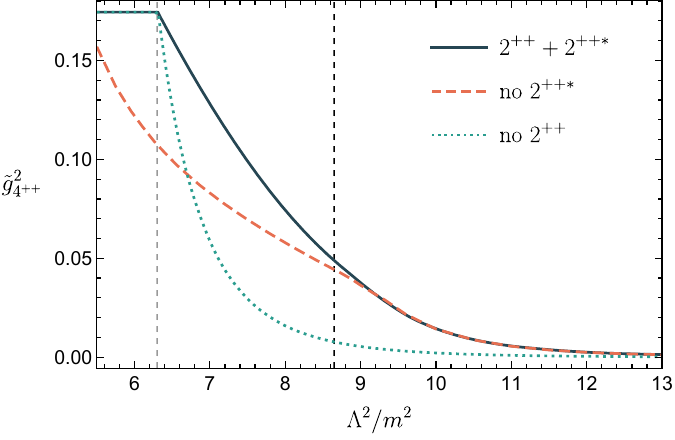}
    \caption{Upper bound on the normalized coupling $\tilde g^2_{4^{++}}$ of the $4^{++}$ glueball at $n_{\max}=16$. $m$ is the mass of the lightest glueball. \textbf{Solid}: spectrum with the $2^{++}$ bound state ($m^2_{2^{++}}=2.24 $), the $2^{++*}$ resonance ($m^2_{2^{++*}}=4.59 $) and the $4^{++}$ ($m^2_{4^{++}}=5.45$). \textbf{Dashed}: without the $2^{++*}$. \text{Dotted}: without the $2^{++}$. Vertical lines mark the spin-6 states on the linear trajectories through the $4^{++}$ and either the lightest spin-two $2^{++}$ state or the $2^{++*}$.}
\label{fig:glue-bound-nmax16}
\end{figure}

\section{Bounds on the $4^{++}$ glueball coupling}
\label{sec:glueballbounds}

We now maximize $\tilde g^2_{4^{++}}$, defined in \eqref{eq:g4tilde}, as a function of the continuum threshold $\Lambda^2\ge m^2_{4^{++}}$. Below $\Lambda^2$ the spectrum contains only the states in \eqref{eq:glueballspec}: the two tensors $2^{++}$ and $2^{++*}$ and the $4^{++}$, with the scalars removed as explained in section~\ref{subsec:spin4}. A useful reference throughout is the mass of the spin-six state on the linear trajectory through a tensor of mass $m_2$ and the $4^{++}$,
\begin{equation}\label{eq:m6}
m_6^2(m_2^2)=2m^2_{4^{++}}-m_2^2\,,
\end{equation}
which gives $m_6^2=6.31$ for the $2^{++*}$ and $m_6^2=8.66$ for the $2^{++}$.

\paragraph{Plateau and kink.}
The bound at $n_{\max}=16$ is the solid black curve in figure~\ref{fig:glue-bound-nmax16}. It converges rapidly with $n_{\max}$ and shows a nearly flat region, $\tilde g^2_{4^{++}}\simeq 0.174$, followed by a sharp decrease. The plateau ends at a kink at $\Lambda^2\simeq 6.31$, which coincides with $m_6^2$ on the trajectory through the $2^{++*}$ and the $4^{++}$.

\begin{figure}[t!]
    \centering
    \includegraphics[width=0.9\linewidth]{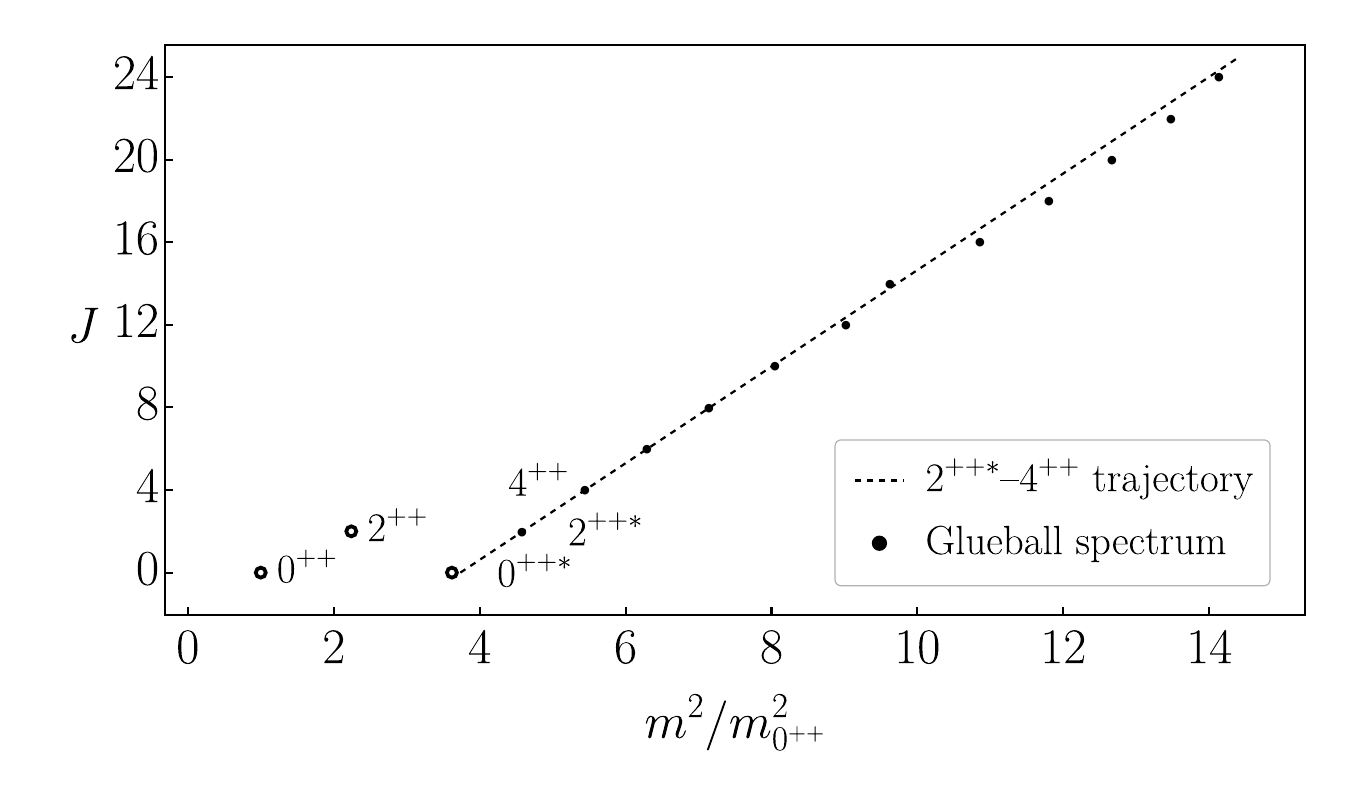}
    \caption{Chew--Frautschi plot of the extremal spectrum at $\Lambda^2=6.30$, just below the kink, at $n_{\max}=16$, in units of $m_{0^{++}}$. The input states are also shown; the $0^{++*}$ and the $2^{++}$ decouple from the extremal solution. The dashed line is the linear trajectory through the $2^{++*}$ and the $4^{++}$. The spin-six state sits at the threshold on this trajectory, and the higher-spin states follow it.}
    \label{fig:glue-cf-kink}
\end{figure}

We extract the spectrum from the extremal functional at $\Lambda^2=6.30$, just below the kink, where the numerics are more stable -- see Figure~\ref{fig:glue-cf-kink}. The first additional state has spin six and sits at the threshold, on the linear trajectory through the $2^{++*}$ and the $4^{++}$. The higher-spin states follow closely the same trajectory up to $J=24$. 
The lighter $2^{++}$ decouples, so the extremal solution is controlled by the $2^{++*}$ and the $4^{++}$ alone. The spectrum is approximately unchanged along the plateau and departs from the linear trajectory once $\Lambda^2$ exceeds the spin-six mass.

This supports the picture of~\cite{Huang:2025icl}, where maximizing higher-spin couplings was found to favour linear Regge trajectories. In \cite{Huang:2025icl} the threshold and the lower masses were fixed and the spin-two mass was varied. The spin-four coupling peaked when the states lay on a linear trajectory and dropped sharply once the threshold exceeded the spin-six mass. Those results were obtained for supersymmetric amplitudes with zero or negative subtractions. We find the same structure in a non-supersymmetric problem with two subtractions, keeping the masses fixed and varying the threshold instead.

\begin{figure}[t!]
    \centering
    \includegraphics[width=0.9\linewidth]{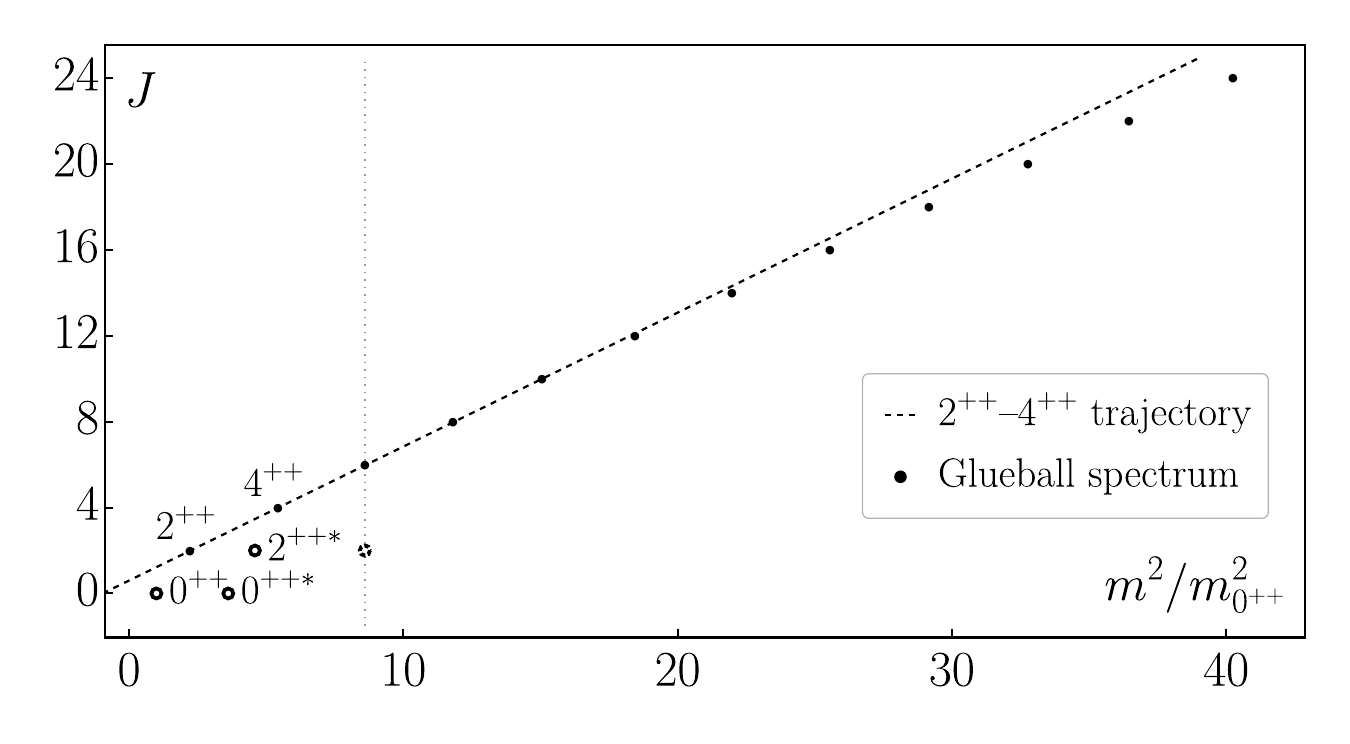}
    \caption{Extremal spectrum at $\Lambda^2=8.66$, the spin-six mass on the linear trajectory through the $2^{++}$ and the $4^{++}$, in the problem without the $2^{++*}$. The states lie on a single linear trajectory. The coupling of the spin-two state at threshold (the dashed circle) vanishes exactly at the kink for every finite $n_{\max}$ and decreases with $n_{\max}$. Since the kink approaches $\Lambda^2=8.66$ as $n_{\max}\to\infty$, we expect this state to decouple in the limit.}
    \label{fig:glueball_spectrum_2}
\end{figure}

\paragraph{The role of the two tensors.}
To disentangle the two spin-two states, we repeat the optimization keeping only one of them -- see Figure~\ref{fig:glue-bound-nmax16}. Without the $2^{++}$ (dotted green curve), the plateau and the kink are unchanged, although the bound falls faster beyond the kink. Without the $2^{++*}$ (dashed red curve), the plateau disappears and a kink appears at a larger threshold. This kink converges more slowly, moving from $\Lambda^2\simeq 9.05$ at $n_{\max}=8$ to $\Lambda^2\simeq 8.85$ at $n_{\max}=16$. Extrapolating the results $n_{\max}=10,\dots,16$ to $n_{\max}\to\infty$ gives
\begin{equation}\label{eq:kinknmaxinfty}
\Lambda^2_{\rm kink} = 8.7 \pm 0.1\,, \qquad
\tilde g^2_{2^{++}} = 0.152 \pm 0.004\,, \qquad
\tilde g^2_{4^{++}} = 0.04 \pm 0.01\,.
\end{equation}
The extrapolated position of the kink above is compatible with the spin-six mass of the linear trajectory through the $2^{++}$ and the $4^{++}$, $m_6^2=8.66$. As a consistency check, we bound $\tilde g_{4^{++}}^2$ at fixed  $\Lambda^2=8.66$ instead of at the kink, and extrapolating, gives compatible couplings, $\tilde g^2_{2^{++}}=0.15\pm0.01$ and $\tilde g^2_{4^{++}}=0.041\pm0.003$.\footnote{All errors reflect the spread among different extrapolations in $1/n_{\max}$.} In the limit $n_{\max}\to\infty$ the kink is then saturated by a single linear trajectory through the $2^{++}$ and $4^{++}$ as in Figure~\ref{fig:glueball_spectrum_2}.

The full bound therefore interpolates between three regimes. On the plateau there is a single linear trajectory through the $2^{++*}$ and the $4^{++}$, and the $2^{++}$ decouples. At intermediate thresholds both tensors contribute, and the bound lies above both single-tensor curves. At large thresholds the $2^{++}$ dominates, and the bound approaches the one without the $2^{++*}$, whose kink is smoothed out.

Both trajectories are fixed by the input masses, $\alpha(t)=\alpha_0+\alpha' t$ with $\alpha'=2/(m^2_{4^{++}}-m_2^2)$ and $\alpha_0=2-\alpha' m_2^2$. The trajectory through the $2^{++}$ and the $4^{++}$, the natural candidate for the leading glueball trajectory at large $N$, has $\alpha_0\simeq0.60$ and $2\pi\sigma\alpha'\simeq0.42$. Its intercept is compatible with the SU(3) lattice estimate $\alpha_0=0.93(24)$ of~\cite{Meyer:2004jc}, whose smaller slope, $2\pi\sigma\alpha'=0.28(2)$, reflects a heavier $J=4$ state than in~\cite{Athenodorou:2021qvs}. The trajectory through the $2^{++*}$ and the $4^{++}$ is instead very steep, $2\pi\sigma\alpha'\simeq1.6$, with $\alpha_0\simeq-8.7$, and is unlikely to describe a physical leading trajectory.

\begin{figure}[t!]
    \centering
    \includegraphics[width=0.7\linewidth]{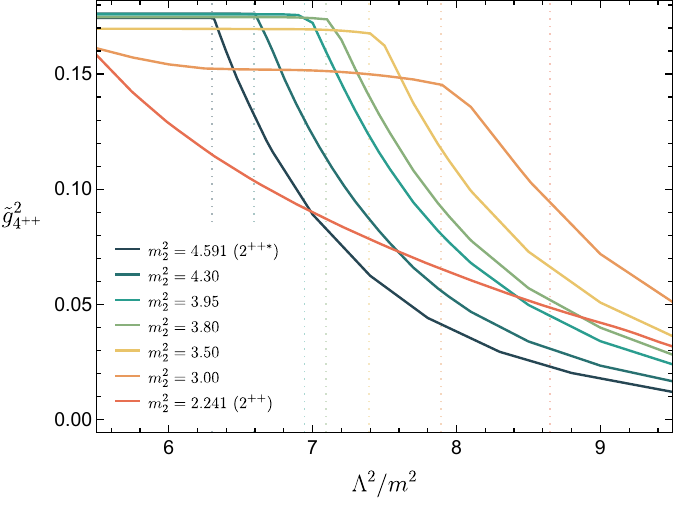}
    \caption{Upper bound on $\tilde g^2_{4^{++}}$ as a function of $\Lambda^2$ at $n_{\max}=12$, with the $4^{++}$ at $m^2_{4^{++}}=5.45$ and a single tensor of mass $m_2^2$. The tensor mass is varied from the $2^{++*}$ at $4.59$ to the $2^{++}$ at $2.241$, across the two-particle threshold at $m_2^2=4$. The dotted vertical lines mark $m_6^2(m_2^2)$, eq.~\eqref{eq:m6}. The kink tracks these values, while the plateau disappears as the tensor becomes lighter.}
    \label{fig:glue-m2scan}
\end{figure}

\paragraph{Dependence on the tensor mass.}
To see how the plateau disappears, we keep a single tensor and vary its mass continuously from $m_2^2=4.59$ to $2.241$, with the $4^{++}$ fixed -- see Figure~\ref{fig:glue-m2scan}, $n_{\max}=12$. The kink tracks $m_6^2(m_2^2)$ throughout, both above and below the two-particle threshold. What changes is the shape of the bound before the kink. For $m_2^2\gtrsim 3.95$ the plateau is flat within numerical precision, and as the tensor becomes lighter the bound develops an increasingly strong dependence on $\Lambda^2$. The change is continuous, with no visible feature when the tensor crosses the threshold.

This dependence comes from the normalization of the coupling itself. Normalizing by $a_2(t)$ \eqref{eq:g4tilde} at $t=2.1$ introduces a new scale. The crossed $t$-channel pole of an explicit tensor exchange of mass $m_2$ contributes\footnote{For a tensor of mass $m_2$, the $t$-channel pole is $g_2^2P_2(z)/(m_2^2-t)$, with $z=(s-u)/(m_2^2-4)$ the cosine of the $t$-channel scattering angle on the pole. Since $P_2(z)=\frac12(3z^2-1)$, it contributes to the coefficient of $(s-u)^2/4$ -- a.k.a $a_2(t)$ -- as in \eqref{eq:Deltaa2}.}
\begin{equation}\label{eq:Deltaa2}
\Delta a_2(t)=g_2^2\,\frac{6}{(m_2^2-4)^2}\,\frac{1}{m_2^2-t}\,,
\end{equation}
which in the fixed-$t$ dispersion relation \eqref{eq:a2sumrule} must be reproduced by the $s$-channel sum, including its high-mass tail. For the $2^{++}$ the normalization point is close to the pole, $m^2_{2^{++}}-t=0.141$, and this crossed term is about six times larger than the direct contribution of the same spin-$2$ exchange $g_2^2K_t(m^2_{2^{++}},2)$ to the $a_2(t)$ sum-rule \eqref{eq:a2sumrule}. Moving the tensor away from the normalization point reduces this sensitivity and produces the plateau seen for heavier masses.\footnote{We checked this by repeating the calculation for $m_2^2=2.241$ at different values of the subtraction point as function of $t$. Between $\Lambda^2=5.5$ and $8.5$ the bound decreases moving farther from the spin-2 pole by a factor of $2.2$ at $t=2.01$, and by a factor of $6$ at $t=2.22$ moving closer.} The plateau is therefore not a generic feature of the coupling bound, but reflects the separation between the tensor mass and the normalization point.

\subsection{The \texorpdfstring{$\tilde g^2_{2^{++}}$--$\tilde g^2_{4^{++}}$}{g2++2-g4++2} plane}\label{sec:glueball-plane}

We have shown in Figure \ref{fig:glue-bound-nmax16} that with both tensors included, the bound on $\tilde g^2_{4^{++}}$ has a single kink, at $\Lambda^2=m_6^2(m^2_{2^{++*}})$, the linear trajectory passing by the $2^{++*}$ and the $4^{++}$ states, where the $2^{++}$ decouples. The trajectory through the $2^{++}$ and the $4^{++}$, which one would naively expect for the leading glueball trajectory, appears only once the $2^{++*}$ is removed. To enforce the presence of the $2^{++}$ state we study a different optimization problem.

\begin{figure}[t!]
    \centering
    \includegraphics[width=0.49\linewidth]{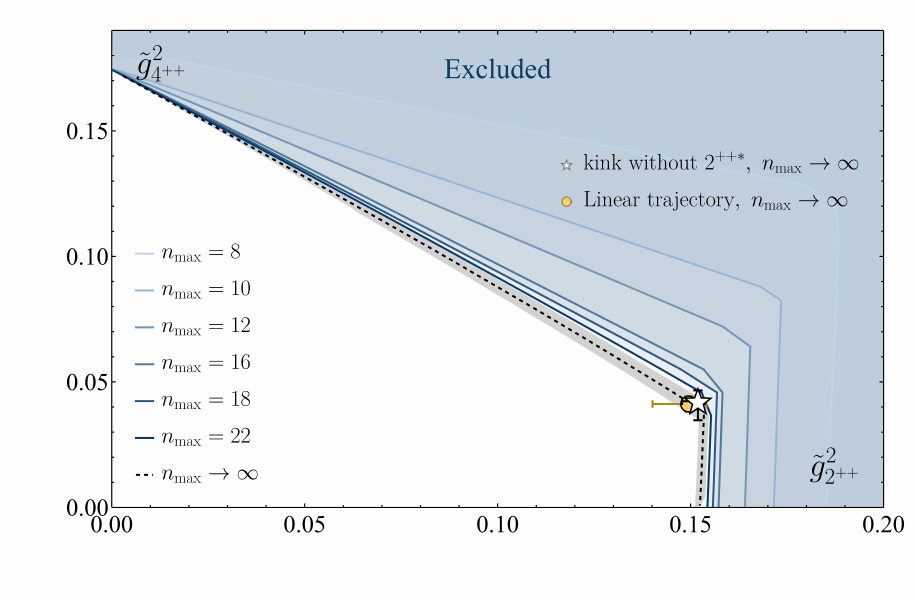}
    \hfill
    \includegraphics[width=0.49\linewidth]{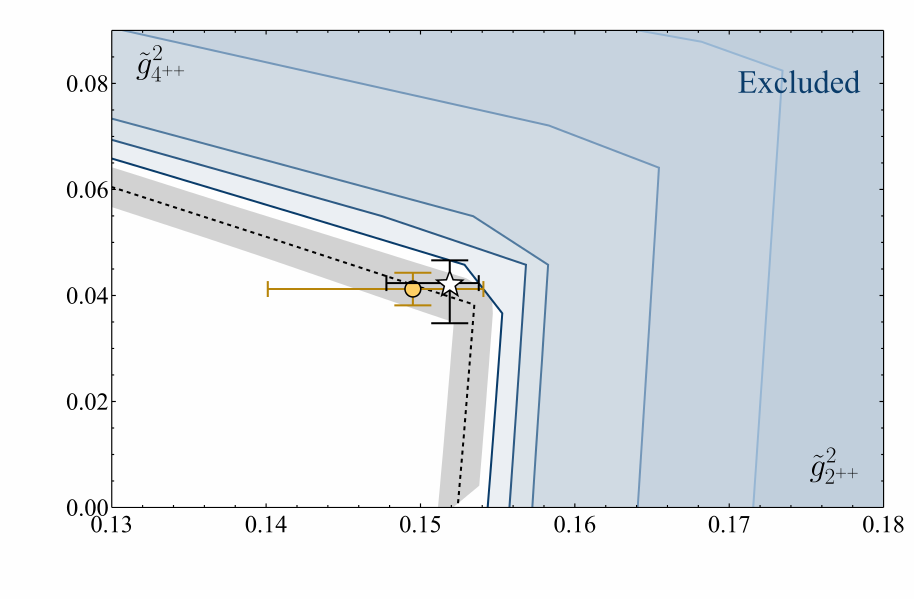}
    \caption{Allowed region in the $(\tilde g^2_{2^{++}},\tilde g^2_{4^{++}})$ plane for increasing $n_{\max}$, and its extrapolation to $n_{\max}\to\infty$; the gray band is the extrapolation uncertainty. \emph{Left:} full region. The yellow dot marks the $n_{\max}\to\infty$ couplings at $\Lambda^2=m_6^2=8.66$ (figure~\ref{fig:glueball_spectrum_2}), and the star the extrapolated kink of the bound without the $2^{++*}$, eq.~\eqref{eq:kinknmaxinfty}. \emph{Right:} zoom of the kink region, where the two marked points are compatible with each other and with the kink of the allowed region.}
    \label{fig:glueball-plane}
\end{figure}

We fix both tensors at their respective physical masses, fix the threshold at a point close in the plateau region near the kink, $\Lambda^2=6.3$, and bound the allowed region of the normalized couplings $(\tilde g^2_{2^{++}},\tilde g^2_{4^{++}})$, shown in Figure~\ref{fig:glueball-plane}. The plot shows a slower convergence than for the single bound problem: the boundary has not visibly converged at $n_{\max}=22$, and therefore we extrapolate it to $n_{\max}\to\infty$. The upper boundary has two distinguished points. At $\tilde g^2_{2^{++}}=0$ it starts from the plateau value $\tilde g^2_{4^{++}}\simeq 0.174$ in Figure~\ref{fig:glue-bound-nmax16}, saturated by the trajectory through the $2^{++*}$ and the $4^{++}$ -- see also Figure~\ref{fig:glue-cf-kink}. At the largest value of $\tilde g^2_{2^{++}}$ it has a clear kink. Because of the slow convergence, extracting the spectrum directly at this kink is difficult. However, we notice that the kink coincides with the solution found without the $2^{++*}$. The values at the kink extracted in \eqref{eq:kinknmaxinfty} already lie close to it at $n_{\max}=16$, and the $n_{\max}\to\infty$ extrapolations of the boundary and of \eqref{eq:kinknmaxinfty} are compatible -- see right panel of Figure~\ref{fig:glueball-plane}. We are then confident to state that the kink of the plane is saturated by the single linear trajectory through the $2^{++}$ and the $4^{++}$ of Figure~\ref{fig:glueball_spectrum_2}.

\begin{figure}[t!]
    \centering
    \includegraphics[width=0.8\linewidth]{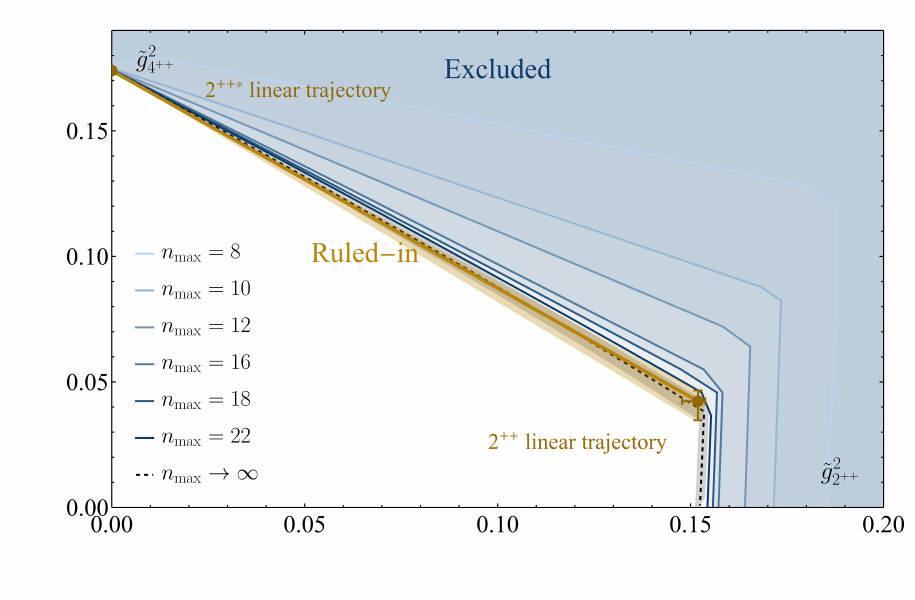}
    \caption{Upper boundary of the $(\tilde g^2_{2^{++}},\tilde g^2_{4^{++}})$ plane and ruled-in segment. Convex combinations of the two extremal amplitudes, the linear trajectory through the $2^{++*}$ and the $4^{++}$ (figure~\ref{fig:glue-cf-kink}) and the one through the $2^{++}$ and the $4^{++}$ (figure~\ref{fig:glueball_spectrum_2}), saturate the upper boundary. The light yellow band is the uncertainty of the $n_{\max}\to\infty$ extrapolation of the kink.}
    \label{fig:glueballs_plane_rulein}
\end{figure}

Between these two points the upper boundary of the allowed region is linear at every fixed $n_\text{max}$ as highlighted in Figure~\ref{fig:glueballs_plane_rulein}. This behavior is expected when the boundary is saturated by convex combinations of particular solutions. Indeed, a convex combination of two feasible amplitudes, that satisfy $a_2(t)=1$, is consistent, and its normalized couplings interpolate linearly between the two endpoints. 
The upper boundary of the plane is thus expected to contain two linear Regge trajectories, one through the $2^{++*}$ and one through the $2^{++}$, each passing through the $4^{++}$.

\paragraph{Tridimensional Space of Couplings}
\begin{figure}
    \centering
    \includegraphics[width=\linewidth]{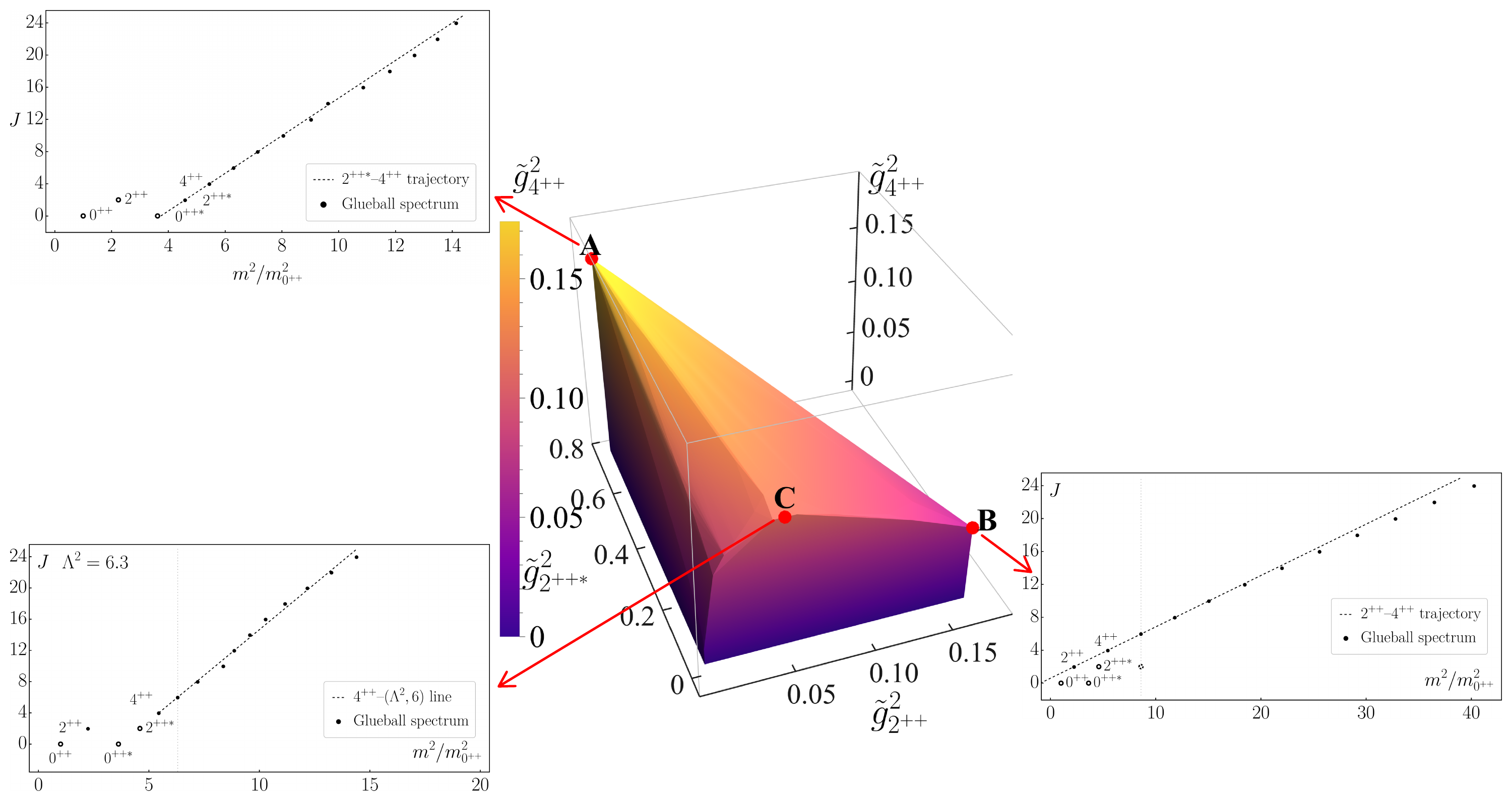}
    \caption{Tridimensional hull of the space of couplings $\{\tilde g_{2^{++}}, \tilde g_{2^{++*}}, \tilde g_{4^{++}}\}$ at $n_\text{max}=12$. At the point $A$ there is only the $2^{++*}-4^{++}$ linear trajectory, at the point $B$ there is only the $2^{++}-4^{++}$ linear trajectory, and at the point $C$ there is the $2^{++}$ and the $4^{++}-(6, \Lambda^2)$ linear trajectory. Convex combinations rule-in the straight ridges between those points.}
    \label{fig:hull_3D}
\end{figure}

Figure~\ref{fig:hull_3D} completes our analysis of the glueball couplings, showing the allowed region in the three-dimensional space $(\tilde g^2_{2^{++}},\tilde g^2_{2^{++*}},\tilde g^2_{4^{++}})$ at $\Lambda^2=6.3$ and $n_{\max}=12$. Its boundary has three cusps, which we denote by $A$, $B$ and $C$, connected by three edges. The edge $AB$ is the straight segment discussed in section~\ref{sec:glueball-plane}. It is saturated by convex combinations of the amplitudes at $A$ and $B$, each containing a single linear trajectory (see the insets of figure~\ref{fig:hull_3D}). The edge $AC$ is also straight, but is visible only in three dimensions. By the same argument, we expect it to be saturated by convex combinations of the amplitudes at $A$ and $C$. The amplitude at $C$ contains the $2^{++}$ and a linear trajectory starting at the $4^{++}$, without the $2^{++*}$. Moving from $C$ to $A$, the $2^{++*}$ turns on and the $2^{++}$ decouples, so that the two tensors exchange their roles.

\section{Discussion and Outlook}
\label{sec:discussion}

We have extended the large $N$ bootstrap of pion scattering to nonzero pion mass, and applied the same dispersive construction to the scattering of the lightest scalar glueball. Our observables are defined directly from the planar amplitude, so the bounds do not rely on the validity of the effective field theory expansion. For pions, the pion mass provides an additional variable with which to test the relation between extremal amplitudes and QCD, and allows comparisons with lattice calculations at the same quark masses.

For pions, both problems retain the main features of the chiral case. The region of normalized Taylor coefficients keeps the straight Skyrme segment and the kink. Its endpoint is fixed analytically by a map to the massless problem, but the deformation of the whole region is not a uniform rescaling. Maximizing the $f_2$ coupling produces a kink at all pion masses studied, from which we extract a candidate leading Regge trajectory. Near the physical pion mass, its states with $J=3,4,5$ are compatible with the experimental ones, after adjusting the input ratio $m_{f_2}^2/m_\rho^2$ so it matches the real world $N=3$ case.

A more direct comparison with large $N$ QCD requires better control of the mass inputs. Our $\rho$ prescription follows the lattice fit of~\cite{Bonanno:2025hzr}. In the absence of a lattice determination of the $f_2$ mass, we used a simple model with a fixed $\rho$--$f_2$ squared-mass spacing. Other choices are equally possible. One could keep $m_{f_2}/m_\rho$ at its physical value, as in~\cite{Albert:2023seb}, or let both states receive the same shift in mass.
A large $N$ lattice determination of $m_{f_2}(m_\pi)$ would remove this ambiguity. Lattice input would also sharpen the comparison of couplings: a large $N$ determination of $L_5$, available through $\bar\ell_4$~\cite{Bonanno:2025hzr}, would replace the $N=3$ value used in our estimate of the matching correction $\epsilon$ (appendix~\ref{app:exp}).

For glueballs, the external state is itself the lightest exchanged particle, and the lightest tensor is a bound state. We found a normalization that remains positive in this situation and is independent of scalar exchanges, so the bound holds for any value of the cubic glueball coupling. Maximizing the $4^{++}$ coupling with the lattice spectrum as input, the extremal solutions are approximately linear Regge trajectories. With both tensors present, the leading trajectory passes through the $2^{++*}$ and the $4^{++}$, and the lightest $2^{++}$ decouples. The trajectory through the $2^{++}$ and the $4^{++}$, with intercept $\alpha_0\simeq0.6$, appears only once the $2^{++*}$ is removed. Understanding a framework selecting that singles out the $2^{++}-4^{++}$ linear trajectory as the leading one is a natural next step. Lattice determinations of three-glueball couplings would provide direct tests of these solutions. 

Two features are common to the meson and glueball problems. In both, the kink appears where the gap reaches the next state of the extremal trajectory, the $\rho_3$ for mesons and the spin-six state for glueballs. In both, the extremal spectrum contains a single leading trajectory without daughters, a feature also noted in other numerical extremal spectra~\cite{Berman:2024kdh}. The meson trajectory is curved, while the glueball one is linear within our accuracy. Meromorphic amplitudes with a finite number of Regge trajectories are, however, incompatible with crossing and unitarity~\cite{Eckner:2024pqt}. The extremal amplitudes must therefore contain further trajectories beyond our numerical reach, for instance, daughters at high mass, as already suggested by the states at finite impact parameter~\cite{Albert:2023seb}. Mixed systems, with external tensor glueballs or $\rho$ mesons, could resolve these states and test whether daughters appear~\cite{Freedman:1966zec}.

There are also open questions about the extremal amplitudes themselves. Stability under increasing the number of null constraints, and control over positivity at large spin and energy, are necessary for assessing the numerical bounds. Reconstructing amplitudes from the candidate spectra also requires reliable spectral weights, which converge much more slowly than the masses. This matters particularly in the bound-state regime, where our numerical problem leaves the sign of the $\rho$ residue unrestricted: the bounds remain valid for the physical problem; the equality of the relaxed and physical optima has been established in our setup by running the problem with the reversed sign and finding out it was unfeasible. This does not hold for example in the glueballs framework where the lightest spin-0 is self coupled or when we have a continuous threshold below the two particle cutoff. The stability of the extracted spectra could be improved by modelling the high-energy tail of the spectral density with its asymptotic form, as done with Tauberian theorems in the thermal bootstrap~\cite{Marchetto:2023xap,Barrat:2025wbi}, where the asymptotics would here be dictated by the Regge behaviour, or by imposing analyticity in spin. We are exploring both directions and hope to report on them soon.

Two extensions seem particularly natural. The first is the coupled pion--kaon system. Including $\pi\pi\to\pi\pi$, $\pi\pi\to K\bar K$ and $K\bar K\to K\bar K$, together with the crossed $\pi K$ channels, would connect the nonstrange and strange sectors. At large $N$, a resonance shared by several channels appears with factorized residues, which ties its couplings in the inelastic amplitudes to those in the elastic ones. The new ingredients are unequal external masses and thresholds, and explicit $SU(3)$ breaking in both the low-energy expansion and the spectrum. Combined with lattice masses, this would constrain strange and nonstrange resonances together, with dispersive analyses of $\pi K$ scattering at $N=3$~\cite{Buettiker:2003pp} as a benchmark.

The second is to include multiparticle amplitudes. At large $N$, amplitudes with any number of mesons must share a common spectrum and factorize consistently on it. Positivity bounds for higher-point tree amplitudes have recently been derived~\cite{Cheung:2025nhw,Berman:2025owb,Kremminger:2026edr}, and applied to the large-$N$ chiral Lagrangian~\cite{Cheung:2026lpv}, where they exclude regions allowed by four-point positivity. The multiparticle flux-tube bootstrap of~\cite{Guerrieri:2024ckc} shows how analyticity, crossing and unitarity can be imposed jointly on processes of different multiplicities. The natural next step is to set up a full multiparticle bootstrap for the coupled system of $2\to2$, $2\to3$ and $3\to3$ amplitudes, constrained jointly by their null constraints and by factorization on a common spectrum.

\section*{Acknowledgements}

We thank Kelian Häring, Denis Karateev, Francesco Riva, Alessandro Vichi for useful discussions. S.B.\ is supported by a PhD studentship funded by the Royal Society through grant URF/R1/241371.
A.G. is
supported by a Royal Society University Research Fellowship, URF/R1/241371.

\appendix

\section{Tree-level matching at order \texorpdfstring{$p^4$}{p4}}\label{app:matching}

In this appendix we derive \eqref{eq:mpi_fpi}, \eqref{eq:M2} and \eqref{eq:c_vs_L}. All the computations are at tree level, which is exact at leading order in $1/N$.

\paragraph{Expansion of the Lagrangian.}
With $U=e^{2i\Phi/F}$, $\Phi=\pi^aT^a$ and $\langle T^aT^b\rangle=\tfrac12\delta^{ab}$, the two terms of \eqref{eq:L2} read, up to fourth order in the pion field and for $\chi=M^2\mathds{1}$,
\begin{equation}\label{eq:L2_expanded}
\begin{aligned}
\frac{F^2}{4}\langle\partial_\mu U\partial^\mu U^\dagger\rangle&=\frac12\,\partial_\mu\pi^a\partial^\mu\pi^a+\frac{1}{3F^2}\big\langle[\Phi,\partial_\mu\Phi][\Phi,\partial^\mu\Phi]\big\rangle+\mathcal{O}(\Phi^6),\\
\frac{F^2}{4}\langle\chi U^\dagger+U\chi^\dagger\rangle&=\frac{N_fF^2M^2}{2}-\frac{M^2}{2}\,\pi^a\pi^a+\frac{M^2}{3F^2}\langle\Phi^4\rangle+\mathcal{O}(\Phi^6).
\end{aligned}
\end{equation}
The operators of \eqref{eq:L4} contribute to the quadratic and quartic Lagrangian as
\small{
\begin{equation}\label{eq:L4_expanded}
\begin{aligned}
\langle\partial_\mu U^\dagger\partial^\mu U\,\partial_\nu U^\dagger\partial^\nu U\rangle&=\frac{16}{F^4}\langle\partial_\mu\Phi\partial^\mu\Phi\,\partial_\nu\Phi\partial^\nu\Phi\rangle+\\&+\frac{16}{3F^6}\left\langle
  2\,[\Phi,\partial_\mu\Phi]\,[\Phi,\partial^\mu\Phi]\,(\partial_\nu\Phi)^2
  + \big[\Phi,(\partial_\nu\Phi)^2\big]\,\big[\Phi,(\partial_\nu\Phi)^2\big]
\right\rangle+\dots,\\
\langle\partial_\mu U^\dagger\partial_\nu U\,\partial^\mu U^\dagger\partial^\nu U\rangle&=\frac{16}{F^4}\langle\partial_\mu\Phi\partial_\nu\Phi\,\partial^\mu\Phi\partial^\nu\Phi\rangle+\\&+\frac{64}{3F^6}\left\langle
  [\Phi,\partial_\mu\Phi]\,[\Phi,\partial_\nu\Phi]\,\partial^\mu\Phi\,\partial^\nu\Phi
  + \frac{1}{2}\,[\Phi,\partial_\mu\Phi]\,\partial_\nu\Phi\,[\Phi,\partial^\mu\Phi]\,\partial^\nu\Phi
\right\rangle+\dots,\\
\langle\partial_\mu U^\dagger\partial^\mu U\,(\chi^\dagger U{+}U^\dagger\chi)\rangle&{=}\frac{4M^2}{F^2}\,\partial_\mu\pi^a\partial^\mu\pi^a+\frac{8M^2}{3F^4}\big\langle[\Phi,\partial_\mu\Phi][\Phi,\partial^\mu\Phi]\big\rangle-\frac{16M^2}{F^4}\langle\partial_\mu\Phi\partial^\mu\Phi\,\Phi^2\rangle+\\
&{+}\frac{16M^2}{45F^6}\left\langle
  15\,\Phi^4\,\partial_\mu\Phi\,\partial^\mu\Phi
  {-} 19\,[\Phi,\partial_\mu\Phi]\,[\Phi,\partial^\mu\Phi]\,\Phi^2
  {+} [\Phi^2,\partial_\mu\Phi]\,[\Phi^2,\partial^\mu\Phi]
\right\rangle+\dots,\\
\langle\chi^\dagger U\chi^\dagger U+U^\dagger\chi U^\dagger\chi\rangle&=2N_fM^4-\frac{8M^4}{F^2}\,\pi^a\pi^a+\frac{64M^4}{3F^4}\langle\Phi^4\rangle-\frac{512\,M^4}{45\,F^6}\langle\Phi^6\rangle+\dots
\end{aligned}
\end{equation}}
where the dots stand for terms with six or more pions. 

\paragraph{Mass and decay constant.}
Collecting the terms quadratic in the pion fields from
\eqref{eq:L2_expanded} and \eqref{eq:L4_expanded}, we obtain
\begin{equation}\label{eq:Lquad}
\mathcal{L}_2^{(2)}+\mathcal{L}_4^{(2)}
=\frac12\Big(1+\frac{8L_5M^2}{F^2}\Big)
\partial_\mu\pi^a\partial^\mu\pi^a
-\frac12\Big(M^2+\frac{16L_8M^4}{F^2}\Big)\pi^a\pi^a,
\end{equation}
where $\mathcal{L}_k^{(n)}$ denotes the terms of chiral order
$p^k$ containing $n$ pion fields.
The coefficient of the kinetic term determines the wave-function
renormalization. Introducing the canonically normalized field
$\pi_r^a$ through $\pi^a=Z^{1/2}\pi_r^a$, we find
$Z=1-\tfrac{8L_5M^2}{F^2}+O(p^4)$.
The physical pion mass is then read from the quadratic term
in $\pi_r^a$,
\begin{equation}
m_\pi^2
=Z\left(M^2+\frac{16L_8M^4}{F^2}\right)
=M^2\left[1+\frac{8M^2}{F^2}(2L_8-L_5)\right]
+O(p^6).
\end{equation}

To determine the decay constant, we restore the external axial
source $a_\mu=a_\mu^aT^a$ in the covariant derivatives,
setting the other external sources to zero. Expanding the
leading-order and $L_5$ operators to first order in both
$a_\mu^a$ and $\pi^a$, we obtain the interaction
\begin{equation}
\mathcal{L}_{a\pi}
=-F\left(1+\frac{8L_5M^2}{F^2}\right)
a_\mu^a\partial^\mu\pi^a+\cdots,
\end{equation}
where we use the convention
$D_\mu U=\partial_\mu U-i\{a_\mu,U\}$.
After the field redefinition $\pi^a=Z^{1/2}\pi_r^a$,
the coefficient of the axial-current coupling gives
\begin{equation}
f_\pi
=F\left(1+\frac{8L_5M^2}{F^2}\right)Z^{1/2}
=F\left(1+\frac{4L_5M^2}{F^2}\right)+O(p^4).
\end{equation}
These reproduce the relations \eqref{eq:mpi_fpi}.

\paragraph{Feynman rule.}
Consider a quartic single-trace monomial $\langle O_1O_2O_3O_4\rangle$, where each $O_i$ is $\Phi$ or $\partial_\mu\Phi$, and four incoming pions with momenta $p_i$ and flavours $a_i$, $\sum_ip_i=0$. Expanding $\Phi=\pi^aT^a$ and contracting the fields with the external legs produces a sum over the $4!$ assignments of legs to slots, each weighted by the trace of the generators in the order of the slots. Grouping them by cyclic classes, the coefficient of the trace $\langle T^{a_1}T^{a_2}T^{a_3}T^{a_4}\rangle$ is the sum over the four cyclic assignments, leg $r+k$ in slot $k$ with $r=0,\dots,3$, where every derivative acting on leg $i$ is replaced by $-ip_i$. By \eqref{eq:flavor_decomposition}, the flavour-ordered amplitude $M(s,t)$ is this coefficient divided by four, with $s=(p_1+p_2)^2$ and $t=(p_2+p_3)^2$; the reversed ordering gives the same function. On the mass shell, $p_i^2=m_\pi^2$, the invariants are
\begin{equation}\label{eq:dots}
p_1\cdot p_2=p_3\cdot p_4=\frac{s-2m_\pi^2}{2},\qquad p_2\cdot p_3=p_4\cdot p_1=\frac{t-2m_\pi^2}{2},\qquad p_1\cdot p_3=p_2\cdot p_4=\frac{u-2m_\pi^2}{2},
\end{equation}
so that, for instance,
\begin{equation}\label{eq:rule_examples}
\begin{aligned}
\langle\partial_\mu\Phi\,\Phi\,\partial^\mu\Phi\,\Phi\rangle&\to-2(u-2m_\pi^2),&\qquad \langle\Phi^4\rangle&\to4,\\
\langle\Phi^2\,\partial_\mu\Phi\partial^\mu\Phi\rangle&\to-(s+t-4m_\pi^2),&\quad
\langle\partial_\mu\Phi\partial_\nu\Phi\,\partial^\mu\Phi\partial^\nu\Phi\rangle&\to(u-2m_\pi^2)^2,\\
\langle\partial_\mu\Phi\partial^\mu\Phi\,\partial_\nu\Phi\partial^\nu\Phi\rangle&\to\tfrac12\big[(s-2m_\pi^2)^2+(t-2m_\pi^2)^2\big].
\end{aligned}
\end{equation}

\paragraph{Leading order.}
Using $\big\langle[\Phi,\partial_\mu\Phi][\Phi,\partial^\mu\Phi]\big\rangle=2\langle\partial_\mu\Phi\,\Phi\,\partial^\mu\Phi\,\Phi\rangle-2\langle\Phi^2\,\partial_\mu\Phi\partial^\mu\Phi\rangle$ and the rules above, the quartic terms of \eqref{eq:L2_expanded} give
\begin{equation}\label{eq:M2_onshell}
M^{(2)}(s,t)=\frac{1}{2F^2}\Big[s+t-\frac{8}{3}m_\pi^2+\frac{2}{3}M^2\Big]=\frac{s+t-2m_\pi^2}{2F^2}-\frac{m_\pi^2-M^2}{3F^2},
\end{equation}
where we have kept the external legs on the physical mass shell. At leading order $m_\pi=M$ and \eqref{eq:M2} follows; at the next order the last term is of order $p^4$ and contributes to $c_{0,0}$.

\paragraph{Next-to-leading order.}
The quartic terms of \eqref{eq:L4_expanded} give the vertex contributions
\begin{equation}\label{eq:NLO_vertices}
\begin{aligned}
L_3'&:\ \frac{2L_3'}{F^4}\Big[(s-2m_\pi^2)^2+(t-2m_\pi^2)^2\Big],&\qquad L_4'&:\ \frac{4L_4'}{F^4}\,(s+t-2m_\pi^2)^2,\\
L_5&:\ \frac{L_5M^2}{F^4}\Big[8(s+t)-\frac{80}{3}m_\pi^2\Big],& L_8&:\ \frac{64L_8M^4}{3F^4},
\end{aligned}
\end{equation}
where $u=4m_\pi^2-s-t$ was used in the $L_4'$ term. The full amplitude at this order is
\begin{equation}\label{eq:M_assembled}
M(s,t)=Z^2\, M^{(2)}(s,t)+\big[\text{vertices \eqref{eq:NLO_vertices}}\big]+... ,
\end{equation}
where $Z^2=1-16L_5M^2/F^2$ is the LSZ factor of the four external legs. Finally $M$ and $F$ are traded for $m_\pi$ and $f_\pi$ with \eqref{eq:mpi_fpi}, which at this order only affects the leading term, $1/(2F^2)=(1+8L_5m_\pi^2/f_\pi^2)/(2f_\pi^2)$ and $m_\pi^2-M^2=8(2L_8-L_5)m_\pi^4/f_\pi^2$. Collecting the coefficients of $s^2+t^2$, $st$, $m_\pi^2(s+t)$ and $m_\pi^4$,
\begin{equation}\label{eq:c_bookkeeping}
\begin{aligned}
c_{2,0}&=2L_3'+4L_4',\\
c_{1,1}&=4L_4',\\
c_{1,0}&=\underbrace{-8L_3'-16L_4'+8L_5}_{\text{vertices}}\ \underbrace{-\,8L_5}_{Z^2}\ \underbrace{+\,4L_5}_{F\to f_\pi}=4\,(L_5-2L_3'-4L_4'),\\
c_{0,0}&=\underbrace{16L_3'+16L_4'-\tfrac{80}{3}L_5+\tfrac{64}{3}L_8}_{\text{vertices}}\ \underbrace{+\,16L_5}_{Z^2}\ \underbrace{-\,8L_5}_{F\to f_\pi}\ \underbrace{+\tfrac{8}{3}L_5-\tfrac{16}{3}L_8}_{\text{on shell}}\\
&=16\,(L_3'+L_4'-L_5+L_8),
\end{aligned}
\end{equation}
which is \eqref{eq:c_vs_L}. Note that the wave-function renormalization and the on-shell condition are essential: dropping them changes $c_{1,0}$ and $c_{0,0}$ but not the quadratic coefficients.

\paragraph{Expansion around the crossing-symmetric point.}
Writing $s=\sigma+\hat s$ and $t=\sigma+\hat t$ in \eqref{eq:M2} and \eqref{eq:M4}, with $\sigma=\tfrac43m_\pi^2$, the leading term gives $(\hat s+\hat t)/(2f_\pi^2)+m_\pi^2/(3f_\pi^2)$, while $s^2+t^2=\hat s^2+\hat t^2+2\sigma(\hat s+\hat t)+2\sigma^2$, $2st=2\hat s\hat t+2\sigma(\hat s+\hat t)+2\sigma^2$ and $s+t=\hat s+\hat t+2\sigma$. Collecting the powers of $\hat s$ and $\hat t$ and comparing with \eqref{eq:Mlow},
\begin{equation}\label{eq:matching_appendix}
\begin{aligned}
M_0&=\frac{m_\pi^2}{3f_\pi^2}+\frac{m_\pi^4}{f_\pi^4}\Big[c_{0,0}+\tfrac83c_{1,0}+\tfrac{32}{9}\big(c_{2,0}+c_{1,1}\big)\Big],\\
g_{1,0}&=\frac{1}{2f_\pi^2}+\frac{m_\pi^2}{f_\pi^4}\Big[c_{1,0}+\tfrac83\big(c_{2,0}+c_{1,1}\big)\Big],\qquad
g_{2,0}=\frac{c_{2,0}}{f_\pi^4},\qquad g_{2,1}=\frac{c_{1,1}}{f_\pi^4},
\end{aligned}
\end{equation}
which is \eqref{eq:matching}; the correction to $g_{1,0}$ defines $\epsilon$, and $1/g_{1,0}=2f_\pi^2(1-2\epsilon)+\dots$ gives the factors in \eqref{eq:gtilde}.

\section{The Dual Formulation}\label{sec:dual}

The problems of Section~\ref{sec:bounds} are linear programs over positive measures, and we solve them through their dual formulation: instead of constructing amplitudes, we construct linear functionals that are non-negative on every amplitude satisfying our assumptions, and each such functional certifies a bound. 
In this section we review the derivation of the dual problems from the primal ones.

\subsection{Dual problem for the Wilson coefficients}\label{subsec:dual_wilson}

We use the shorthand \eqref{eq:average} for the linear functionals of the positive densities $\rho_J$, 
and we denote the null-constraint kernels of \eqref{eq:null_XY} collectively by $f_n(z,J)$, with $n$ running over the pairs $(n,k)$ of both families with $n+k\leq n_{\max}$. In units $\Lambda=1$, the primal problem for the upper boundary of the $\tilde g_2'$--$\tilde g_2$ plane reads (see figure \ref{fig:g20g21nmax})
\begin{equation}\label{eq:primal}
\begin{aligned}
&\text{maximize}\quad g_{2,1}=\langle K_{2,1}\rangle \quad\text{over}\quad \rho_J\geq0,\\
&\text{subject to}\quad \langle K_{1,0}\rangle = g^*_{1,0}\equiv1,\qquad \langle K_{2,0}\rangle = g^*_{2,0},\qquad \langle f_n\rangle=0 \quad\forall\,n ,
\end{aligned}
\end{equation}
where $g^*_{2,0}=\tilde g_2$ is the value at which we slice the plane,
and twice the maximum of $g_{2,1}$ gives the upper-boundary value
of $\tilde g_2'$. The same factor of two converts the lower bound.

The standard step is to introduce a Lagrange multiplier for every equality constraint,
\begin{equation}\label{eq:primal_lagrangian}
    \mathcal{L}=\langle K_{2,1}\rangle-\lambda_{1,0} \big( \langle K_{1,0}\rangle-g_{1,0}^*\big)-\lambda_{2,0} \big( \langle K_{2,0}\rangle-g_{2,0}^*\big)-\sum_n \lambda_n \langle f_{n}\rangle .
\end{equation}
The multipliers are unconstrained 
since every constraint is an equality: the only inequality of the problem is $\rho_J\geq0$. Their overall sign is a convention, that we fixed so that the dual condition below takes the form of a positivity. 
On any feasible density $\mathcal{L}$ coincides with the objective $g_{2,1}$. Exchanging the roles of the densities and of the multipliers, we rewrite the same expression as
\begin{equation}\label{eq:dual_lagrangian}
    \mathcal{L}=\lambda_{1,0} \, g_{1,0}^*+\lambda_{2,0} \, g_{2,0}^* - \Big\langle\, \lambda_{1,0} K_{1,0}+\lambda_{2,0} K_{2,0}-K_{2,1}+\sum_n \lambda_n f_{n} \,\Big\rangle.
\end{equation}
The first term only depends on the multipliers, which becomes the dual objective. The second is linear in the densities: if the kernel inside the average is non-negative pointwise, the average is non-negative for every admissible $\rho_J$, and \eqref{eq:dual_lagrangian} implies $g_{2,1}\leq \lambda_{1,0} \, g_{1,0}^*+\lambda_{2,0} \, g_{2,0}^*$ for every feasible density. Collecting
\begin{equation}\label{eq:dual_vectors}
    \vec{f}(z,J)=\left( K_{1,0}, K_{2,0}, K_{2,1}, \{f_n\} \right), \qquad \vec{\lambda}=\left( \lambda_{1,0}, \lambda_{2,0}, \lambda_{2,1}, \{\lambda_n\} \right),
\end{equation}
the dual problem reads
\begin{equation}\label{eq:dual}
\begin{aligned}
    &\text{minimize}\quad \lambda_{1,0} \, g_{1,0}^*+\lambda_{2,0} \, g_{2,0}^* ,\\
    &\text{subject to}\quad \vec{\lambda}\cdot\vec{f}(z,J)\geq 0 \quad \forall \, z\geq \Lambda^2, \ \forall \, J, \qquad \lambda_{2,1}=-1 .
\end{aligned}
\end{equation}
Every feasible $\vec\lambda$ gives a rigorous upper bound on $g_{2,1}$, and the minimization over $\vec\lambda$ gives the best one; for a linear problem of this type no duality gap is expected, and the dual optimum coincides with the primal maximum. The lower boundary of the region is obtained in the same way with $\lambda_{2,1}=+1$, in which case $-(\lambda_{1,0} \, g_{1,0}^*+\lambda_{2,0} \, g_{2,0}^*)$ is a lower bound on $g_{2,1}$, to be maximized. Scanning $g^*_{2,0}$ over its allowed range \eqref{eq:simple_bounds} traces the whole boundary of figure~\ref{fig:g20g21nmax}.

\subsection{Dual problem for the couplings}\label{subsec:dual_coupling}

For the coupling problem the $\rho$ and $f_2$ poles are resolved explicitly, and every linear functional is replaced by the extended average $\langle h\rangle'$ of \eqref{eq:average_poles}. We first assume for simplicity that both states lie above the two-pion threshold; the bound-state case is discussed in the next section~\ref{app:Bound_state}. 
With the normalization
\[
g_{1,0}=\langle K_{1,0}\rangle'=1,
\]
necessary to obtain finite bounds, the observable studied in Section \ref{subsec:f2_coupling} is defined as the fraction of $g_{1,0}$ coming from the $f_2$ pole,
\begin{equation}
\tilde g^2_{\pi\pi f_2}
=
g^2_{\pi\pi f_2}\,
m_{f_2}^2\,
\frac{K_{1,0}(m_{f_2}^2,2)}{g_{1,0}}.
\label{eq:f2coupling}    
\end{equation}

The primal problem is therefore
\begin{equation}\label{eq:primal_coupling}
\begin{aligned}
&\text{maximize}\quad
g^2_{\pi\pi f_2}\,
m_{f_2}^2\,
K_{1,0}(m_{f_2}^2,2)
\quad \text{over}\quad
g^2_{\pi\pi\rho}\geq0,\quad
g^2_{\pi\pi f_2}\geq0,\quad
\rho_J\geq0,
\\
&\text{subject to}\quad
\langle K_{1,0}\rangle'=1,
\qquad
\langle f_n\rangle'=0
\quad\forall\,n .
\end{aligned}
\end{equation}
Proceeding as in the previous subsection, define $\Phi(z,J)
\equiv
 \lambda_{1,0}K_{1,0}(z,J)
 +\sum_n\lambda_n f_n(z,J).$
The Lagrangian can then be written as
\begin{equation}\label{eq:dual_lagrangian_coupling}
\begin{aligned}
\mathcal{L}
=&\;
\lambda_{1,0}
-g^2_{\pi\pi\rho}\,m_\rho^2\,\Phi(m_\rho^2,1)
-
g^2_{\pi\pi f_2}\,m_{f_2}^2
\Big[
\Phi(m_{f_2}^2,2)
-K_{1,0}(m_{f_2}^2,2)
\Big]
\\
&-
\sum_J\int_{\Lambda^2}^{\infty}
dz\,16(2J+1)\rho_J(z)\Phi(z,J).
\end{aligned}
\end{equation}
On a feasible configuration $\mathcal L=\tilde g^2_{\pi\pi f_2}$. Since all spectral weights are non-negative, the conditions
\[
\Phi(m_\rho^2,1)\geq0,
\qquad
\Phi(m_{f_2}^2,2)\geq K_{1,0}(m_{f_2}^2,2),
\qquad
\Phi(z,J)\geq0
\]
imply
\[
\tilde g^2_{\pi\pi f_2}\leq\lambda_{1,0}.
\]
The dual problem is therefore
\begin{equation}\label{eq:dual_coupling}
\begin{aligned}
&\text{minimize}\quad \lambda_{1,0},
\\
&\text{subject to}\quad
\Phi(m_\rho^2,1)\geq0,
\qquad
\Phi(m_{f_2}^2,2)
\geq K_{1,0}(m_{f_2}^2,2),
\\
&\hspace{34mm}
\Phi(z,J)\geq0
\qquad
\forall\,z\geq\Lambda^2,\ \forall\,J .
\end{aligned}
\end{equation}
The inhomogeneous condition at the $f_2$ mass is the analogue of fixing the coefficient of the objective kernel in the Wilson-coefficient problem. At fixed $m_\pi$, $m_\rho$ and $m_{f_2}$, the resulting bound is a function of $\Lambda$. When the bound is saturated, complementary slackness implies that the unresolved spectral density can have support only at zeros of the optimal functional $\Phi(z,J)$; similarly, a nonzero $\rho$ residue requires $\Phi(m_\rho^2,1)=0$, while at the $f_2$ pole we have
\[
\Phi(m_{f_2}^2,2)
=
K_{1,0}(m_{f_2}^2,2).
\]

\subsection{Bound states}\label{app:Bound_state}

When the $\rho$ becomes lighter than two pions, its pole lies below
the two-pion threshold. We consider the mass range
\[
    \sigma<m_\rho^2<4m_\pi^2<m_{f_2}^2<\Lambda^2,
    \qquad \sigma=\frac{4}{3}m_\pi^2,
\]
so that the crossing-symmetric point remains below the $\rho$ pole,
while the $f_2$ and the continuum remain above threshold.
In the convention of \eqref{eq:coupling_def}, the physical coupling
condition \eqref{eq:physical_coupling} gives
\[
    g_{\pi\pi\rho}^2m_\rho^2
    =\hat g_\rho^2(m_\rho^2-4m_\pi^2)\leq0,
    \qquad \hat g_\rho^2\geq0.
\]
The coefficient $g_{\pi\pi\rho}^2$ therefore has the opposite sign
to that of a state above threshold. From the Lagrangian
\eqref{eq:dual_lagrangian_coupling}, the corresponding dual condition is
\begin{equation}\label{eq:dual_bound_state}
    \Phi(m_\rho^2,1)\leq0,
\end{equation}
replacing the above-threshold condition $\Phi(m_\rho^2,1)\geq0$.
The conditions at the $f_2$ pole and on the continuum are those of
\eqref{eq:dual_coupling}.

With the physical residue sign, positivity of the normalization
$g_{1,0}$ continues to hold term by term. Although the $\rho$ kernel
is negative,
\[
    K_{1,0}(m_\rho^2,1)
    =\frac{1}{(m_\rho^2-\sigma)(m_\rho^2-4m_\pi^2)}<0,
\]
its contribution to $g_{1,0}$ is non-negative:
\begin{equation}\label{eq:g10_bound_state}
    g_{\pi\pi\rho}^2m_\rho^2K_{1,0}(m_\rho^2,1)
    =\frac{\hat g_\rho^2}{m_\rho^2-\sigma}\geq0.
\end{equation}
The $\rho$ contributions to $g_{2,0}$ and $g_{2,1}$ are likewise
$\hat g_\rho^2/(m_\rho^2-\sigma)^2\geq0$.
Since the $f_2$ and the continuum also contribute non-negatively
to $g_{1,0}$, the normalization is strictly positive for every
nonzero physical spectrum.

For the numerical results of section~\ref{sec:bounds} in the
bound-state regime, we leave the sign of $g_{\pi\pi\rho}^2$
unrestricted. In the dual problem this replaces
\eqref{eq:dual_bound_state} by
\[
    \Phi(m_\rho^2,1)=0.
\]
Allowing either residue sign enlarges the primal feasible set.
The resulting upper bounds therefore remain valid for the physical
problem, although, in principle, they may be weaker than those obtained by imposing
the physical sign condition. Here if we instead set the explicit opposite sign condition
\[
    K_{1,0}(m_\rho^2,1)
    =\frac{1}{(m_\rho^2-\sigma)(m_\rho^2-4m_\pi^2)}>0,
\]
the problem becomes unfeasible. 
Thus our bounds are equivalent to those of the physical setup.

In this relaxed problem, the individual $\rho$ contribution to
$g_{1,0}$ need not be positive. Positivity of the total normalization
instead follows from the null constraint $\mathcal X^{(2,1)}$.
Appendix~\ref{app:g10positivity} uses this constraint to eliminate
the $\rho$ contribution and obtain \eqref{eq:g10_final}, which
expresses $g_{1,0}$ as the unchanged $f_2$ contribution plus a
continuum integral with strictly positive kernels.
If $g_{1,0}=0$, the $f_2$ residue and the continuum must both vanish;
the same null constraint then forces the $\rho$ residue to vanish.
Thus $g_{1,0}>0$ for every nonzero spectrum satisfying the relaxed
constraints, and the normalization $g_{1,0}=1$ remains justified.
Since the $f_2$ contribution is unchanged in this representation,
its normalized weight \eqref{eq:norm} also remains between zero
and one.

\section{Numerical Implementation}\label{sec:numerics}

Every dual problem asks for a finite vector $\vec\lambda$ that is non-negative against an infinite family of constraints: one for every spin $J\geq0$ and every $z$ above the cutoff $\Lambda^2$. No solver accepts infinitely many conditions; we impose some of them and we check the convergence afterwards. Here we describe the procedure for the pion problems; the glueball bound is treated in the same way.

For fixed $J$ the kernels \eqref{eq:kernels} and the null constraints \eqref{eq:null_XY} are rational in $z$. Multiplying them by $(3z-4m_\pi^2)^{n_{\max}+1}(z-4m_\pi^2)^J$, which is positive above the cutoff, turns every component into a polynomial of degree $J+n_{\max}-1$. For $J\leq J_{\rm poly}$ we impose positivity on the whole half-line $z\geq \Lambda^2$, as a semidefinite condition. Since the degree grows with $J$, above $J_{\rm poly}$ this becomes too expensive, so we impose positivity only at a finite set of points $(z,J)$. This is a relaxation: the resulting bounds approach the true one from below.

To approach the true bound, we choose the points iteratively. Each iteration solves the current problem, evaluates the resulting functional at many new points, and adds those where it is negative. We stop when the scan finds no violation, or when the bound changes by less than $10^{-6}$ in relative terms for three consecutive iterations.\footnote{We are grateful to Kelian Häring for suggesting us to implement this strategy.}

At $n_{\max}=16$ a point takes typically twenty iterations and about 2500 constraints from the scan, and the last increment is of order $10^{-8}$.\footnote{These values are measured on the coupling problem; the runs for the Wilson coefficients behave similarly, with between four and fifteen iterations.} For the glueball bound a point takes between 7 and 10 iterations and about 600 constraints from the scan, with last increments between $10^{-11}$ and $10^{-7}$.
The parameters of the three problems are collected in table~\ref{tab:numerics}.
\begin{table}[tbp]
\centering
\begin{tabular}{llll}
\toprule
 & Wilson coefficients & coupling & glueball \\
\midrule
$J_{\rm poly}$ & $20$ & $20$ & $40$ \\
$J_{\max}$ & $10^4$ & $10^{10}$ & $10^{10}$ \\
working precision & $110$ digits & $110$ digits  &$110$ digits \\
\texttt{SDPB duality gap threshold} & $10^{-10}$ & $10^{-10}$& $10^{-10}$ \\
\bottomrule
\end{tabular}
\caption{Numerical parameters of the three problems.}
\label{tab:numerics}
\end{table}
As an a posteriori check, we reconstruct the optimal $\vec\lambda$ and evaluate the dual functional where the scan never looks: at $J>J_{\max}$, above the maximal energy of the scan, and as $z\to\infty$. For $0.03\leq m_\pi^2/m_{\rho,0}^2\leq0.40$ (Figures~\ref{fig:gpipi_nmax} and~\ref{fig:gpipi_masses}) it is positive there, up to numerical noise.

\subsection{Spectrum extraction}
\label{subsec:spectrum-extraction}
The spectrum is extracted as a series of zeros in the dual formulation.
We extract the spectrum in two steps: the zeros of the optimal dual functional, as returned by \texttt{spectrum} of \texttt{SDPB}, give a first and incomplete picture; we then refine it by locating the local minima of the functional directly.

In this section we focus on the spectra of the $\tilde g_{\pi\pi f_2}^2$ problem extracted in section~\ref{subsec:f2_coupling}; analogous considerations hold for the glueball problem.
The spectrum at the kink in figure~\ref{fig:spectrum_at_the_kink} is read off the zeros of the optimal dual functional, at each point of the kink line and each truncation order $n_{\max}$. The \texttt{SDPB} zero search is performed with an aggressive threshold (\texttt{--threshold=1e-10}). As a consequence, some physical resonances are cut, and the spectrum at fixed $n_{\max}$ has gaps in spin. At the same time, spurious poles appear, as also observed in~\cite{Albert:2022oes,Albert:2026xyz,Caron-Huot:2021rmr,us_to_appear}. Those poles carry small residues, and we discard every state with a normalized residue below $10^{-4}$.

We first build a reference trajectory. Starting from $n_{\max}=16$ and moving down to $n_{\max}=9$, for each spin we keep the states that pass the residue cut. The gaps left at higher $n_{\max}$ are therefore filled from lower truncation orders. Together with the $\rho$ and the $f_2$, this yields a single trajectory for every value of $m_\pi^2$.

\begin{figure}
    \centering
    \includegraphics[width=\linewidth]{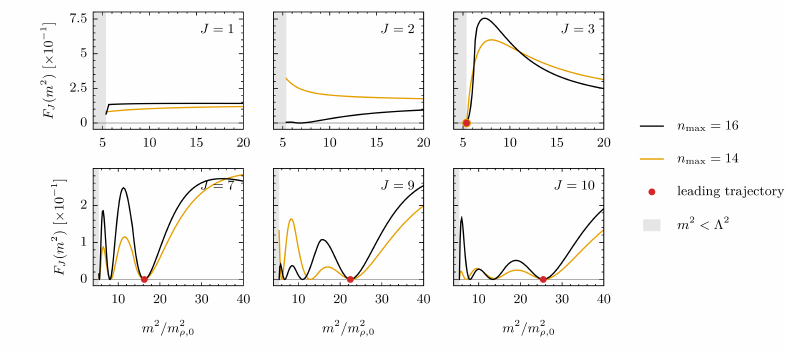}
    \caption{Dual functionals for different spins of the $\tilde g_{\pi\pi f_2}^2$ maximization problem at the kink. The orange line is the dual functional at $n_\text{max}=14$, the black one at $n_\text{max}=16$, and the red dots are the minima on the leading trajectory. This dual functional differs from the one seen by \texttt{spectrum} of \texttt{SDPB} by the prefactor normalization $e^{-x}$ and the positive polynomials discussed in section~\ref{sec:numerics}.}
    \label{fig:dual_functionals}
\end{figure}

As we can see from figure~\ref{fig:dual_functionals}, the dual functional is full of other zeros that usually appear between the threshold and the reference trajectory, and are more or less stable in $n_\text{max}$ (e.g. stable for $J=7$, unstable for $J=9$); they appear as a consequence of numerical truncation in spin and crossing \cite{us_to_appear}, and we expect them to disappear at higher $n_\text{max}$.
Repeating the optimization problem with a spin dependent threshold, that removes most of those zeros, does not change either the optimal solution or the position of the physical resonances~\cite{Albert:2023seb}. We have explicitly checked it for the kinks found in Figures~\ref{fig:gpipi_nmax}, \ref{fig:glue-bound-nmax16} and \ref{fig:glueball-plane}.

As the plots show, physical states sit at local minima of the functional. We therefore extract the trajectory at the highest fixed $n_{\max}$: for each spin we look for a local minimum of the optimal dual functional in a window around the expected position, given by the reference trajectory, by linear interpolation between neighbouring spins or, above the highest reference spin, by linear extrapolation from the highest-spin resonances. A minimum is not guaranteed to exist (numerically) inside the window: if none is found, the spin is left out. Where both are available, the minima found in this way agree with the zeros of the original extraction of \texttt{SDPB}.

The quoted masses are those at $n_{\max}=16$. Their uncertainty is the sum in quadrature of two contributions. The first is the width of the minimum, namely the range in which the functional stays compatible with its minimum value. The second is the convergence in $n_{\max}$, which we model as a uniform distribution of width twice the shift between the minima at $n_{\max}=16$ and $n_{\max}=14.$\footnote{The former error is usually negligible compared to the latter.} The truncation in $J_{\max}$ gives a further error, which is negligible.

\subsection{The meson trajectory}\label{appsubsec:the_meson_trajectory}
We extract the meson trajectory at the kink in Figure~\ref{fig:gpipi_masses} for different pion masses and, at fixed spin, we study the evolution of the masses with the pion mass. The squared mass of each resonance is well described by a linear function of $m_\pi^2$: $m_J^2 = A_J + B_J\, m_\pi^2$, in units of $m_{\rho,0}^2$. The spin-3 state sits at the continuum threshold, $m_3^2=\Lambda^2$, and its line is the fit of the kink positions of figure~\ref{fig:kinkfits}, eq.~\eqref{eq:rho3_mass}; the masses with $J\geq4$ are extracted from the extremal functionals at $n_{\max}=16$, setting the threshold along that line.

Besides the two errors above, each mass carries a third one, due to the location of the kink. The width is independent from point to point and weights the fit. The truncation in $n_{\max}$ and the kink shift all the points of a given spin together, so they are not averaged out by the fit: we evaluate their effect on $A_J$ and $B_J$ separately and add it in quadrature \textit{a posteriori}.

\begin{figure}[t!]
    \centering
    \includegraphics[width=\textwidth]{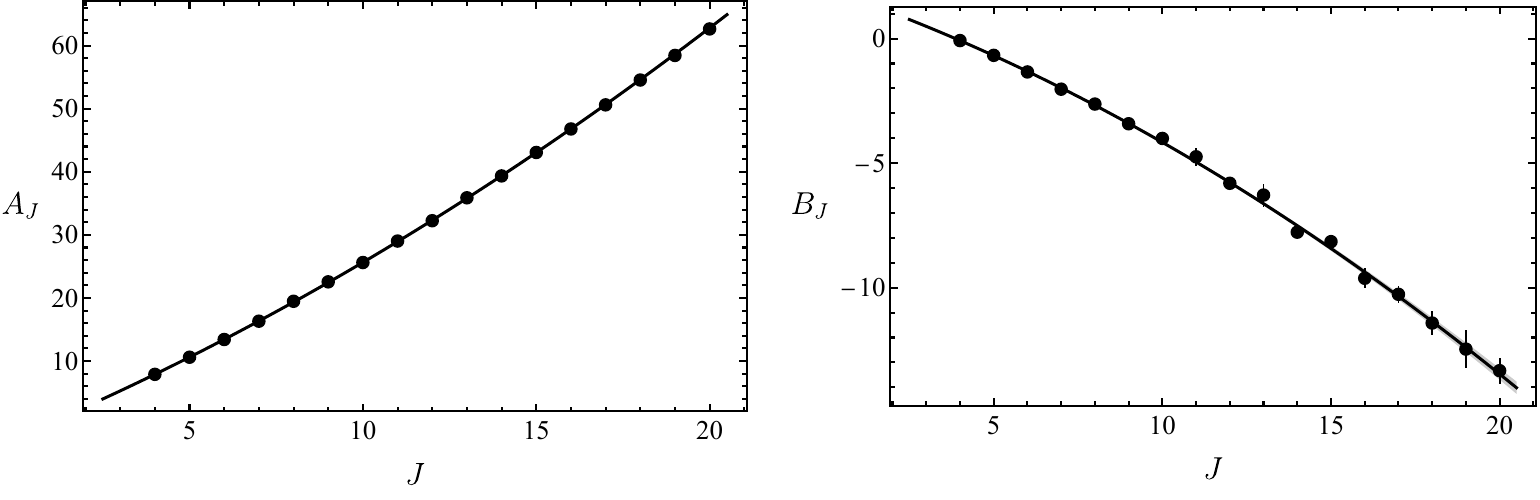}
    \caption{Spin dependence of the fitted pion-mass evolution \eqref{eq:trajectory_pion_mass_fit}. Left: the extrapolated chiral-limit squared masses $A_J$. Right: the pion-mass slopes $B_J$. Error bars are the numerical uncertainty estimates described in the text. Solid lines are quadratic fits for $4\leq J\leq20$ using \eqref{eq:quadratic_regge_params}. }
    \label{fig:quadratic_fits}
\end{figure}

The coefficients $A_J$ and $B_J$ are then described by quadratic polynomials in $J$ over $4\leq J\leq20$, as shown in figure~\ref{fig:quadratic_fits}. These two coefficients at a given spin are correlated. In these fits the width error is independent between spins, while the kink and truncation error are common to all of them.

Combining the fits gives the interpolation
\begin{equation}\label{eq:quadratic_regge}
    \frac{m_J^2}{m_{\rho,0}^2}
    =\left(\alpha_A+\alpha_B \frac{m^2_\pi}{m^2_{\rho,0}}\right)+\left(\beta_A+\beta_B \frac{m^2_\pi}{m^2_{\rho,0}}\right)J+\left(\gamma_A+\gamma_B \frac{m^2_\pi}{m^2_{\rho,0}}\right)J^2,
    \qquad 4\leq J\leq20,
\end{equation}
with
\begin{equation}\label{eq:quadratic_regge_params}
    \begin{aligned}
        \alpha_A&=-2.1(1), & \beta_A&=2.32(3), & \gamma_A&=0.046(2),\\
        \alpha_B&=1.99(7), & \beta_B&=-0.46(2), & \gamma_B&=-0.016(1).
    \end{aligned}
\end{equation}
In particular, their central values give $B_J=\alpha_B+\beta_BJ+\gamma_BJ^2<0$ throughout the fitted spin range. The decreasing higher-spin masses thus characterize the numerical extrema obtained with the prescribed $\rho$ and $f_2$ inputs.

Since $\gamma_A+\gamma_B \frac{m^2_\pi}{m^2_{\rho,0}}>0$ over the sampled pion masses, $m_J^2$ is convex in $J$: the trajectory $J(m^2)$ is concave, and its slope decreases with spin. At fixed spin, increasing the pion mass decreases $\partial m_J^2/\partial J$, that is, it increases the slope of the trajectory.

\section{Determination of the experimental value of \texorpdfstring{$\tilde g^2_{\pi\pi f_2}$}{g2pipif2}}
\label{app:exp}

In this appendix we extract $\tilde g^2_{\pi\pi f_2}$, defined in \eqref{eq:norm}, from real-world data. The coupling entering the sum rules is the residue of the flavour-ordered amplitude at the $f_2$ pole, Eq.~\eqref{eq:coupling_def}. We relate it to the measured $f_2\to\pi\pi$ width through unitarity, for $N_f=2$ and in the normalization of Eq.~\eqref{eq:partial_waves}.

\paragraph{From the residue to the singlet density.}
At $N=\infty$ the $f_2$ pole gives $\mathrm{Im}\,M(s,t)=\pi g^2_{\pi\pi f_2}m_{f_2}^2\,\delta(s-m_{f_2}^2)\,P_2(x_t)$. Projecting it into partial waves
\begin{equation}
\rho_2(s)=\frac{1}{32\pi}\int_{-1}^{1}dx\,P_2(x)\,\mathrm{Im}\,M
=\frac{g^2_{\pi\pi f_2}m_{f_2}^2}{80}\,\delta(s-m_{f_2}^2)\,,
\end{equation}
in agreement with \eqref{eq:rho_meromorphic}. The $f_2$ is an isospin singlet. Writing $\mathrm{Im}\,M_{\rm singlet}=16\pi\sum_J(2J+1)\,\rho^{\rm singlet}_J P_J(x_t)$, Eqs.~\eqref{eq:nice_channels} and \eqref{eq:proj_nice} give $\rho^{\rm singlet}_J=6\rho_J$ for even $J$, hence
\begin{equation}\label{eq:rhosinglet}
\rho^{\rm singlet}_2(s)=\frac{3}{40}\,g^2_{\pi\pi f_2}m_{f_2}^2\,\delta(s-m_{f_2}^2)\,.
\end{equation}

\paragraph{Unitarity and the width.}
Define $f_J(s)=\frac{1}{32\pi}\int_{-1}^1dx\,P_J(x)\,M_{\rm singlet}(s,t)$, so that $\rho^{\rm singlet}_J=\mathrm{Im}\,f_J$. For identical pions, elastic two-pion unitarity in this normalization reads
\begin{equation}
\mathrm{Im}\,f_J=\frac{\beta}{2}\,|f_J|^2\,,\qquad \beta(s)=\sqrt{1-\frac{4m_\pi^2}{s}}\,,
\end{equation}
where the factor $\beta/2$ comes from the two-body phase space, $\int d\Phi_2=\frac{\beta}{8\pi}\int\frac{d\Omega}{4\pi}$, together with the symmetry factor $1/2$. At finite $N$ the $f_2$ pole moves to $s=m_{f_2}^2-i\,m_{f_2}\Gamma_{f_2}$ and, near the pole, $f_2\simeq R/(m_{f_2}^2-s-i m_{f_2}\Gamma_{f_2})$. Unitarity then fixes the $\pi\pi$ contribution to the width in terms of the residue, $m_{f_2}\Gamma_{\pi\pi}=\beta_{f_2}R/2$, with $\beta_{f_2}\equiv\beta(m_{f_2}^2)$. In the narrow-width limit
\begin{equation}
\mathrm{Im}\,f_2\;\to\;\pi R\,\delta(s-m_{f_2}^2)=\frac{2\pi m_{f_2}\Gamma_{\pi\pi}}{\beta_{f_2}}\,\delta(s-m_{f_2}^2)\,.
\end{equation}
Comparing with \eqref{eq:rhosinglet},
\begin{equation}\label{eq:widthf2}
\Gamma(f_2\to\pi\pi)=\frac{3\,g^2_{\pi\pi f_2}}{80\pi}\,m_{f_2}\,\beta_{f_2}\,.
\end{equation}
This differs from Eq.~\eqref{eq:M2_onshell} of~\cite{Albert:2023seb} by a factor $\beta_{f_2}^4$. There the coupling, which we denote $g^2_{\rm v}$, is defined through the three-point vertex normalized for massless pions, so the width contains the full centrifugal factor $\beta_{f_2}^{2J+1}$. In the notation of \eqref{eq:physical_coupling}, $g^2_{\rm v}=\hat g^2 m_{f_2}^{2J-2}$, and $g^2_{\pi\pi f_2}=\beta_{f_2}^{4}\,g^2_{\rm v}$. The two definitions coincide in the chiral limit. At finite pion mass only the residue enters the sum rule for $g_{1,0}$.

\paragraph{Numerical value.}
With the inputs of Table~\ref{tab:exp_value_N3} we have $\beta_{f_2}=0.9758$ and $\Gamma_{\pi\pi}=\Gamma_{f_2}\,{\rm BR}(f_2\to\pi\pi)=156.6^{+5.7}_{-2.6}$~MeV, and \eqref{eq:widthf2} gives
\begin{equation}
g^2_{\pi\pi f_2}=10.54^{+0.38}_{-0.17}\,.
\end{equation}
Using $g_{1,0}=1/(2f_\pi^2)$, with $x_0(m_{f_2}^2)=1.0335$ and $P_2(x_0)=1.1023$, we obtain
\begin{equation}\label{eq:D7}
\tilde g^{2\,({\rm exp})}_{\pi\pi f_2}=0.125^{+0.005}_{-0.003}\,.
\end{equation}
The result increases monotonically with $\Gamma_{f_2}$, ${\rm BR}$ and $f_\pi$. We therefore combine the upper and lower uncertainties separately in quadrature, treating the three inputs as uncorrelated; the uncertainties on the masses are negligible. 
 
\begin{center}
\begin{tabular}{|l|c|l|}
\hline
Quantity & Value & Source \\
\hline
$m_{\pi^\pm}$ & $139.57039 \pm 0.00018$ MeV & PDG 2026~\cite{PDG2026} \\
$m_\rho$ & $775.26 \pm 0.23$ MeV & PDG 2026~\cite{PDG2026} (neutral only, $e^+e^-$) \\
$f_\pi = f_{\pi^+}/\sqrt{2}$ & $92.07 \pm 0.85$ MeV & PDG 2024, Eq.~(72.14)~\cite{PDG2024} \\
$m_{f_2}$ & $1275.4 \pm 0.8$ MeV & PDG 2026~\cite{PDG2026} \\
$\Gamma_{f_2}$ & $185.8^{+2.8}_{-2.1}$ MeV & PDG 2026~\cite{PDG2026} \\
$\mathrm{BR}(f_2\to\pi\pi)$ & $\left(84.3^{+2.8}_{-1.0}\right)\%$ & PDG 2026~\cite{PDG2026} \\
\hline
\end{tabular}
\captionof{table}{PDG inputs and derived values for the $f_2\pi\pi$ coupling at $N=3$. Uncertainties on derived quantities are combined in quadrature, treating the $f_2$ width and branching ratio as uncorrelated.}
\label{tab:exp_value_N3}
\end{center}

\paragraph{Theoretical uncertainties.}
The quoted uncertainty is experimental only. The narrow-width extraction neglects effects of order $\Gamma_{f_2}/m_{f_2}\simeq0.15$. In addition, \eqref{eq:D7} uses the leading-order relation $g_{1,0}=1/(2f_\pi^2)$, while \eqref{eq:gtilde} gives $g_{1,0}=(1+2\epsilon)/(2f_\pi^2)$, with
\begin{equation}
\epsilon=\frac{m_\pi^2}{f_\pi^2}\Big[c_{1,0}+\frac83\big(c_{2,0}+c_{1,1}\big)\Big].
\end{equation}
We now estimate $\epsilon$ at the physical point. The coefficients $c_{n,k}$ are given in \eqref{eq:c_vs_L} in terms of the low-energy constants $L_3'$, $L_4'$ and $L_5$. Since $c_{2,0}=2L_3'+4L_4'$, we can write $c_{1,0}=4(L_5-2L_3'-4L_4')=4L_5-4c_{2,0}$, so that
\begin{equation}
\epsilon=\frac{4L_5m_\pi^2}{f_\pi^2}+\frac{m_\pi^2}{f_\pi^2}\Big[\frac83\,c_{1,1}-\frac43\,c_{2,0}\Big].
\end{equation}
The two terms have a different origin, and we estimate them separately.

The first term contains the constant $L_5$, which multiplies the quark mass in the chiral Lagrangian. By \eqref{eq:mpi_fpi} it equals $f_\pi/F-1$, where $F$ is the pion decay constant in the chiral limit: it is the relative change of the decay constant due to the quark mass. Since $L_5$ is leading in $1/N$, we use its phenomenological value at $N=3$, $L_5\simeq1.0\times10^{-3}$~\cite{Bijnens:2014lea}. With $m_\pi=139.6$~MeV and $f_\pi=92.1$~MeV this gives $4L_5m_\pi^2/f_\pi^2\simeq0.01$.\footnote{The full ratio at $N=3$, $f_\pi/F\simeq1.06$, also includes chiral logarithms and the constant $L_4$, both suppressed at large $N$.}

The second term contains only the four-derivative couplings $c_{2,0}$ and $c_{1,1}$, which are related to the normalized coefficients bounded in section~4.1. By \eqref{eq:g2defs} and \eqref{eq:g2eg2prime}, at leading order, $\tilde g_2=2\Lambda^2c_{2,0}/f_\pi^2$ and $\tilde g_2'=4\Lambda^2c_{1,1}/f_\pi^2$, so that
\begin{equation}
\frac{m_\pi^2}{f_\pi^2}\Big[\frac83\,c_{1,1}-\frac43\,c_{2,0}\Big]=\frac23\,\frac{m_\pi^2}{\Lambda^2}\,\big(\tilde g_2'-\tilde g_2\big)\,,
\end{equation}
up to corrections of order $m_\pi^4/\Lambda^4$. At the physical point, with $\Lambda=m_\rho$, the prefactor is $\frac23\,m_\pi^2/m_\rho^2\simeq0.02$. In the allowed region of Figure~\ref{fig:g20g21masses} at $m_\pi^2/\Lambda^2=0.03$, $\tilde g_2'-\tilde g_2$ ranges from $\simeq-1.04$, at the upper-left corner, to $\simeq2.3$, at the upper-right corner, so this term lies between $-0.02$ and $0.05$ for any amplitude compatible with our assumptions. For a $\rho$-dominated amplitude, at the endpoint of the Skyrme segment, $\tilde g_2'-\tilde g_2\simeq1$ and the term is $\simeq0.02$.

Adding the two contributions, a $\rho$-dominated amplitude gives $\epsilon\simeq0.03$, so that the correction to \eqref{eq:D7} is $2\epsilon\simeq0.06$.

\section{Positivity of the normalization in the bound-state problem}
\label{app:g10positivity}
We show that $g_{1,0}>0$ for every nonzero spectrum satisfying the sum rules, even when the sign of the $\rho$ residue is left free, as in the relaxed bound-state problem of section~\ref{app:Bound_state}. We work in the mass range $\sigma<m_\rho^2<4m_\pi^2<m_{f_2}^2<\Lambda^2$, with $\sigma=\frac43m_\pi^2$, and assume $g_{\pi\pi f_2}^2\geq0$ and $\rho_J(z)\geq0$ above the cutoff. No sign condition on $g_{\pi\pi\rho}^2$ is used.

With the $\rho$ and the $f_2$ below the gap, $g_{1,0}$ \eqref{eq:kernels} and the first null constraint $\mathcal X^{(2,1)}$ \eqref{eq:null_const_rhof2} read
\begin{align}
g_{1,0}&=g_{\pi\pi\rho}^2m_\rho^2K_{1,0}(m_\rho^2,1)+g_{\pi\pi f_2}^2m_{f_2}^2K_{1,0}(m_{f_2}^2,2)+16\sum_J(2J+1)\int_{\Lambda^2}^{\infty}dz\,\rho_J(z)K_{1,0}(z,J),
\label{eq:g10_rhof2}\\
0&=g_{\pi\pi\rho}^2m_\rho^2\mathcal X^{(2,1)}_1(m_\rho^2)+g_{\pi\pi f_2}^2m_{f_2}^2\mathcal X^{(2,1)}_2(m_{f_2}^2)+16\sum_J(2J+1)\int_{\Lambda^2}^{\infty}dz\,\rho_J(z)\mathcal X^{(2,1)}_J(z).
\label{eq:X21_rhof2}
\end{align}
With $\zeta=z-\sigma$, $d=z-4m_\pi^2$ and $x=\zeta/d$, the kernels are
\begin{equation}
K_{1,0}(z,J)=\frac{P_J(x)}{\zeta^2},
\qquad
\mathcal X^{(2,1)}_J(z)=\frac{4}{\zeta^2d}\left(\frac{P'_J(x)}{\zeta}-\frac{P''_J(x)}{d}\right).
\label{eq:X21_kernel}
\end{equation}
Two cases are special. For spin one, $K_{1,0}(z,1)=1/(\zeta d)$ and
\begin{equation}
\mathcal X^{(2,1)}_1(z)=\frac{4}{\zeta^2}K_{1,0}(z,1);
\end{equation}
for spin two: $\mathcal X^{(2,1)}_2(z)=\frac{12}{\zeta^2d}\left(\frac{x}{\zeta}-\frac1d\right)=0$.

Solving \eqref{eq:X21_rhof2} in terms of \eqref{eq:g10_rhof2}, the $\rho$ term cancels in $g_{1,0}$:
\begin{equation}
g_{1,0}=g_{\pi\pi f_2}^2m_{f_2}^2K_{1,0}(m_{f_2}^2,2)+16\sum_J(2J+1)\int_{\Lambda^2}^{\infty}dz\,\rho_J(z)\,C_J(z),
\label{eq:g10_final}
\end{equation}
where 
\begin{equation}
C_J(z)\equiv K_{1,0}(z,J)-\frac{D_\rho^2}{4}\mathcal X^{(2,1)}_J(z)
=\frac{1}{\zeta^2}\left[P_J(x)+\frac{D_\rho^2}{d^2}\left(P''_J(x)-\frac{P'_J(x)}{x}\right)\right],
\label{eq:C_kernel_def}
\end{equation}
with $D^2_\rho=m^2_\rho-\sigma^2$.

We now show that $C_J(z)>0$ for every spin and every $z>m_{f_2}^2$. In this range $x=1+2\sigma/d>1$, so $P_J\geq1$ and, since $P_J(1+y)$ has positive coefficients in $y$, all its derivatives are non-negative. For $J=0$, $C_0=1/\zeta^2>0$. For $J=1$,
\begin{equation}
C_1(z)=\frac{1}{\zeta d}\left[1-\left(\frac{D_\rho}{\zeta}\right)^2\right]>0,
\end{equation}
because $z>m_{f_2}^2>m_\rho^2$ implies $\zeta>D_\rho$. For $J\geq2$, differentiating $P'_{J-1}=xP'_J-JP_J$ gives
\begin{equation}
xP''_J-P'_J=P''_{J-1}+(J-2)P'_J\geq0,
\end{equation}
so $P''_J-P'_J/x\geq0$ and $C_J >0$.

Every term in \eqref{eq:g10_final} is therefore non-negative, and any nonzero $f_2$ residue makes $g_{1,0}$ strictly positive.

\section{Fully crossing symmetric amplitudes}
\label{app:stu}

We can extend our method to the double-subtracted fully crossing symmetric amplitude of a
scalar of mass $m$, symmetric under all permutations of $s$, $t$
and $u=4m^2-s-t$: this is the massive analogue of \cite{Caron-Huot:2020cmc}. In this case only even spins contribute to the spectral
density.

\paragraph{Dispersion relation and null constraints.}
We subtract at the crossing symmetric point $s=t=\sigma=\tfrac43 m^2$ and define the kernel
\begin{equation}
K(z;s,t;t_0)\equiv\frac{1}{z-s}+\frac{1}{z-4m^2+s+t}
-\frac{1}{z-t_0}-\frac{1}{z-4m^2+t+t_0}\,.
\end{equation}
Dispersing at fixed $t$ and imposing the full $s\leftrightarrow t\leftrightarrow u$ symmetry, the crossing equations become a condition on the spectral density alone,
\begin{equation}
\sum_{J\,\text{even}}\int_{\Lambda^2}^\infty dz\,16(2J+1)\,\rho_J(z)\,
\mathcal{F}_J(z;s,t)=0\,,\qquad \forall\, s,t\,,
\end{equation}
with
\begin{equation}
\label{eq:crossing_symmetric_ncs}
\mathcal{F}_J(z;s,t)=P_J(x_t)\,K(z;s,t;\sigma)-P_J(x_u)\,K(z;s,u;\sigma)
+P_J(x_0)\Big[K(z;t,\sigma;\sigma)-K(z;u,\sigma;\sigma)\Big],
\end{equation}
where $x_t=1+\frac{2t}{z-4m^2}$, $x_u=1+\frac{2u}{z-4m^2}$ and $x_0=x_t|_{t=\sigma}$. Taylor expanding around the crossing symmetric point gives the null constraints
\begin{equation}
\label{eq:ncsforreal}
\mathcal{N}^{(n,k)}_J(z)\equiv\partial_s^n\partial_t^k\,
\mathcal{F}_J(z;s,t)\Big|_{s=t=\sigma}\,,\qquad n+k\le n_{\max}\,.
\end{equation}
Unlike the two families \eqref{eq:XY_def},
the set $\{(n,k)\}$ is highly redundant. At $n_{\max}=9$ there are 55
derivatives: 25 vanish identically, and the remaining 30 span a space of
dimension only 9; we keep a linearly independent subset.

\paragraph{Observables.}
The low energy coefficients are the Taylor coefficients of the amplitude
around the crossing symmetric point:
\begin{align}
g_2&=\sum_J\int dz\,16(2J+1)\rho_J(z)\,\frac{p_0}{(z-\sigma)^3}\,,\\
g_3&=\sum_J\int dz\,16(2J+1)\rho_J(z)\left[\frac{3\,p_0}{(z-\sigma)^4}
-\frac{2\,p_1}{(z-\sigma)^3}\right],\\
g_4&=\sum_J\int dz\,16(2J+1)\rho_J(z)\,\frac{p_0}{(z-\sigma)^5}\,,
\end{align}
where
\begin{equation}\label{eq:p_def}
\left.\partial_t^kP_J\bigl(x_t(z)\bigr)\right|_\sigma
=p_k\equiv
\left(\frac{2}{z-4m_\pi^2}\right)^kP_J^{(k)}\bigl(x_0(z)\bigr),
\qquad k\geq0.
\end{equation}
We define the dimensionless ratios
\begin{equation}
\tilde g_3\equiv\frac{g_3}{g_2}\,\Lambda^2\,,\qquad
\tilde g_4\equiv\frac{g_4}{2\,g_2}\,\Lambda^4\,,
\end{equation}
which for a massless Goldstone boson reduce to the variables of~\cite{Caron-Huot:2020cmc}.

\paragraph{Numerical setup.}
The boundary is obtained by maximizing/minimising $\tilde g_3$ at fixed $\tilde g_4$
with SDPB, at 100 digits of precision.
\begin{itemize}
\item \emph{Positivity.} It is imposed exactly as polynomial positivity for
$J_\text{poly}\le20$. Spins $20<J\le J_\text{max}=10^4$ are included on a grid in $z$, refined by a
cutting-plane search for violations restricted to even spins.
\item \emph{Convergence check.} Raising the spin cutoff $J_\text{max}$ from $10^4$ to $10^5$ changes
the bound by $10^{-7}$ relative. Going from $n_{\max}=10$ to $12$ changes it by
at most $10^{-3}$ relative, and by $10^{-5}$ close to $\tilde g_4^{\max}$.
\end{itemize}
All results below are at $n_{\max}=12$.

\begin{figure}[H]
    \centering
    \includegraphics[width=\linewidth]{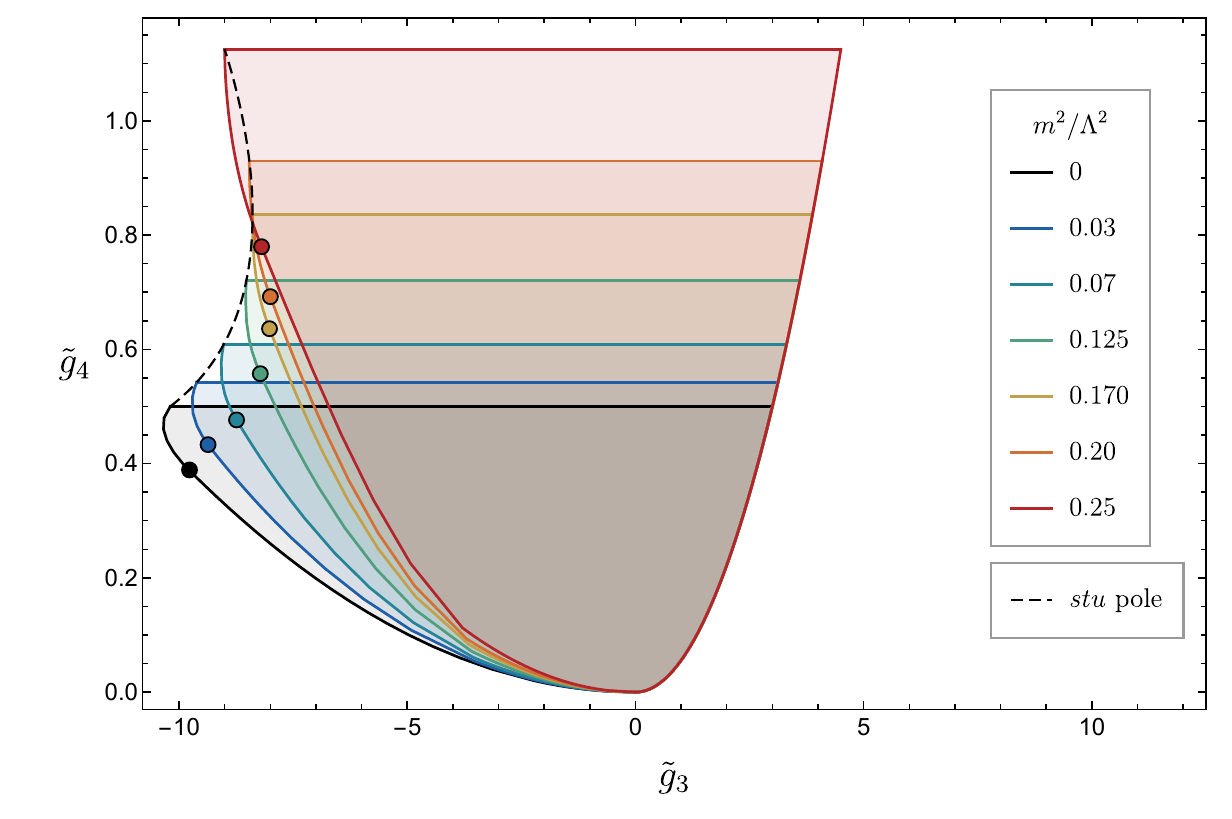}
    \caption{Allowed region in the $(\tilde g_3,\tilde g_4)$ plane for the
    $stu$ crossing symmetric problem, for several values of the scalar mass $m^2/\Lambda^2$. The left boundaries are numerical, at
    $n_{\max}=12$ for $m^2>0$; the massless curve is from
    \cite{Caron-Huot:2020cmc,us_to_appear} at $n_{\max}=10$. The upper line is
    $\tilde g_4^{\max}$ and the right boundary is the scalar-exchange parabola
    $\tilde g_4=\tilde g_3^2/18$. Dots mark the kinks of
    table~\ref{tab:stu-kink}. The dashed curve is the $stu$ pole, which traces
    the upper-left corner as $m^2/\Lambda^2$ varies. }
    \label{fig:stu}
\end{figure}

\paragraph{Analytic boundaries.}
Three parts of the boundary of the allowed region are known analytically.
\begin{itemize}
\item \emph{Upper boundary.} Since $\rho_J\ge0$, the ratio $g_4/g_2$ is
maximised by putting all the spectral weight at the gap, as in
section~\ref{subsec:g20_g21_plane}:
\begin{equation}
\tilde g_4\le\tilde g_4^{\max}=\frac{\Lambda^4}{2(\Lambda^2-\sigma)^2}
=\frac{9\Lambda^4}{2\,(3\Lambda^2-4m^2)^2}\,.
\end{equation}
\item \emph{Right boundary.} A single scalar of mass $z\ge\Lambda^2$ gives
$\tilde g_3=3\Lambda^2/(z-\sigma)$ and $\tilde g_4=\Lambda^4/\big(2(z-\sigma)^2\big)$. It
therefore saturates the parabola
\begin{equation}
\tilde g_4=\frac{\tilde g_3^2}{18}\,,\qquad
0\le\tilde g_3\le\frac{3\Lambda^2}{\Lambda^2-\sigma}\,,
\end{equation}
whose shape does not depend on the scalar mass.
\item \emph{Upper-left corner.} The left end of the upper boundary is fixed by
the $stu$ pole amplitude:
\begin{equation}\begin{split}
&\tilde g_3^{\,\rm corner}=
\frac{18\Lambda^2\Big[3\Lambda^2-12m^2+(6\Lambda^2-8m^2)\log\frac{\Lambda^2}{2\Lambda^2-4m^2}\Big]}
{(3\Lambda^2-4m^2)\Big[9\Lambda^2-36m^2+(12\Lambda^2-16m^2)\log\frac{\Lambda^2}{2\Lambda^2-4m^2}\Big]}\,,\\
&\tilde g_4=\tilde g_4^{\max}=\frac{9\Lambda^4}{2\,(3\Lambda^2-4m^2)^2}\,.
\end{split}
\end{equation}
It reproduces the massless value $-10.192$~\cite{Caron-Huot:2020cmc} at
$m=0$. At $m^2/\Lambda^2\to1/4$, where the two-particle threshold
$4m^2$ reaches the gap, it tends to $-9$. It is minimal in $\tilde g_3$ at
$m^2/\Lambda^2\simeq0.170$, where $\tilde g_3 \simeq-8.388$.
\end{itemize}

\paragraph{The kink.}
As in the massless case, the left boundary develops a kink in the derivative of the boundary \cite{Berman:2024eid,us_to_appear}: the derivative
$\partial\tilde g_3/\partial\tilde g_4$ is continuous, but its slope jumps. The position of the kink as a function of the mass of the scalar are reported in
table~\ref{tab:stu-kink}.

\begin{table}[H]
\centering
\begin{tabular}{c|ccc}
$m^2/\Lambda^2$ &
$\tilde g_4^{\rm kink}$ & $\tilde g_3^{\rm kink}$\\\hline
$0$~\cite{Caron-Huot:2020cmc,us_to_appear}  & $0.3889$ & $-9.77$\\
$0.03$ &  $0.433$ & $-9.36$\\
$0.07$ &  $0.476$ & $-8.74$\\
$0.125$ &  $0.557$ & $-8.22$\\
$0.170$ & $0.636$ & $-8.02$\\
$0.20$ & $0.692$ & $-8.00$\\
$0.25$ & $0.780$ & $-8.19$
\end{tabular}
\caption{Position of the kink on the left boundary at $n_{\max}=12$, as a
function of the scalar mass.}
\label{tab:stu-kink}
\end{table}

\section{The spectrum at the kink as a function of the pion mass}
\label{app:kink_spectrum_mpi}

In figure~\ref{fig:spectrum_masses} the resonances with $J\geq 4$ become lighter as the pion mass increases, although the string tension is kept fixed and the $\rho$ and the $f_2$ become heavier. We show that this is not an effect due to the pion mass itself, but to the ratio
\begin{equation}
	\label{eq:r_def}
	r\equiv\frac{m_{f_2}^2}{m_\rho^2}
	=1+\frac{\Delta}{m_\rho^2},
	\qquad
	\Delta\equiv m_{f_2}^2-m_\rho^2=\frac{1}{\alpha'} ,
\end{equation}
which by \eqref{eq:rho_mass} and \eqref{eq:f2_mass} decreases with $m_\pi$ at fixed $\Delta$. At fixed $r$, but with a mass $m_\pi^2 \simeq 7 \, m_\pi^2(\text{phys})$, our spectrum at the kink agrees with the massless one of \cite{Albert:2023seb}, within $1 \%$, for every spin up to $J=12$. The extraction of the spectra is described in appendix~\ref{subsec:spectrum-extraction}.

\paragraph{Why the trajectory bends.}
The null constraints \eqref{eq:null_XY} and the normalisation are homogeneous in the masses, so the extremal problem depends only on $r$, on $\mu\equiv m_\pi^2/m_\rho^2$ and on the cutoff $\Lambda^2/m_\rho^2$. The position of the kink fixes the last one in terms of the first two, and the spectrum at the kink takes the form
\begin{equation}
	\label{eq:F_J}
	m_J^2=m_\rho^2\,\Theta_J(r,\mu),
\end{equation}
for some function $\Theta_J$. For the heavy states, the pion mass enters $\Theta_J$ only through the argument of the Legendre polynomials: $x_t=1+2t/(s-4m_\pi^2)$. For a resonance of spin $J$, the impact parameter $b=2J/\sqrt{m_J^2-4m_\pi^2}$ is affected by a relative amount $2m_\pi^2/m_J^2$: for instance, at $m_\pi^2/m_{\rho,0}^2=0.2$, this is about $4\%$ for the $\rho_3$ and below $1\%$ at $J=12$. The dependence on $r$ is instead always of order one.

Neglecting the dependence of $\Theta_J$ from $\mu$, the masses $m^2_J$ of the higher spin particles can become lighter with an heavier pion mass if $\partial_r \Theta_J>0$. This is exactly what we observe and the due shift to the pion mass is not enough to compensate this effect.

\begin{figure}[H]
	\centering
	\includegraphics[width=\textwidth]{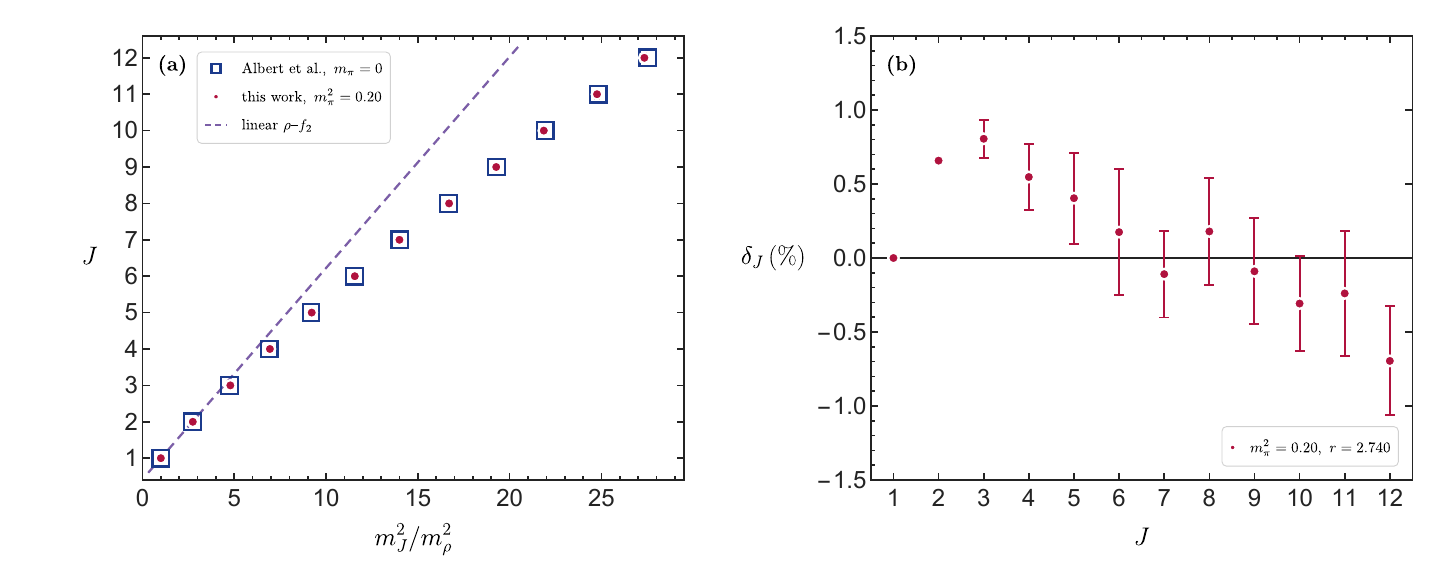}
	\caption{Spectrum at the kink at $m_\pi^2/m_{\rho,0}^2=0.20$ ($r \approx 2.74$) against the massless spectrum of \cite{Albert:2023seb} ($r\approx2.72$), in units of the $\rho$ mass of each theory. (a)~Chew--Frautschi plot: blue squares from \cite{Albert:2023seb}, red dots from the zeros of the extremal functional at the kink; the dashed line is the linear trajectory through the $\rho$ and the $f_2$ at $r\approx2.72$. (b)~Relative difference $\delta_J=m_J^2/m_{J}^{2}\big|_{\text{\cite{Albert:2023seb}}}-1$; the $f_2$ at $+0.66\%$ reflects the difference in $r$.}
	\label{fig:kink_vs_AHRV}
\end{figure}

\paragraph{Comparison at fixed $r$.}
To isolate the effect of $r$, we compare with a theory with the same $r$ and no pion mass: the spectrum of \cite{Albert:2023seb}, which has $r \approx 2.72$. At $m_\pi^2/m_{\rho,0}^2=0.20$ our value is $r \approx 2.74$ ($m_{f_2}/m_\rho=1.655$ against $1.65$). Figure~\ref{fig:kink_vs_AHRV} shows the two spectra, each in units of the $\rho$ mass of its own theory. They agree within $1\%$ for every spin up to $J=12$. Therefore, the lightening of the high-spin states is due to the decrease of $r$ and not to the pion mass.

The differences are compatible with zero within one standard deviation for $6\leq J\leq11$. There is a trend in the deviations: the light states lie above the massless ones, while the heaviest lie below. This effect can be explained both by the discrepancy in $r$ and by the subleading corrections given by the pion mass.

\section{Endpoint of the Skyrme segment and the Massless to Massive Map}
\label{app:skyrme_endpoint}
We prove the upper bound \eqref{eq:massive_skyrme_endpoint_bound}, which, together with the vector completion \eqref{eq:vector_uv_completion}, fixes the endpoint of the Skyrme line.
The argument maps every massive amplitude into an auxiliary amplitude satisfying the corresponding massless assumptions. We assume $\Lambda^2 \geq 4m_\pi^2$, which guarantees that all poles lie above the two-particle threshold.~\footnote{When $\Lambda^2=4m_\pi^2$, the parameter $\lambda$ diverges at a pole sitting exactly at threshold. The physical residue nevertheless has a finite limit. For a finite physical spin-$J$ coupling,
\[
    \lim_{z\downarrow4m_\pi^2}
    (z-4m_\pi^2)^J P_J(\lambda y)
    =
    \frac{(2J)!}{2^J(J!)^2}
    (4m_\pi^2-\sigma)^J y^J.
\]
The monomial $y^J$ has nonnegative Legendre coefficients with the same
spin parity. Positivity of the shifted residue therefore persists at
threshold, and the upper bound continues to hold.}

Recall that $\sigma=4m_\pi^2/3$ and define
\begin{equation}\label{eq:shifted_massless_amplitude}
    A(\hat s,\hat t)
    =M(\sigma+\hat s,\sigma+\hat t)-M_0,
    \qquad
    \hat u=-\hat s-\hat t.
\end{equation}
The shifted variables obey massless kinematics,
$\hat s+\hat t+\hat u=0$, and crossing is unchanged,
$A(\hat s,\hat t)=A(\hat t,\hat s)$. The massless Regge
assumptions for $\hat t\leq0$ correspond to $t\leq\sigma$, and therefore
follow from the massive ones provided the latter hold up to $t=\sigma$.
Moreover,
$A(0,0)=0$, its poles have shifted squared masses
$\zeta=z-\sigma\geq\Lambda^2-\sigma>0$, and all Taylor coefficients of
positive order around $\hat s=\hat t=0$ coincide with those of the original
amplitude around the crossing-symmetric point $s=t=u=\sigma$. It remains to
show that the shifted pole residues admit a positive massless partial-wave
expansion.

For a pole at $z\geq\Lambda^2>4m_\pi^2$, with $\zeta=z-\sigma$, we define
\[
    \lambda(z)=\frac{z-\sigma}{z-4m_\pi^2}\geq1,
    \qquad
    y_s=1+\frac{2\hat s}{\zeta},
    \qquad
    y_t=1+\frac{2\hat t}{\zeta},
    \qquad
    y_u=1+\frac{2\hat u}{\zeta}.
\]
The massive angular variables factorize as
\[
    x_t(z)=1+\frac{2t}{z-4m_\pi^2}=\lambda(z)\,y_t ,
    \qquad
    x_s(z)=\lambda(z)\,y_s ,
    \qquad
    x_u(z)=\lambda(z)\,y_u ,
    \qquad
    x_0(z)=\lambda(z),
\]
where the last relation is the value at $\hat t=0$, i.e.\ $t=\sigma$.
The expansion in the massless Legendre basis takes
the following form:
\begin{equation}\label{eq:positive_legendre_dilation}
    P_J(\lambda y)
    =
    \sum_{\substack{\ell\leq J\\J-\ell\ {\rm even}}}
       c_{J\ell}(\lambda)P_\ell(y),
    \qquad
    c_{J\ell}(\lambda)\geq0,
    \qquad \lambda\geq1 .
\end{equation}
To prove the positivity, write
\begin{equation}
    P_J(\lambda y)=e^{r\,y\partial_y}P_J(y),
    \qquad \lambda \equiv e^r .
\end{equation}
Its action on Legendre polynomials is
\[
    y\partial_yP_J(y)
    =
    JP_J(y)
    +
    \sum_{\substack{\ell<J\\J-\ell\ {\rm even}}}
       (2\ell+1)P_\ell(y).
\]
In each finite-spin sector, this generator has nonnegative matrix
elements in the Legendre basis. Its exponential therefore has
nonnegative matrix elements for $r\geq0$. The transformation also
preserves spin parity: only $\ell$ with $J-\ell$ even appear, so that
$(-1)^J=(-1)^\ell$.

Every kernel becomes a massless
kernel with the pole at $\zeta$:
\begin{equation}\label{eq:shifted_kernels}
\begin{gathered}
    K_1(z;s,\sigma)=\frac{1}{\zeta-\hat s}-\frac{1}{\zeta},
    \qquad
    K_1(z;t,\sigma)=\frac{1}{\zeta-\hat t}-\frac{1}{\zeta},
    \qquad
    K_1(z;s+t-\sigma,\sigma)=\frac{1}{\zeta+\hat u}-\frac{1}{\zeta},
    \\
    K_{\rm fold}(z;s,u,\sigma)
    =\frac{1}{\zeta-\hat s}+\frac{1}{\zeta-\hat t}
     -\frac{1}{\zeta}-\frac{1}{\zeta+\hat u}.
\end{gathered}
\end{equation}
The pion mass has disappeared from these expressions. It remains only in the angular variables,
through $x=\lambda(z)\,y$, and \eqref{eq:positive_legendre_dilation} removes it
pole by pole:
\begin{equation}\label{eq:shifted_legendre}
    P_J(x_t)=\sum_\ell c_{J\ell}\,P_\ell(y_t),
    \qquad
    P_J(x_0)=\sum_\ell c_{J\ell}\,P_\ell(1),
    \qquad
    (-1)^J P_J(x_u)=\sum_\ell c_{J\ell}\,(-1)^\ell P_\ell(y_u),
\end{equation}
with $c_{J\ell}\equiv c_{J\ell}\big(\lambda(z)\big)$. The last relation uses
$(-1)^J=(-1)^\ell$.

Let $h_J(z;s,t)$ be a generic kernel and $h^{(0)}_\ell(\zeta;\hat s,\hat t)$ denote the same function at $m_\pi=0$,
with spin $\ell$ and pole at $\zeta$. Each $h_J$ is a sum of products of one
Legendre polynomial and one kernel, so
\eqref{eq:shifted_kernels} and \eqref{eq:shifted_legendre} give
\begin{equation}\label{eq:kernel_dictionary}
    h_J(z;\sigma+\hat s,\sigma+\hat t)
    =\sum_{\substack{\ell\leq J\\ J-\ell\ {\rm even}}}
     c_{J\ell}\big(\lambda(z)\big)\,
     h^{(0)}_\ell(z-\sigma;\hat s,\hat t).
\end{equation}
Because $\partial_s=\partial_{\hat s}$, $\partial_t=\partial_{\hat t}$ and
$s=t=\sigma$ corresponds to $\hat s=\hat t=0$, the identity holds separately at every order of the
Taylor expansion:
\begin{equation}\label{eq:dictionary_taylor}
    K_{n,k}(z,J)=\sum_\ell c_{J\ell}\,K^{(0)}_{n,k}(\zeta,\ell),
    \quad
    \mathcal X^{(n,k)}_J(z)=\sum_\ell c_{J\ell}\,\mathcal X^{(0)(n,k)}_\ell(\zeta),
    \quad
    \mathcal Y^{(n,k)}_J(z)=\sum_\ell c_{J\ell}\,\mathcal Y^{(0)(n,k)}_\ell(\zeta).
\end{equation}
For example:
\[
    K_{2,1}(z,J)=\frac{P_J'(x_0)}{(z-4m_\pi^2)(z-\sigma)^2}
    =\sum_\ell c_{J\ell}\,\frac{P'_\ell(1)}{\zeta^3}
    =\sum_\ell c_{J\ell}\,K^{(0)}_{2,1}(\zeta,\ell).
\]

Now, defining:
\begin{equation}\label{eq:shifted_spectral_densities}
    (2\ell+1)\rho^{(0)}_\ell(\zeta)
    {\equiv}
    \sum_{\substack{J\geq\ell\\J-\ell\ {\rm even}}}
       (2J+1)\,\rho_J(\zeta+\sigma)\,
       c_{J\ell}\big(\lambda(\zeta+\sigma)\big),
    \quad
    \langle h\rangle_0\equiv\sum_\ell\int_{\Lambda^2-\sigma}^\infty d\zeta\,
       16(2\ell+1)\,\rho^{(0)}_\ell(\zeta)\,h(\zeta,\ell).
\end{equation}
every massive average then becomes a massless one:
\begin{equation}\label{eq:functional_map}
    \big\langle h_J\big\rangle=\big\langle h^{(0)}_\ell\big\rangle_0 .
\end{equation}
This has four consequences.
\begin{enumerate}
\item \emph{Positivity and gap.} Since $\rho_J\geq0$ and $c_{J\ell}\geq0$, we
      have $\rho^{(0)}_\ell\geq0$. Its support is $\zeta\geq\Lambda'^2\equiv\Lambda^2-\sigma>0$.
\item \emph{Dispersion relations.} Applied to the integrand of
      \eqref{eq:DR_fixed_t}, \eqref{eq:functional_map} gives
      \begin{equation}\label{eq:shifted_amplitude_massive_rho}
          A(\hat s,\hat t)
          =\Big\langle
             P_\ell(y_t)\Big(\frac{1}{\zeta-\hat s}-\frac{1}{\zeta}\Big)
            +P_\ell(1)\Big(\frac{1}{\zeta-\hat t}-\frac{1}{\zeta}\Big)
           \Big\rangle_0 .
      \end{equation}
      This is the once-subtracted fixed-$t$ dispersion relation of a massless
      planar amplitude~\cite{Albert:2022oes}. Applied to
      \eqref{eq:DR_fixed_u}, it gives the massless fixed-$u$ relation.
      The signature factor $(-1)^\ell$ is inherited from $(-1)^J$.
\item \emph{Taylor coefficients.} By \eqref{eq:dictionary_taylor},
      $g_{n,k}=\langle K_{n,k}\rangle=\langle K^{(0)}_{n,k}\rangle_0$, so the
      shifted amplitude has \emph{the same} $g_{n,k}$. The simplest case is
      \begin{equation}\label{eq:shifted_normalization_moment}
          g_{1,0}
          =\Big\langle\frac{P_J(x_0)}{(z-\sigma)^2}\Big\rangle
          =\Big\langle\frac{1}{\zeta^2}\sum_\ell c_{J\ell}P_\ell(1)\Big\rangle
          =\Big\langle\frac{P_\ell(1)}{\zeta^2}\Big\rangle_0 .
      \end{equation}
\item \emph{Null constraints.} By \eqref{eq:dictionary_taylor},
      $\langle\mathcal X^{(0)(n,k)}\rangle_0=\langle\mathcal X^{(n,k)}\rangle=0$,
      and likewise for $\mathcal Y$.
\end{enumerate}
The shifted amplitude \eqref{eq:shifted_massless_amplitude} is therefore an
admissible massless amplitude with gap $\Lambda'^2=\Lambda^2-\sigma$ and the same
Taylor coefficients.

\begin{figure}[t!]
    \centering
    \includegraphics[width=0.49\linewidth]{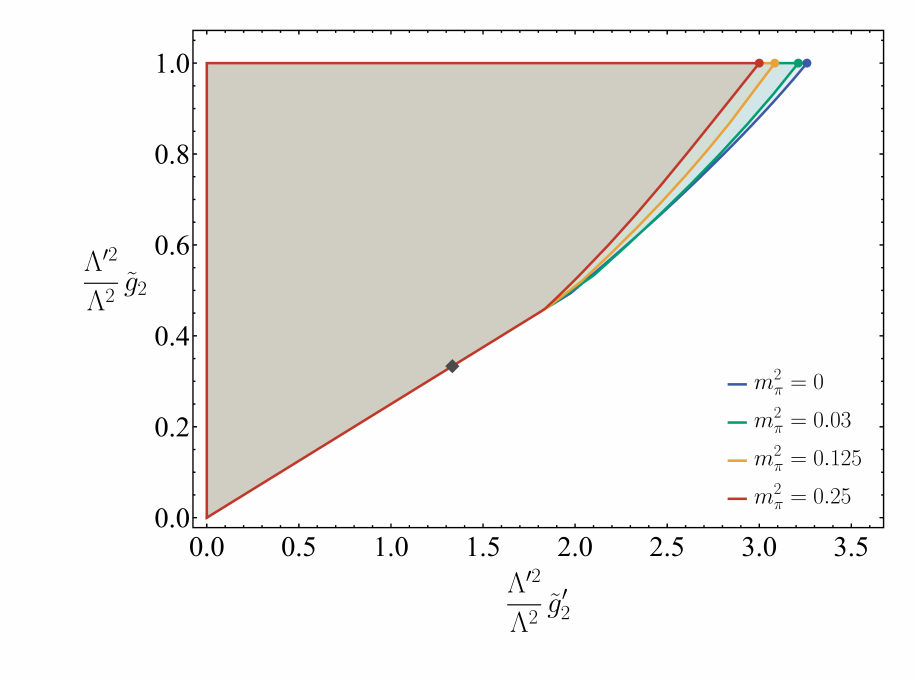}
    \hfill
    \includegraphics[width=0.49\linewidth]{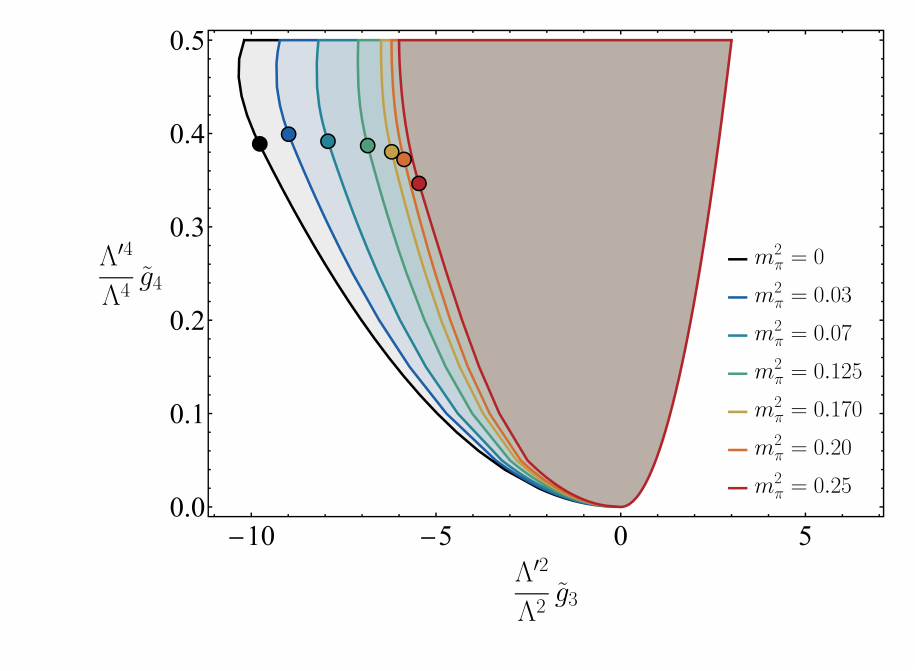}
    \caption{Rescaled version of the plots in figures~\ref{fig:g20g21masses}
    (\emph{Left}) and \ref{fig:stu} (\emph{Right}). After rescaling, the allowed
    region at $m_2$ lies inside the one at $m_1<m_2$: bounds obtained at $m_1$
    are also bounds at $m_2$, at the same $n_{\rm max}$. The right plot applies
    the map \eqref{eq:general_mass_map} to the fully crossing-symmetric problem
    of appendix~\ref{app:stu}, using the pole structure
    \eqref{eq:shifted_amplitude_stu}. There the coefficients are normalized by
    $g_2$.}
    \label{fig:rescaled_plots}
\end{figure}

The numerical bounds use only the null constraints with $n+k\leq n_{\rm max}$.
Let $\vec V_J(z)$ be the vector of kernels that enter at a given $n_{\rm max}$:
$K_{1,0}$, $K_{2,0}$, $K_{2,1}$, and $\mathcal X^{(n,k)}_J$, $\mathcal Y^{(n,k)}_J$
with $n+k\leq n_{\rm max}$. Let $\vec V^{(0)}_\ell(\zeta)$ be its massless counterpart,
which contains the same list of constraints. By \eqref{eq:dictionary_taylor}, component by
component,
\begin{equation}\label{eq:vector_dictionary}
    \vec V_J(z)=\sum_\ell c_{J\ell}\big(\lambda(z)\big)\,
    \vec V^{(0)}_\ell(z-\sigma).
\end{equation}
Take any dual functional $\vec\alpha$ that is valid for the massless problem with
gap $\Lambda'^2$ at this $n_{\rm max}$, i.e.\ any $\vec\alpha$ with
$\vec\alpha\cdot\vec V^{(0)}_\ell(\zeta)\geq0$ for all $\ell$ and all $\zeta\geq\Lambda'^2$. For every $J$
and every $z\geq\Lambda^2$,
\[
    \vec\alpha\cdot\vec V_J(z)
    =\sum_\ell c_{J\ell}\,\vec\alpha\cdot\vec V^{(0)}_\ell(z-\sigma)\geq0,
\]
since $c_{J\ell}\geq0$ and $z-\sigma\geq\Lambda'^2$. The same $\vec\alpha$ is therefore a
valid functional for the massive problem at the \emph{same} $n_{\rm max}$, and
it gives the same bound on the $g_{n,k}$. 

Therefore, at every finite $n_{\rm max}$, the
massive allowed region lies inside the massless one with gap $\Lambda'^2$.

The massive and the massless problems normalize the coefficients with different gaps:
\[
    \tilde g_2=\Lambda^2\,\frac{g_{2,0}}{g_{1,0}},\quad
    \tilde g_2'=2\Lambda^2\,\frac{g_{2,1}}{g_{1,0}},
    \qquad\qquad
    \tilde g_2^{(0)}=\Lambda'^2\,\frac{g_{2,0}}{g_{1,0}},\quad
    \tilde g_2^{\prime(0)}=2\Lambda'^2\,\frac{g_{2,1}}{g_{1,0}}.
\]
Since the $g_{n,k}$ are the same,
\begin{equation}\label{eq:massive_massless_rescaling}
    \big(\tilde g_2^{\prime(0)},\,\tilde g_2^{(0)}\big)
    =\frac{\Lambda^2-\sigma}{\Lambda^2}\,
    \big(\tilde g_2',\,\tilde g_2\big).
\end{equation}
This is a dilation of the plane, so the Skyrme line $\tilde g_2'=4\tilde g_2$
is mapped into itself.

The vector completion \eqref{eq:vector_uv_completion}
was admissible for every $M_\infty$, so it lies inside the allowed region at every
$n_{\rm max}$, and it sits on the Skyrme line. Hence, for every $n_{\rm max}$,
\begin{equation}\label{eq:endpoint_sandwich}
    \frac{\Lambda^2}{3(\Lambda^2-\sigma)}
    \;\leq\;
    \tilde g_2^{\max}(n_{\rm max})
    \;\leq\;
    \frac{\Lambda^2}{\Lambda^2-\sigma}\,
    \tilde g_2^{(0)\max}(n_{\rm max}).
\end{equation}
In the massless problem $\tilde g_2^{(0)\max}\to1/3$ as $n_{\rm max}\to\infty$
\cite{Fernandez:2022kzi}. The two sides of \eqref{eq:endpoint_sandwich} then
coincide, and
\[
    \lim_{n_{\rm max}\to\infty}\tilde g_2^{\max}
    =\frac{\Lambda^2}{3\Lambda^2-4m_\pi^2},
\]
which is \eqref{eq:massive_skyrme_endpoint_bound} in units $\Lambda=1$.

The converse map is not positive, indeed, the massive boundary generally lies strictly inside the rescaled massless one.

Nothing in the first part of the proof requires the target mass to vanish. Shift instead by
$\sigma_2-\sigma_1$. This maps an admissible amplitude with mass $m_2$ and gap $\Lambda$,
$\Lambda^2\geq4m_2^2$, into one with mass $m_1\leq m_2$ and gap
\begin{equation}
    \Lambda'^2=\Lambda^2-\sigma_2+\sigma_1 ,
    \qquad
    \Lambda'^2-4m_1^2=\Lambda^2-4m_2^2+2(\sigma_2-\sigma_1)>0 .
\end{equation}
The denominators again shift rigidly and the angular variables are related by the
dilation factor
\[
    \lambda'(z)=\frac{z-\sigma_2+\sigma_1-4m_1^2}{z-4m_2^2}
    =1+\frac{2(\sigma_2-\sigma_1)}{z-4m_2^2}\geq1 .
\]
The Taylor coefficients $g_{n,k}$ are unchanged, so the normalized coefficients
$\tilde g_{n,k}=\Lambda^{2(n-1)}g_{n,k}/g_{1,0}$ obey
\begin{equation}\label{eq:general_mass_map}
    \tilde g_{n,k}\big|_{m_1/\Lambda'}
    =\left(\frac{\Lambda'^2}{\Lambda^2}\right)^{n-1}
    \tilde g_{n,k}\big|_{m_2/\Lambda}.
\end{equation}
If the coefficients are normalized by $g_{n_0,k_0}$ instead of $g_{1,0}$, the
exponent becomes $n-n_0$.

The same map applies to the amplitudes of appendix~\ref{app:stu},
including the statement at fixed $n_{\rm max}$.


 \bibliographystyle{JHEP}
 \bibliography{biblio.bib}

\end{document}